\documentclass[a4paper,twoside]{memoir}

\usepackage[T1]{fontenc}
\usepackage{mathpazo}
\usepackage[semibold,scale=0.85]{sourcecodepro}
\usepackage{microtype}

\usepackage{amssymb}
\usepackage{amsmath}
\usepackage{amsthm}
\usepackage{amsfonts}
\usepackage{mathtools}

\usepackage[hidelinks,draft=false]{hyperref}
\usepackage{doi}
\usepackage[capitalise]{cleveref}
\usepackage{enumitem}
\usepackage[semicolon,square,sort]{natbib}
\usepackage[final]{listings}
\usepackage{siunitx}

\usepackage{jhdefs}

\setsecnumdepth{subsection}
\maxtocdepth{subsection}

\chapterstyle{dash}
\renewcommand{\chapnumfont}{\huge\bfseries\scshape}

\setsecheadstyle{\LARGE\bfseries\scshape}
\setsubsecheadstyle{\Large\bfseries\scshape}

\copypagestyle{chapter}{plain}
\makeevenfoot{chapter}{}{\textbf{\thepage}}{}
\makeoddfoot{chapter}{}{\textbf{\thepage}}{}

\createmark{chapter}{left}{shownumber}{Chapter~}{\hspace{4mm}}
\createmark{section}{right}{shownumber}{}{\hspace{4mm}}
\makepagestyle{mypage}
\makeheadrule{mypage}{\textwidth}{\normalrulethickness}
\makeevenhead{mypage}{\llap{\textbf{\thepage}\hspace{3\marginparsep}}\scshape\leftmark}{}{}
\makeoddhead{mypage}{}{}{\scshape\rightmark\rlap{\hspace{3\marginparsep}\textbf{\thepage}}}

\makeatletter
\g@addto@macro{\appendix}{%

\createmark{chapter}{left}{shownumber}{Appendix~}{\hspace{4mm}}
}
\makeatother

\makeatletter
\g@addto@macro{\backmatter}{%
\createmark{chapter}{left}{shownumber}{}{\hspace{4mm}}
\makeoddhead{mypage}{}{}{\rlap{\hspace{3\marginparsep}\textbf{\thepage}}}
}
\makeatother

\newcommand{\thetitlepage}{{%
  \clearpage
  \thispagestyle{empty}
  \centering
  \LARGE\textsc{Master's Thesis}
  \vspace{0.4 cm}
  \hrule height 1.5 pt
  \vspace{0.4 cm}
  \Huge\textbf{\textsc{\thetitle}}
  \vspace{0.4 cm}
  \hrule height 1.5 pt
  \vspace{0.4 cm}
  \LARGE\textsc{\theauthor}

  \vfill
  \includegraphics[width=0.3\textwidth]{\thelogo}
  \vfill

  \Large{\textbf{\textsc{Supervisors}} \\ \smallskip
  \textsc{\thesupervisors}

  \vspace{0.8 cm}
  \textbf{\textsc{\theuniversity}} \\ \smallskip
  \textsc{\thegroup}

  \vspace{0.8 cm}
  \textsc{\thedate}}
  \cleardoublepage
}}

\newcommand{\university}[1]{\def\theuniversity{#1}}
\newcommand{\group}[1]{\def\thegroup{#1}}
\newcommand{\supervisors}[2]{\def\thesupervisors{#1 \\ #2}}
\newcommand{\logo}[1]{\def\thelogo{#1}}
\newcommand{\place}[1]{\def\theplace{#1}}

\renewcommand{\cite}[1]{\citep{#1}}

\newtheorem{theorem}{Theorem}[chapter]
\newtheorem{lemma}[theorem]{Lemma}
\newtheorem{corollary}[theorem]{Corollary}
\theoremstyle{definition}
\newtheorem{definition}[theorem]{Definition}
\newtheorem{remark}[theorem]{Remark}
\newtheorem{example}[theorem]{Example}

\newtheorem*{continuedex}{Example~\continuedexref~\textit{(continued)}}
\newenvironment{continuedexample}[1]{\newcommand\continuedexref{\ref{#1}}\continuedex}{\endcontinuedex}
\newtheorem*{recalllem}{Lemma~\recalllemref~\textit{(recalled)}}
\newenvironment{recalllemma}[1]{\newcommand\recalllemref{\ref{#1}}\recalllem}{\endrecalllem}
\newtheorem*{recalldef}{Definition~\recalldefref~\textit{(recalled)}}
\newenvironment{recalldefinition}[1]{\newcommand\recalldefref{\ref{#1}}\recalldef}{\endrecalldef}

\newcommand{\twosub}[2]{\substack{#1, \\ \mathclap{#2}}}
\newcommand{\nospace}[1]{\mathclap{#1}}

\newlist{assumptions}{enumerate}{1}
\setlist[assumptions,1]{label=\alph*., ref=\alph*}
\crefname{assumptionsi}{}{}
\Crefname{assumptionsi}{Assumption}{Assumptions}
\creflabelformat{assumptionsi}{(#2#1#3)}

\crefname{lemma}{Lemma}{Lemmata}

\newcommand{\tiff}{\quad\text{iff}\quad}
\newcommand{\tand}{\quad\text{and}\quad}

\newcommand{\cons}[1]{Cons(#1)}
\newcommand{\comprule}[1]{\overleftarrow{Comp}(#1)}
\newcommand{\compdef}[1]{\overrightarrow{Comp}(#1)}
\newcommand{\comp}[1]{Comp(#1)}
\newcommand{\oc}[1]{OC(#1)}
\newcommand{\trans}[1]{Trans(#1)}
\newcommand{\irref}[1]{Irref(#1)}
\newcommand{\ocax}[1]{\irref{#1} \land \trans{#1}}
\newcommand{\ocomp}[1]{\overrightarrow{OC}(#1)}
\newcommand{\lvl}[1]{lvl_{#1}}
\newcommand{\ncomp}[1]{\overrightarrow{OC}^{*}(#1)}
\newcommand{\nc}[1]{OC^{*}(#1)}
\newcommand{\nat}[1]{Nat(#1)}
\newcommand{\ocomps}[1]{\overrightarrow{OC}^{s}(#1)}
\newcommand{\ocs}[1]{OC^{s}(#1)}
\newcommand{\ord}[2]{ord(#1,#2)}

\newcommand{\terms}{\tuple{t}}
\newcommand{\vars}{\tuple{V}}
\newcommand{\lang}[1]{\mathcal{L}(#1)}
\newcommand{\at}[1]{\mathcal{A}(#1)}
\newcommand{\inst}[1]{Inst(#1)}
\newcommand{\rs}{\tuple{r}}

\newcommand{\headlit}[1]{Head_{L}(#1)}
\newcommand{\bodylitp}[1]{Body^+_{L}(#1)}

\newcommand{\uc}{\widetilde{\forall}}

\newcommand{\values}[1]{[#1]}
\newcommand{\taub}{\tau^B}
\newcommand{\otaub}{\tau^B_<}
\newcommand{\form}[2]{F_{#1}(#2)}
\newcommand{\oform}[2]{F^{<}_{#1}(#2)}
\newcommand{\pform}[1]{F_{#1}}

\newcommand{\sigsort}{\sigma_{0}}
\newcommand{\sigord}{\sigma_{0}^{<}}
\newcommand{\siglvl}{\sigma_{0}^{lvl}}
\newcommand{\ext}[2]{#1 \mathinner{\uparrow} #2}
\newcommand{\red}[2]{#1 \mathinner{\downarrow} #2}

\newcommand{\step}[1]{T_{#1}}
\newcommand{\stepi}[2]{T_{#1}^{#2}}
\newcommand{\stepf}[1]{T_{#1}^{\infty}}

\newcommand{\round}[1]{round(#1)}
\newcommand{\abs}[1]{|#1|}

\newcommand{\glr}[2]{#1^#2}
\newcommand{\glrp}[2]{(#1)^#2}

\lstdefinelanguage{mg}{
  morekeywords={not},
  sensitive,
}
\lstdefinelanguage{formula}{
  morekeywords={and, or, not, exists, forall, true, false},
  sensitive,
}

\newcommand{\code}[1]{\lstinline{#1}}

\title{Ordered Completion for Non-Locally Tight \mg Programs}
\author{Jan Heuer}
\date{April 2025}
\place{Berlin}
\university{University of Potsdam}
\group{Knowledge Processing and \\ Information Systems}
\supervisors{Tobias Stolzmann, M.\ Sc.}{Prof.\ Dr.\ Torsten Schaub}
\logo{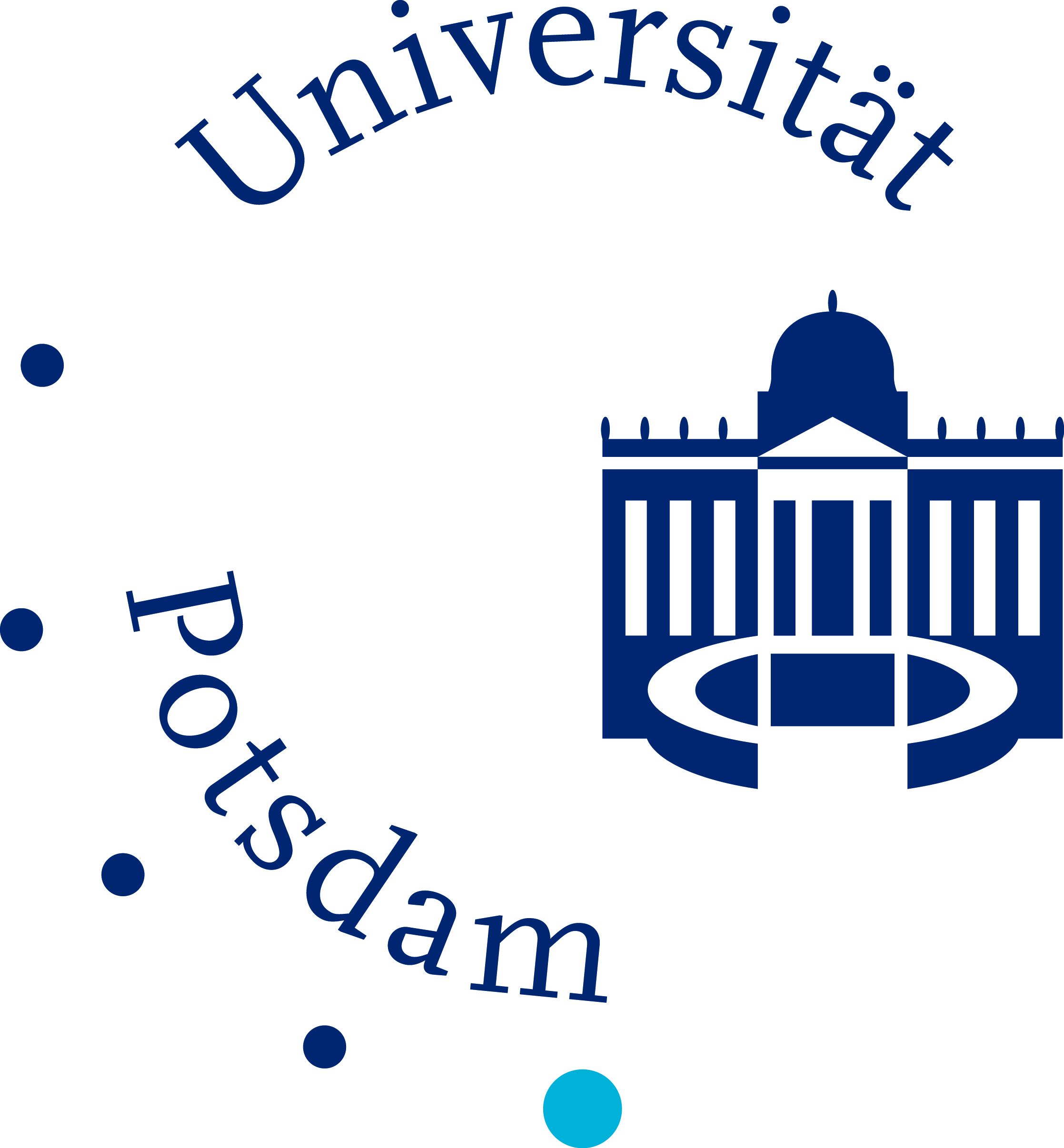}

\hypersetup{
  pdftitle = {Jan Heuer},
  pdfauthor = {Ordered Completion for Non-Locally Tight mini-gringo Programs},
}

\begin{document}
\frontmatter
\thetitlepage
\pdfbookmark{\contentsname}{toc}
\tableofcontents*
\chapter{Acknowledgements}

First and foremost, I would like to thank my thesis supervisors, Tobias Stolzmann and Torsten Schaub, for their guidance, support, and insightful feedback throughout the course of this work.
I am also grateful to Jorge Fandinno, Zach Hansen, and Christoph Wernhard for the valuable discussion and feedback they provided.
Special thanks go to Javier Romero for numerous engaging conversations and suggestions that have greatly contributed to the improvement of this thesis.

Last but not least, I am deeply grateful for the constant support of my family and friends during my studies and the writing of this thesis.
Most importantly, I want to thank my fianc{\'e}, Akmal, for his encouragement and belief in me.

\chapter{Abstract}

Completion is a well-known transformation to capture the stable model semantics of logic programs.
It turns a logic program into a set of definitions, where the disjunctions of its rules defines each predicate.
The completion of a logic program is then just a first-order theory.
Stable models of a program are also models of the completion, but the converse does not hold in general.
On the class of tight logic programs, i.e., programs that do not use positive recursion, the two semantics coincide.
More recent results extended this exact correspondence to the class of locally-tight programs (i.e., programs that do not use non-terminating recursion).
However, local-tightness is a property that can not be tested with simple syntactic methods, as is the case for tightness.
Completion plays an important role in the verification of answer set programs.
It is used in the verification of so-called external equivalence, a form of equivalence under certain inputs where only subsets of the stable models of the programs are relevant (the output predicates).
Ordinary answer set equivalence as well as checking that a program adheres to a first-order specification are special cases of external equivalence.
With the currently existing methods for the verification tool \anthem, the checking of external equivalence is limited in two ways:
(1)~there is currently no way to check local tightness automatically, and (2)~it is not possible to verify programs that are not locally tight.
Thus, it is of interest to consider alternatives to completion that can capture the stable model semantics of arbitrary (i.e., not necessarily locally tight) programs.
This thesis investigates an extension of completion called ordered completion that accomplishes this goal.
Ordered completion was introduced by~\cite{AsuncionEtAl2012} to provide a transformation capturing the stable model semantics of arbitrary logic programs.
The transformation is limited in that it can only accurately capture finite stable models.
This thesis extends their work to the language of \mg (a subset of the language used by the \clingo solver).
To do so, the concept of well-supported models is generalised to programs in the language of \mg.
In the same manner, definitions and theorems of the ordered completion are extended for the \mg language.
Furthermore, a modification of ordered completion is introduced, which lifts the limitation that ordered completion only captures finite stable models.
This modification would usually require a stronger logic (first-order logic with arithmetic), but in the \mg context this is already necessary.
Ordered completion for \mg is implemented as a translation in the verification tool \anthem.
Some first practical experiments for verifying the ordered completion of simple logic programs are presented.

\mainmatter
\chapter{Introduction}
\label{chp:introduction}

Answer Set Programming (ASP) is a declarative programming approach that has been successfully applied to complex search and optimisation problems.
In this thesis, we assume some familiarity with ASP.
For introductions on ASP see~\cite{Baral2010,GebserEtAl2012,Lifschitz2019}.

In this thesis, we will focus on the formal definition of stable models or answer sets.
Originally, stable models were defined for propositional logic programs using minimal models and the reduct operation~\cite{GelfondLifschitz1988}.
The stable model semantics has since been generalised to more expressive languages.
Furthermore, many alternative definitions have been proposed in the literature.
For an overview, see for example~\cite{Lifschitz2010}.

A notable alternative definition is the so-called completion.
Completion was first introduced in~\cite{Clark1978} for databases and later defined for different classes of logic programs~\cite{Fages1994,ErdemLifschitz2003,HarrisonEtAl2017a}.
The idea of completion is to transform a logic program into a theory in classical logic.
This is done by taking all rules talking about some predicate as the definition of this predicate.
For example, let us consider the logic program
\begin{align}
  \label{eq:tight}
  \begin{split}
     & p(X) \revimp q(X),       \\
     & p(X) \revimp \pnot r(X), \\
     & r(1),                    \\
     & q(1).
  \end{split}
\end{align}
Its completion is the theory
\begin{align*}
  & \forall X (p(X) \equi q(X) \lor \lnot r(X)), \\
  & \forall X (r(X) \equi X = 1),                \\
  & \forall X (q(X) \equi X = 1).
\end{align*}
The $\revimp$ direction of each equivalence ensures that the rules for the predicate are satisfied.
The $\imp$ direction makes sure that a predicate is only true if it is supported by one of its rules.
I.e., only if one of its rules provides a justification for the truth.
This means we take the rules of the program to exactly define each predicate, and we assume that there are no other reasons why a predicate can be true.

In this example, the stable models of the program are exactly the models of its completion.
But this exact correspondence only holds for so-called tight logic programs~\cite{Fages1994}.
Tight logic programs do not contain any recursion over positive predicates.
For instance, the program
\begin{align}
  \label{eq:non-tight}
  \begin{split}
     & p(X) \revimp q(X), \\
     & q(X) \revimp p(X),
  \end{split}
\end{align}
is not tight.
Indeed, its completion
\begin{align*}
  &\forall X (p(X) \equi q(X)),
\end{align*}
has two models: the empty model and the set consisting of $p(X)$ and $q(X)$ for all values of $X$.
Only the former is a stable model of the program.

Completion is of particular interest as an alternative semantics for ASP as: (1) it connects ASP to classical logic, (2) it provides the basis of modern ASP solvers, and (3) its usage in the verification of ASP.
(1) The connection to classical logic provides us with some interesting results, for example, on the expressivity of ASP: ASP is more expressive than standard first-order logic.
For example, it is easy to write a logic program computing the transitive closure of a relation, while transitive closure can not be expressed in first-order logic~\cite{EbbinghausFlum1995}.
(2) For the design of an ASP solver such as \clingo~\cite{GebserEtAl2019}, completion is supplemented with loop formulas~\cite{LinZhao2004} in order to capture the stable models of non-tight programs.
(3) In verification of ASP (using the verification tool \anthem~\cite{LifschitzEtAl2018,LifschitzEtAl2019,LifschitzEtAl2020,FandinnoEtAl2020,FandinnoEtAl2023}) completion is used as the basis for verifying external equivalence of programs~\cite{FandinnoEtAl2023}.
This is a form of equivalence in which the inputs and outputs of the logic program are taken into account.
The verification of answer set equivalence as well as verifying that a logic program adheres to some first-order specification are special cases of external equivalence.

As completion is used to establish this equivalence, the approach is limited to tight logic programs.
A more recent result~\cite{FandinnoEtAl2024a} extends the equivalence of completion and stable models to the class of locally tight logic programs.
The idea of local tightness is that positive recursion is allowed, but it must be terminating.
I.e., a program like the non-tight example from above (\cref{eq:non-tight}) is still not allowed, but the program
\begin{align}
  \label{eq:locally-tight}
  \begin{split}
     & p(X+1) \revimp p(X) \land X > 0, \\
     & p(1),
  \end{split}
\end{align}
is locally tight as in the positive recursion the argument decreases (the head contains $p(X+1)$ while the body contains $p(X)$) and this decrease terminates ($X > 0$).
However, while tightness is a simple syntactic property that can be efficiently checked, local tightness is more complex, and so far, no algorithms to check it have been developed.
Besides, many interesting logic programs are not locally tight, such as the following program
\begin{align}
  \label{eq:transitive}
  \begin{split}
     & t(X,Y) \revimp e(X,Y),              \\
     & t(X,Y) \revimp e(X,Z) \land t(Z,Y),
  \end{split}
\end{align}
that computes the transitive closure of a graph.

It is thus of interest to investigate extensions of completion for usage in verification.
As mentioned above, in solving completion is supplemented by loop formulas.
In the solving case, only loop formulas of propositional programs are used (due to the grounding and solving approach), but in the verification scenario we work on first-order programs (here, grounding is not possible as we want to verify properties for all inputs).
While loop formulas were generalised to the first-order case~\cite{ChenEtAl2006}, they are not suitable for use in automated verification: a logic program can have infinitely many loop formulas.

Therefore, we are interested in alternative extensions or modifications of completion.
One such approach is the concept of ordered completion~\cite{AsuncionEtAl2012}.
In ordered completion the formulas of the completion are modified by adding formulas that talk about the derivation order of atoms.
This makes it possible to rule out models in which cyclic derivations occur, such as in the completion of program \cref{eq:non-tight}.
But ordered completion does not fully capture the stable model semantics: it only accurately captures finite stable models.
(Note that this is a general limitation of standard first-order logic, as ASP is more expressive as mentioned above.)

However, in order to use ordered completion in the verification of ASP as an alternative to completion, we have to generalise the work of~\cite{AsuncionEtAl2012} to the \mg language.
\mg is a subset of the input language of the ASP solver \clingo.
The language considered in~\cite{AsuncionEtAl2012} is simpler as it, for example, does not contain integer arithmetic.
This is the goal of this thesis: we want to give a definition of ordered completion for the language of \mg and prove its equivalence to the stable model semantics.

The structure of this thesis is as follows.
\cref{chp:mini-gringo} reviews the \mg language: we start by reviewing the syntax and semantics of the language before reviewing how to express \mg programs as first-order formulas.
In \cref{chp:support} we define the support semantics for \mg.
The support semantics is another alternative semantics of ASP that we will use in the equivalence proof of ordered completion.
In this chapter, we take definitions from literature, generalise them to the \mg language, and prove the correctness of this alternative characterisation of stable models.
\cref{chp:completion} starts by reviewing completion of \mg programs before defining ordered completion for \mg programs and formalising the relation to the stable model semantics.
We furthermore give an alternative definition of ordered completion that makes use of a first-order logic with arithmetic to remove the restriction that ordered completion can only capture finite stable models.
We finish the chapter by considering how we can simplify the definition of ordered completion to obtain simpler but equivalent formulas.
In \cref{chp:verification} we provide some details of an implementation of the ordered completion transformation that we then use in some practical verification experiments.
We also explore the limitations of using ordered completion in verification.
Finally, \cref{chp:conclusion} concludes this thesis by summarising its contributions, reviewing related work, and providing ideas for future work.

\chapter{Review: The \texorpdfstring{\mg}{mini-gringo} Language}
\label{chp:mini-gringo}

This chapter reviews the language of \mg~\cite{LifschitzEtAl2018,LifschitzEtAl2019,FandinnoEtAl2020,FandinnoEtAl2023,FandinnoEtAl2024a}\footnote{Note that in some of the cited publications the language was not yet called \mg.}, a subset of the \ag~\cite{GebserEtAl2015} language.
In particular, this chapter reviews the syntax of \mg and its semantics, in \cref{sec:mg-syntax,sec:mg-semantics} respectively.
Furthermore, we review the $\tau^{*}$ transformation in \cref{sec:tau-star}.
Using this transformation, we can express logic programs as theories in a two-sorted logic.
We also give a quick review of this logical language in \cref{sec:sorted}.

While readers familiar with the literature on \mg may want to skip this chapter, we want to note that we modified and added some definitions that differ compared to other publications.
Namely, see \cref{rem:rule-parts,rem:standard,rem:form}.

\section{The Syntax of \texorpdfstring{\mg}{mini-gringo} Programs}
\label{sec:mg-syntax}

We start by defining so-called program terms.
As the building blocks for program terms, we assume three countably infinite sets of symbols: numerals, symbolic constants, and program variables.
We further assume that a 1-to-1 correspondence between numerals and integers is chosen.
For a given integer~$n$, we denote its corresponding numeral by $\num{n}$.

\begin{definition}[Program Terms]
  \emph{Program terms} are recursively defined as follows:
  \begin{itemize}
    \item Numerals, symbolic constants, program variables, and the symbols $\myinf$ and $\mysup$ are program terms,
    \item if $t$ is a program term then $\abs{t}$ is a program term,
    \item if $t_{1}$ and $t_{2}$ are program terms and $\circ$ is one of the \emph{operation names}
          \begin{equation*}
            + \quad - \quad \times \quad / \quad \backslash \quad \twodots
          \end{equation*}
          then $(t_{1} \circ t_{2})$ is a program term.
  \end{itemize}
\end{definition}

We write $-t$ as a shorthand for $\num{0}-t$.
Program terms (or other expressions) that do not contain variables are called \emph{ground}.
If a ground expression does not contain operation names, it is called \emph{precomputed}.
We assume a total order on the precomputed terms such that $\myinf$ is its least element, $\mysup$ is its greatest element, and for any integers $m$ and $n$, $\num{m} < \num{n}$ if and only if $m < n$.

Using program terms, we can then form literals and comparisons.

\begin{definition}[Literals]
  An \emph{atom} is an expression of the form $p(\terms)$, where $p$ is a symbolic constant and $\terms$ is a tuple of program terms.
  (If $\terms$ is empty, we can omit the parentheses.)
  A \emph{literal} is an atom possibly preceded by one or two occurrences of \emph{negation} $\pnot$.
\end{definition}

\begin{definition}[Comparisons]
  A \emph{comparison} is an expression of the form ${(t_{1} \prec t_{2})}$, where $t_{1}$ and $t_{2}$ are program terms and $\prec$ is on of the \emph{relation names}
  \begin{equation*}
    = \quad \neq \quad < \quad > \quad \leq \quad \geq.
  \end{equation*}
\end{definition}

Finally, we can combine literals and comparisons to form rules and programs.

\begin{definition}[Rules and Programs]
  A \emph{rule} is an expression of the form
  \begin{equation}
    \label{eq:gringo-rule}
    Head \revimp Body,
  \end{equation}
  where
  \begin{itemize}
    \item $Body$ is a (possibly empty\footnote{When $Body$ is empty $\revimp$ can be dropped from \cref{eq:gringo-rule}.}) conjunction of literals and comparisons,
    \item $Head$ is either an atom $p(\terms)$ (then \cref{eq:gringo-rule} is a \emph{basic rule}), or an atom in braces $\{p(\terms)\}$ (then \cref{eq:gringo-rule} is a \emph{choice rule}), or empty (then \cref{eq:gringo-rule} is a \emph{constraint}).
  \end{itemize}

  A \emph{program} is a finite set of rules.
\end{definition}

\begin{remark}\label{rem:rule-parts}
  We introduce some notations for accessing parts of rules.
  For a rule $R$ of form \cref{eq:gringo-rule}, $\body{R}$ is the formula $Body$, $\bodylitp{R}$ stands for the set of positive literals in $Body$.
  Note here the distinction that $\body{R}$ denotes a formula, whereas $\bodylitp{R}$ denotes a set of literals.

  $\head{R}$ is the element $Head$, i.e., $p(\terms)$, $\{p(\terms)\}$, or $\false$ to represent the empty head.
  $\headlit{R}$ is the literal in the head, i.e., $p(\terms)$ if $\head{R} = p(\terms)$ or $\head{R} = \{p(\terms)\}$ and $\false$ otherwise.
\end{remark}

Finally, for a program $\Pi$, $\lang{\Pi}$ denotes the set of predicates occurring in $\Pi$.
For simplicity, we assume that a predicate $p$ only occurs in a single arity.\footnote{To remove this restriction, $\lang{\Pi}$ would have to be modified to be the set of predicates together with their arity, and subsequent definitions would have to be modified accordingly.}

\section{The Semantics of \texorpdfstring{\mg}{mini-gringo} Programs}
\label{sec:mg-semantics}

The semantics of \mg programs is based on the semantics of \ag~\cite{GebserEtAl2015} as \mg is a subset of the language of \ag.
The definition of the semantics in~\cite{GebserEtAl2015} uses a transformation $\tau$ to transform \ag rules into infinitary propositional formulas.
However, $\tau$ is only defined on ground rules.
I.e., we first have to replace rules by all the possible ways in which we can instantiate their variables.
The stable models of \ag programs are then defined in terms of the stable models of infinitary propositional formulas according to~\cite{Truszczynski2012}.

Below we will reproduce the transformation $\tau$ for \mg.
In our case, the resulting formulas will always be finitary propositional formulas.
Thus, the definition of stable models according to~\cite{Ferraris2005} is sufficient for our purpose.
This follows the treatment in other publications concerning \mg, such as~\cite{FandinnoEtAl2024a}.
For the definition of rounding operations in division and modulo operations, we follow the most recent publication~\cite{FandinnoEtAl2024a} that corrected a minor error from the earlier publications regarding the division and modulo operations.

First, we define a transformation that, given a \mg term, produces a set of possible values of this term.
This is necessary as \mg terms can have zero (e.g., arithmetic operations involving symbolic constants), one, or multiple values (interval terms), whereas a term in (propositional) logic has exactly one value.

\begin{definition}[Values of a Term]
  For every ground term $t$, we define the set~$\values{t}$ of its values as follows:
  \begin{itemize}
    \item if $t$ is a numeral, a symbolic constant, $\myinf$ or $\mysup$, then $\values{t} = \{t\}$,
    \item if $t$ is $|t_{1}|$, then $\values{t}$ is the set of numerals $\num{|n|}$ for all integers $n$ such that $\num{n} \in \values{t_{1}}$,
    \item if $t$ is $(t_{1} \circ t_{2})$ with $\circ \in \{+,-,\times\}$, then $\values{t}$ is the set of numerals $\num{n_{1} \circ n_{2}}$ for all integers $n_{1}, n_{2}$ such that $\num{n_{1}} \in \values{t_{1}}$ and $\num{n_{2}} \in \values{t_{2}}$,
    \item if $t$ is $(t_{1} / t_{2})$, then $\values{t}$ is the set of numerals $\num{\round{n_{1} / n_{2}}}$ for all integers $n_{1}, n_{2}$ such that $\num{n_{1}} \in \values{t_{1}}$ and $\num{n_{2}} \in \values{t_{2}}$ and $n_{2} \neq 0$,
    \item if $t$ is $(t_{1} \backslash t_{2})$, then $\values{t}$ is the set of numerals $\num{n_{1} - n_{2} \times \round{n_{1} / n_{2}}}$ for all integers $n_{1}, n_{2}$ such that $\num{n_{1}} \in \values{t_{1}}$ and $\num{n_{2}} \in \values{t_{2}}$ and $n_{2} \neq 0$,
    \item if $t$ is $(t_{1} \twodots t_{2})$, then $\values{t}$ is the set of numerals $\num{m}$ for all integers $m$ such that, for some integers $n_{1}$ and $n_{2}$, $n_{1} \in \values{t_{1}}$, $n_{2} \in \values{t_{2}}$, $n_{1} \leq m \leq n_{2}$,
  \end{itemize}
  where the rounding function is defined as follows:
  \[
    \round{n} =
    \begin{cases}
      \lfloor n \rfloor & \text{if}\ n \geq 0, \\
      \lceil n \rceil   & \text{if}\ n < 0.
    \end{cases}
  \]

  For ground terms $t_{1}, \dots, t_{n}$, $\values{t_{1}, \dots, t_{n}}$ denotes the set of tuples $r_{1}, \dots, r_{n}$ for all $r_{1} \in \values{t_{1}}, \dots, r_{n} \in \values{t_{n}}$.
\end{definition}

Next, we define $\tau$ for the bodies of rules, i.e., for literals and comparisons.

\begin{definition}[Transformation of Literals]
  For ground literals $Lit$, we define $\tau(Lit)$ as
  \begin{itemize}
    \item $\bigvee_{\rs \in \values{\terms}} p(\rs)$, if $Lit$ is $p(\terms)$,
    \item $\bigvee_{\rs \in \values{\terms}} \lnot p(\rs)$, if $Lit$ is $\pnot p(\terms)$,
    \item $\bigvee_{\rs \in \values{\terms}} \lnot \lnot p(\rs)$, if $Lit$ is $\pnot \pnot p(\terms)$.
  \end{itemize}

  For any ground comparison $t_{1} \prec t_{2}$, we define $\tau(t_{1} \prec t_{2})$ as
  \begin{itemize}
    \item $\true$, if the relation $\prec$ holds for some $r_{1} \in \values{t_{1}}$ and some $r_{2} \in \values{t_{2}}$,
    \item $\false$ otherwise.
  \end{itemize}

  If each $C_{1}, \dots, C_{n}$ is a ground literal or a ground comparison, then the formula $\tau(C_{1} \land \dots \land C_{n})$ is the conjunction $\tau(C_{1}) \land \dots \land \tau(C_{n})$.
\end{definition}

We then extend $\tau$ to \mg rules as follows.

\begin{definition}[Transformation of Rules]
  For ground rules $R$, we define $\tau(R)$ as
  \begin{itemize}
    \item $\tau(Body) \imp \bigwedge_{\rs \in \values{\terms}} p(\rs)$, if $R$ is a basic rule $p(\terms) \revimp Body$,
    \item $\tau(Body) \imp \bigwedge_{\rs \in \values{\terms}} \big(p(\rs) \lor \lnot p(\rs)\big)$, if $R$ is a choice rule $\{ p(\terms) \} \revimp Body$,
    \item $\lnot \tau(Body)$, if $R$ is a constraint $\revimp Body$.
  \end{itemize}
\end{definition}

To extend $\tau$ to non-ground \mg programs, we use the concept of instances of rules/programs.
An \emph{instance} of a rule is a ground rule obtained by replacing all of its variables by precomputed terms.
For a rule $R$, the set~$\inst{R}$denotes the set of instances of $R$.
For a program $\Pi$, $\inst{\Pi}$ denotes the set of instances of all rules of $\Pi$.

We can then define the $\tau$ transformation for \mg programs.

\begin{definition}[Transformation of Programs]
  For any program $\Pi$, $\tau(\Pi)$ is the set of formulas $\tau(R)$ for all instances $R$ of the rules of $\Pi$.
\end{definition}

Next, we define stable models of propositional theories.
As stated before, for the case of \mg, we only have to consider finite propositional formulas.
However, we will, in general, still have infinite sets of (finite) propositional formulas as the instantiation of a program will, in general, result in an infinite set.
To define stable models of propositional theories, we reproduce the definition from~\cite{Ferraris2005}.
This definition follows the usual approach of building a reduct with respect to some stable model candidate $I$ and checking whether the minimal consequences of the reduct match our candidate $I$.

We first define the reduct operation.
\clearpage

\begin{definition}[Reduct of a Theory]\label{def:reduct}
  Let $F$ be a propositional formula and let $I$ be a set of precomputed atoms.
  The reduct of $F$ with respect to $I$, denoted by $\glr{F}{I}$, is defined as
  \begin{enumerate}
    \item if $I \not\models F$, then $\glr{F}{I} = \false$,
    \item if $I \models p$ where $p$ is an atom, then $\glr{p}{I} = p$,
    \item if $I \models G \otimes H$ where $\otimes \in \{\land, \lor, \imp\}$, then $\glrp{G \otimes H}{I} = \glr{G}{I} \otimes \glr{H}{I}$.
  \end{enumerate}

  For a propositional theory $\Gamma$, the reduct $\glr{\Gamma}{I}$ is the set of formulas $\glr{F}{I}$ for all $F \in \Gamma$.
\end{definition}

Before turning to the definition of stable models, we state a property of reducts that will be useful in subsequent chapters.

\begin{lemma}\label{lem:sat-reduct}
  Let $F$ be a propositional formula and let $I$ be a set of precomputed atoms.
  $I$ is a model of $F$ iff $I$ is a model of $\glr{F}{I}$.
\end{lemma}

\begin{proof}
  We prove this lemma by induction on the structure of $F$.

  First, let $F$ be an atomic formula, i.e., $F = p$ for some atom $p$.
  Then if $I \models p$ we have $\glr{p}{I} = p$ and thus also $I \models \glr{p}{I}$.
  If $I \models \glr{p}{I}$ then $\glr{p}{I}$ must be $p$ (otherwise $\glr{p}{I} = \false$ which can not be satisfied by $I$) and thus also $I \models p$.

  For the case that $F = \false$ we have $\glr{\bot}{I} = \bot$ so that no $I$ can be a model of $F$ nor $\glr{F}{I}$.

  Next, we consider the case that $F = G \otimes H$ for $\otimes \in \{\land, \lor, \imp\}$.
  By the induction hypothesis, we know that our claim holds for $G$ and $H$.
  We start by considering the case that $\otimes = \land$.
  We then have the following equivalence:
  \begin{align*}
    I \models F
    &\tiff I \models G \land H \\
    &\tiff I \models G \tand I \models H \\
    &\tiff I \models \glr{G}{I} \tand I \models \glr{H}{I} \tag{by induction hypothesis} \\
    &\tiff I \models \glrp{G \land H}{I} \\
    &\tiff I \models \glr{F}{I}
  \end{align*}
  We can show the remaining two cases in the same manner.

  Thus, we have shown that our claim holds for all $F$.
\end{proof}

Using the reduct of a theory, we then have the following definition of stable models for propositional theories.

\begin{definition}[Stable Models of Propositional Theories]
  Let $\Gamma$ be a propositional theory and let $I$ be a set of precomputed atoms.
  $I$ is a stable models of $\Gamma$ iff $I$ is the minimal model of $\glr{\Gamma}{I}$.
\end{definition}

Finally, we define the stable models of \mg programs in terms of stable models of propositional theories.

\begin{definition}[Stable Models of \mg Programs]\label{def:sm-tau}
  Let $\Pi$ be a \mg program and let $I$ be a set of precomputed atoms.
  $I$ is a stable model of $\Pi$ iff $I$ is a stable model of the propositional theory $\tau(\Pi)$.
\end{definition}

\section{Two-Sorted Formulas}
\label{sec:sorted}

While the grounding approach is sufficient for defining the semantics of \mg it is not always a helpful approach as grounding in general produces infinite sets.
If we, for example, want to verify properties of \mg programs using automated theorem provers, we can not use any semantics that uses infinite sets.
Thus, there is an alternative to the $\tau$ transformation called the $\tau^{*}$ transformation.
The difference is that $\tau^{*}$ does not require the grounding step but instead produces first-order formulas rather than propositional formulas.

However, to define $\tau^{*}$ (below in \cref{sec:tau-star}), a standard first-order logic is not sufficient.
Instead, we need a first-order logic that has two types of variables to account for the implicit typing in \mg rules.
For example, in the rule $p(X) \revimp q(X)$, the variable $X$ is of the so-called general type that includes both symbolic constants and integers.
On the other hand, in the rule $p(X+1) \revimp q(X)$, the variable $X$ is of the integer type, a sub-type of the general type.

We review this two-sorted logical language following~\cite{LifschitzEtAl2019,FandinnoEtAl2024a}.
First, we have the signature $\sigsort$ of this logic.

\begin{definition}[Signature]
  The signature $\sigsort$ includes the following
  \begin{itemize}
    \item the sort \emph{general} and its sub-sort \emph{integer},
    \item all precomputed terms of \mg as object constants; an object constant is assigned the sort integer iff it is a numeral,
    \item the symbol $||$ as a unary function constant whose argument and value have the sort integer,
    \item the symbols $+,-,\times$ as binary functions constants whose arguments and values have the sort integer,
    \item predicate symbols $p/n$ as $n$-ary predicate symbols whose arguments have the sort general,
    \item the comparison symbols $=,\neq,\leq,\geq,<,>$ as binary predicate constants whose arguments have the sort general.
  \end{itemize}
\end{definition}

Next, we define formulas over this signature.

\begin{definition}[Formulas]
  Formulas over $\sigsort$ are built as follows
  \begin{itemize}
    \item Variables of two types: program variables (ranging over precomputed program terms) and integer variables (ranging over numerals/integers)
    \item Arithmetic terms: arithmetic terms formed from numerals and integer variables using the operations $||,+,-,\times$
    \item Formula terms: arithmetic terms, program variables, symbolic constants, and the symbols $\myinf$ and $\mysup$
    \item Atomic formulas:
          \begin{itemize}
            \item $p(\terms)$, where $p/n$ is a predicate symbol and $\terms$ is a tuple of $n$ formula terms
            \item $t_{1} \prec t_{2}$, where $t_{i}$ are formula terms and $\prec$ is one of the comparisons
          \end{itemize}
    \item Formulas are formed from atomic formulas using the logical connectives $\false,\land,\lor,\imp$ and quantifiers $\forall,\exists$ in the usual way
  \end{itemize}
\end{definition}

Finally, we consider \emph{standard} interpretations of the signature.
Standard interpretations are certain well-behaved interpretations in that they interpret certain functions and predicates in the usual (expected) way.

\begin{definition}[Standard Interpretations]
  An interpretation of the signature $\sigsort$ is \emph{standard} if
  \begin{enumerate}
    \item its domain of the sort general is the set of precomputed terms,
    \item its domain of the sort integer is the set of numerals,
    \item every object constant represents itself,
    \item the unary and binary function constants are interpreted as usual in arithmetic,
    \item the comparison predicate constants are interpreted in accordance to the \mg semantics.
  \end{enumerate}
\end{definition}

\begin{remark}\label{rem:standard}
  Let $I$ be a set of precomputed atoms.
  $I$ can be uniquely extended to a standard interpretation that maps all members of $I$ to true.
  We denote this standard interpretation by\footnote{In earlier work on \anthem this is simply denoted as $I^{\uparrow}$. We use a slightly more complex notation here as we will later also extend sets of precomputed atoms to other signatures.} $\ext{I}{\sigsort}$.
\end{remark}

\section{\texorpdfstring{\mg}{mini-gringo} Programs as Two-Sorted Formulas}
\label{sec:tau-star}

In this section, we define the transformation $\tau^{*}$, which transforms logic programs into closed first-order formulas.
The formulas produced by $\tau^{*}$ are strongly equivalent to the propositional formulas produced by $\tau$.
However, the formulas resulting from applying $\tau^{*}$ can be used by automated reasoning tools.
In the literature, there are slightly different definitions of $\tau^{*}$ due to some minor errors in the treatment of the division and modulo operations in earlier publications.
This review follows the most recent publication~\cite{FandinnoEtAl2024a}.

First, we have a transformation to express that a variable is the value of a term.
This corresponds to the function $\values{\cdot}$ used in the $\tau$ transformation.

\begin{definition}[$\val{t}{V}$]
  Let $t$ be a term and $V$ be a program variable.
  The formula $\val{t}{V}$ expresses that $V$ is one of the values of $t$ and is defined as follows
  \begin{itemize}
    \item if $t$ is a precomputed term or a variable, then $\val{t}{V}$ is $V = t$,
    \item if $t$ is $|t_{1}|$, then $\val{t}{V}$ is $\exists I (\val{t_{1}}{I} \land V = |I|)$,
    \item if $t$ is $t_{1} \circ t_{2}$ for $\circ \in \{+,-,\times\}$, then $\val{t}{V}$ is
          \[
            \exists I,J (\val{t_{1}}{I} \land \val{t_{2}}{J} \land V = I \circ J),
          \]
    \item if $t$ is $t_{1} / t_{2}$, then $\val{t}{V}$ is
          \begin{alignat*}{3}
            \exists I,J,K ( & \val{t_{1}}{I} \land \val{t_{2}}{J} \land {} && K \times |J| \leq |I| < (K + \num{1}) \times |J| \land {} \\
            & &&( (I \times J \geq \num{0} \land V = K) \lor {} \\
            & &&\phantom{(} (I \times J < \num{0} \land V = - K))),
          \end{alignat*}
    \item if $t$ is $t_{1} \backslash t_{2}$, then $\val{t}{V}$ is
          \begin{alignat*}{3}
            \exists I,J,K ( & \val{t_{1}}{I} \land \val{t_{2}}{J} \land {} && K \times |J| \leq |I| < (K = \num{1}) \times |J| \land {} \\
            & &&((I \times J \geq \num{0} \land V = I - K \times J) \lor {} \\
            & &&\phantom{(}(I \times J < \num{0} \land V = I + K \times J))),
          \end{alignat*}
    \item if $t$ is $t_{1} \twodots t_{2}$, then $\val{t}{V}$ is
          \[
            \exists I,J,K (\val{t_{1}}{I} \land \val{t_{2}}{J} \land I \leq K \leq J \land V = K),
          \]
  \end{itemize}
  where $I,J,K$ are fresh integer variables.

  \begin{sloppypar}
    If $\terms$ is a tuple $t_{1}, \dots, t_{n}$ of terms and $\vars$ is a tuple $V_{1}, \dots, V_{n}$ of distinct program variables, then the formula $\val{\terms}{\vars}$ stands for the conjunction ${\val{t_{1}}{V_{1}} \land \dots \land \val{t_{n}}{V_{n}}}$.
  \end{sloppypar}
\end{definition}

The following lemma connects this definition to the definition of $\values{\cdot}$.
It is a consequence of~\citep[Proposition~1]{LifschitzEtAl2019}.

\begin{lemma}\label{lem:val-values}
  Let $\terms$ be a tuple of terms and let $\rs$ be a tuple of precomputed terms.
  The formula $\val{\terms}{\rs}$ is equivalent to $\true$ iff $\rs \in \values{\terms}$.
\end{lemma}

Next is the transformation $\tau^{B}$ applied to rule bodies.

\begin{definition}[$\tau^{B}$]
  Let $B$ be a literal or a comparison, then we define $\tau^{B}(B)$ as follows
  \begin{itemize}
    \item $\exists \vars (\val{\terms}{\vars} \land p(\vars))$ if $B$ is $p(\terms)$,
    \item $\exists \vars (\val{\terms}{\vars} \land \lnot p(\vars))$ if $B$ is $\pnot p(\terms)$,
    \item $\exists \vars (\val{\terms}{\vars} \land \lnot \lnot p(\vars))$ if $B$ is $\pnot \pnot p(\terms)$,
    \item $\exists V_{1}, V_{2} (\val{t_{1}}{V_{1}} \land \val{t_{2}}{V_{2}} \land V_{1} \prec V_{2})$ if $B$ is $t_{1} \prec t_{2}$.
  \end{itemize}

  If $Body$ is a conjunction $B_{1} \land \ldots \land B_{n}$ of literals and comparisons, then $\tau^{B}(Body)$ stands for the conjunction $\tau^{B}(B_{1}) \land \dots \land \tau^{B}(B_{n})$.
\end{definition}

The following lemma states the relationship to $\tau$ applied to a rule body.
It is a consequence of~\citep[Proposition~2]{LifschitzEtAl2019}.

\begin{lemma}\label{lem:tau-taub}
  Let $Body$ be a conjunction of literals and comparisons, and let $I$ be a set of precomputed atoms.
  Then $I \models \tau(Body)$ iff $\ext{I}{\sigsort} \models \tau^{B}(Body)$.
\end{lemma}

Finally, we define the $\tau^{*}$ transformation applied to rules and programs.

\begin{definition}[$\tau^{*}$]
  Let $R$ be a rule, then we define $\tau^{*}$ as follows
  \begin{itemize}
    \item $\uc (\val{\terms}{\vars} \land \tau^{B}(Body) \imp p(\vars))$ if $R$ is a basic rule,
    \item $\uc (\val{\terms}{\vars} \land \tau^{B}(Body) \land \lnot\lnot p(\vars) \imp p(\vars))$ if $R$ is a choice rule,
    \item $\uc (\lnot \tau^{B}(Body))$ if $R$ is a constraint,
  \end{itemize}
  where $\uc F$ denotes the universal closure of $F$.

  For a logic program $\Pi$, $\tau^{*}(\Pi)$ denotes the set of sentences $\tau^{*}(R)$ for all $R \in \Pi$.
\end{definition}

\begin{remark}\label{rem:form}
  We introduce a helpful notation for basic and choice rules here.
  For both types of rules, the result of applying the $\tau^{*}$ transformation has the same form $\uc (F \imp p(\vars))$ where $F$ is as above.
  We represented this common part by $\form{R}{\vars}$ where we define $\form{R}{\vars}$ as follows
  \begin{itemize}
    \item if $R$ is a basic rule, then $\form{R}{\vars} = \val{\terms}{\vars} \land \taub(\body{R})$,
    \item if $R$ is a choice rule, then $\form{R}{\vars} = \val{\terms}{\vars} \land \taub(\body{R}) \land \lnot \lnot p(\vars)$.
  \end{itemize}
\end{remark}

Indeed, for basic or choice rules $R$ we then have $\tau^{*}(R) = \uc (\form{R}{\vars} \imp p(\vars))$.
This notation enables us to treat basic and choice rules in a uniform way later when defining completion.

Finally, we can state the relationship between $\tau$ and $\tau^{*}$.
The following lemma is a consequence of~\citep[Proposition~3]{LifschitzEtAl2019}.

\begin{lemma}\label{lem:tau-tau*}
  Let $\Pi$ be a \mg program and let $I$ be a set of precomputed atoms.
  Then $I \models \tau(\Pi)$ iff $\ext{I}{\sigsort} \models \tau^{*}(\Pi)$.
\end{lemma}

\chapter{Support Semantics for \texorpdfstring{\mg}{mini-gringo}}
\label{chp:support}

This chapter introduces an alternative semantics for \mg programs based on well-supported models.
Supported models were introduced in~\cite{AptEtAl1988}.
Their extension well-supported models (sometimes called supported models with well-ordering) has, for example, been considered in~\cite{Elkan1990,Fages1991}.
In this chapter, we use their definitions and generalise them to the \mg language.
The main idea of this generalisation is to adapt the definitions from~\cite{Elkan1990,Fages1991} to include an application of the $\tau$ transformation.
Similarly, the proofs of equivalence of the well-support semantics and the stable model semantics using reducts are generalised for the \mg case.

We start by introducing supported models for \mg in \cref{sec:support}.
Afterwards, in \cref{sec:well-support} we introduce well-supported models for \mg.
We then show that the well-supported models of a program and its reduct are the same.
Finally, we prove the equivalence to the stable model semantics of \mg in terms of reducts (\cref{def:sm-tau}).

\section{Supported Models for \texorpdfstring{\mg}{mini-gringo}}
\label{sec:support}

Intuitively, we call a model supported if for every true atom in the model, there is some justification or support of its truth.
That is, there is some rule in the logic program that contains the atom in the head and whose body is true in the model.

For example, let us consider a logic program containing the single rule
\[
  p \revimp q.
\]
The set $\{p\}$ is a model of this program, however, it is not supported, as the only possible support for $p$ is the rule $p \revimp q$, but the body of this rule is not true in the model.
If we add $q$ to our model, $p$ will be supported, but now we do not have a support for $q$.
Note that this example has no supported model.

If we instead consider the example program
\[
  p \revimp q \qquad q \revimp p
\]
we can see that the set $\{p,q\}$ is a supported model: $p$ is supported by the rule $p \revimp q$ and $q$ is supported by the rule $q \revimp p$.

Supported models are equivalent to the models of the completion of a logic program (see \cref{sec:completion} for details on completion).
Both semantics are limited in that they only capture the stable models semantics of (locally) tight programs.
However, we will not focus on this relation here.
We only consider supported models as a first step to defining well-supported models.

Formally, we define supported interpretations as follows.

\begin{definition}[Supported Interpretation]
  Let $\Pi$ be a \mg program and let $I$ be a set of precomputed atoms.
  $I$ is a \emph{supported interpretation} of $\Pi$ iff for every precomputed atom $p(\rs) \in I$ there exists an instance $R$ of a rule\footnote{Note that this can only be a basic rule or a choice rule. $R$ can not be a constraint as in this case the first condition can not be satisfied.} in $\Pi$ such that:
  \begin{enumerate}
    \item $\headlit{R} = p(\terms)$ and $\rs \in \values{\terms}$,
    \item $I \models \tau(\body{R})$.
  \end{enumerate}
\end{definition}

A set $I$ of precomputed atoms is a \emph{supported model} of a logic program $\Pi$ if and only if $I$ is a supported interpretation of $\Pi$ and $I$ is a model of $\tau(\Pi)$.
That is, $I$ is a supported interpretation and it satisfies all the rules of $\Pi$.

\section{Well-Supported Models for \texorpdfstring{\mg}{mini-gringo}}
\label{sec:well-support}

\subsection{Definition of Well-Supported Models}
Well-supported models extend supported models by taking into account the order in which atoms are derived.
They additionally require a well-behaved order on the set of true atoms, which corresponds to their derivation order.
The support of an atom can then only use atoms that are derived earlier with regard to this order.

Let us again consider the example logic program
\[
  p \revimp q \qquad q \revimp p.
\]
To check whether the supported model $\{p,q\}$ is also well-supported, we need an order $\prec$ on $\{p,q\}$.
We choose the order $p \prec q$ here.
Now $q$ is well-supported: the supporting rule is still $q \revimp p$ and the body of the rule only uses $p$ and we have $p \prec q$.
On the other hand, $p$ is not well-supported: the only possible rule is $p \revimp q$ but the body of this rule is $q$, and we have $q \not\prec p$.

Instead of the order~$p \prec q$, we could have chosen $q \prec p$, but this leads to the inverse issue: $p$~is well-supported while $q$ is not.
Thus, the set $\{p,q\}$ is not a well-supported model.
Note that this corresponds to the fact that $\{p,q\}$ is also not a stable model of the program.

Before formally defining well-supported models, we first look at the conditions on the ordering $\prec$.
We want a derivation order that corresponds to our understanding of stable models.
Our derivation order should have the following properties: it should be transitive, non-cyclic and finite.
We want to rule out cases where $p$ derives $q$ and vice versa as well as ensure that the derivation of an atom $p$ only contains finitely many steps.\footnote{Note that this does not rule out infinite models. It only means that in an infinite model, each atom only has finitely many derivation steps.}
Orders of this kind are called strict well-founded partial orders.

Let us review what the formal criteria of a strict well-founded partial order are.
We start by defining strict partial orders.

\begin{definition}[Strict partial order]
  A relation $\prec$ on $I$ is a \emph{strict partial order} if it satisfies the following properties:
  \begin{enumerate}
    \item Irreflexivity: for all $p \in I$, $\lnot (p \prec p)$,
    \item Transitivity: for all $p, q, r \in I$, $p \prec q \land q \prec r \imp p \prec r$.
  \end{enumerate}
\end{definition}

Note that a transitive relation is irreflexive if and only if it is asymmetric ($\prec$ is asymmetric if for all $p,q \in I$ it holds that $p \prec q \imp \lnot (q \prec p)$).
Thus, we could replace irreflexivity by asymmetry in the above definition.
Next, we define well-founded relations.

\begin{definition}[Well-founded relation]
  A relation $\prec$ on $I$ is \emph{well-founded} if there is no infinitely decreasing chain.
  I.e., there is no infinite sequence $p_{0}, p_{1}, p_{2}, \dots$ with $p_{i} \in I$ such that $p_{i+1} \prec p_{i}$ for all $i$.
\end{definition}

In the special case that $I$ is a finite set, the condition of well-foundedness is automatically satisfied by every strict partial order as stated in the following lemma.

\begin{lemma}\label{lem:fin-well-founded}
  Let $I$ be a \emph{finite} set and $\prec$ be a \emph{strict partial order} on $I$.
  Then $\prec$ is also a well-founded relation on $I$.
\end{lemma}

If there was an infinitely decreasing chain $p_{0}, p_{1}, p_{2}, \dots$ on $I$ then there would be some $i < j$ such that $p_{i} = p_{j}$ as $I$ only has finitely many elements.
By transitivity, we then have $p_{j} \prec p_{i}$ which is in contradiction to the assumption that $\prec$ is a strict partial order and thus, in particular, irreflexive.

As an example, the usual order~$<$ on the integers is a strict partial order.
However, it is not a well-founded order.
Starting at any integer~$n$ we have an infinitely decreasing chain as the integers do not have a smallest element.
If we restrict ourselves to natural numbers instead, $<$~is well-founded, as we now have $0$ as the smallest element.
Therefore, any decreasing chain starting at $n$ will terminate when it reaches $0$.
On the other hand, the order $\leq$ (on either naturals or integers) is not even a strict partial order as it is reflexive.

Combining these two definitions, we get the strict well-founded partial orders that match our intuition of the properties of a derivation order.
Using these definitions, we can now turn to defining well-supported interpretations.

\begin{definition}[Well-Supported Interpretation]\label{def:wsm}
  Let $\Pi$ be a \mg program and let $I$ be a set of precomputed atoms.
  $I$ is a well-supported interpretation of $\Pi$ iff there exists a strict well-founded partial ordering $\prec$ on $I$ so that for every precomputed atom $p(\rs) \in I$ there exists an instance $R$ of a rule in $\Pi$ such that:
  \begin{enumerate}
    \item $\headlit{R} = p(\terms)$ and $\rs \in \values{\terms}$,
    \item $I \models \tau(\body{R})$,
    \item for all $q(\terms^{\prime}) \in \bodylitp{R}$ there is some $\rs^{\prime} \in \values{\terms^{\prime}}$ with $q(\rs^{\prime}) \in I$ and $q(\rs^{\prime}) \prec p(\rs)$.
  \end{enumerate}
\end{definition}

Note that the first two conditions are exactly those of a supported interpretation.
We then add the condition that the positive body literals used to derive $p(\rs)$ have to be smaller than $p(\rs)$ with respect to the order $\prec$.

A set of precomputed atoms is a well-supported model of a logic program $\Pi$ if and only if $I$ is a well-supported interpretation of $\Pi$ and $I$ is a model of $\tau(\Pi)$.

\subsection{Well-Supported Models of Reducts}
Before considering the relation of well-supported and stable models, we want to state that well-supported models of a program $\Pi$ are also well-supported models of its reduct.
However, we can not directly state this as the reduct operation (\cref{def:reduct}) is only defined on $\tau(\Pi)$ and the reduct itself is not a logic program.%
\footnote{The theory $\glr{\tau(\Pi)}{I}$ essentially has the form of a positive \mg program (see \cref{lem:red-theory}) but it is in general an infinite set whereas \mg programs are finite sets. By introducing variables to the ``rules'' in $\glr{\tau(\Pi)}{I}$, it may be possible to convert it to a finite set of rules in some cases.}
Whereas our definition of well-supported models is in terms of programs.

Thus, we will first characterise reducts of $\tau(\Pi)$ as theories of a certain form and then extend the definition of well-supported models to these theories.
We call these theories fittingly \emph{reduct theories} as they are the result of the reduct operation as shown below in \cref{lem:red-theory}.

\begin{definition}[Reduct Theory]\label{def:red-theory}
  We call a theory $\Gamma$ a \emph{reduct theory} if it only contains formulas of the following form:
  \[
    B_{1} \land \dots \land B_{n} \imp (q(\rs_{1}) \land \dots \land q(\rs_{m})),
  \]
  where each $B_{i} = (p_{i}(\rs_{i,1}) \lor \dots \lor p_{i}(\rs_{i,m_{i}}))$.
  If $F \in \Gamma$ is such a formula we define
  \begin{alignat*}{1}
    \head{F} &= \{ q(\rs_{1}), \dots, q(\rs_{m}) \}, \\
    \body{F} &= B_{1} \land \dots \land B_{n}.
  \end{alignat*}
  Note that both $\body{F}$ and the formula $q(\rs_{1}) \land \dots \land q(\rs_{m})$ can also be the empty conjunction, i.e., $\true$.
\end{definition}

The following lemma justifies the usage of the name \emph{reduct theory}.

\begin{lemma}\label{lem:red-theory}
  Let $\Pi$ be a \mg program and let $I$ be a set of precomputed atoms.
  The theory $\glr{\tau(\Pi)}{I}$ is either
  \begin{enumerate}
    \item a theory that is not satisfied by $I$ or,
    \item a reduct theory.
  \end{enumerate}
\end{lemma}

We prove this lemma in \cref{chp:red-theories} by analysing all the cases of what a \mg rule can be and what its reduct is.
The case that $I \not\models \glr{\tau(\Pi)}{I}$ will not be relevant in the following, as we will always assume that $I$ is a model of the reduct $\glr{\tau(\Pi)}{I}$.

Now we can state an alternative definition of well-supported models for reduct theories.

\begin{definition}[Well-supported Model of a Reduct Theory]\label{def:wsm-red-theory}
  Let $\Gamma$ be a reduct theory and let $I$ be a set of precomputed atoms.
  $I$ is a well-supported interpretation of $\Gamma$ iff there exists a strict well-founded partial ordering $\prec$ on $I$ so that for every precomputed atom $p(\rs) \in I$ there exists a formula $F \in \Gamma$ such that:
  \begin{enumerate}
    \item $p(\rs) \in \head{F}$,
    \item $I \models \body{F}$,
    \item for every conjunct $p_{i}(\rs_{i,1}) \lor \dots \lor p_{i}(\rs_{i,m_{i}})$ in $\body{F}$ there is some $\rs_{i,j}$ such that $p_{i}(\rs_{i,j}) \in I$ and $p_{i}(\rs_{i,j}) \prec p(\rs)$.
  \end{enumerate}
  A set $I$ of precomputed atoms is a well-supported model of $\Gamma$ iff it is a well-supported interpretation of $\Gamma$ and it is a model of $\Gamma$.
\end{definition}

This definition is essentially the same as \cref{def:wsm} but adjusted to reduct theories.
Now we can state that well-supported models of a program are also well-supported models of its reduct.

\begin{lemma}\label{lem:wsm-prog-reduct}
  Let $\Pi$ be a \mg program and let $I$ be a set of precomputed atoms.
  We denote by $\Gamma$ the reduct of $\tau(\Pi)$ with respect to $I$.
  If $I$ is a well-supported model of $\Pi$, then $I$ is also a well-supported model of $\Gamma$.
\end{lemma}

The idea of the proof is that building the reduct of a supporting rule $R$ only removes the negative body of $R$.
Thus, the reduct of $R$ still fulfils conditions 1~and~3 as they only talk about the head and positive body.
The second condition also still holds: if $I$ satisfies the whole body of $R$, it in particular satisfies the positive body of $R$.

\begin{proof}
  As $I$ is a model of $\tau(\Pi)$ it is also a model of $\Gamma = \glr{\tau(\Pi)}{I}$ by \cref{lem:sat-reduct}.
  To show the well-supportedness of $I$ on $\Gamma$, we use the same order $\prec$ that shows the well-supportedness on $\Pi$.
  Now let $p(\rs) \in I$ and let $R \in \inst{\Pi}$ be the supporting rule of $p(\rs)$ (regarding the well-support on $\Pi$), i.e., we have:
  \begin{assumptions}
    \item\label{ass:wsm-prog-reduct-head} $\headlit{R} = p(\terms)$ and $\rs \in \values{\terms}$,
    \item\label{ass:wsm-prog-reduct-body} $I \models \tau(\body{R})$,
    \item\label{ass:wsm-prog-reduct-order} for all $q(\terms^{\prime}) \in \bodylitp{R}$ there is some $\rs^{\prime} \in \values{\terms^{\prime}}$ with $q(\rs^{\prime}) \in I$ and $q(\rs^{\prime}) \prec p(\rs)$.
  \end{assumptions}

  We show that the properties stated in \cref{def:wsm-red-theory} hold.
  We know that $I \models \tau(R)$ (as $I \models \tau(\Pi)$).
  By \cref{ass:wsm-prog-reduct-body} (i.e., $I \models \tau(\body{R})$) we thus also have $I \models \tau(\head{R})$.
  Then (according to \cref{chp:red-theories})
  \begin{equation}\label{eq:wsm-red}
    \glr{\tau(R)}{I} = \bigwedge_{q(\terms^{\prime}) \in \bodylitp{R}} \Big(\bigvee_{\twosub{\rs^{\prime} \in \values{\terms^{\prime}}}{q(\rs^{\prime}) \in I}} q(\rs^{\prime}) \Big) \imp \bigwedge_{\twosub{\rs \in \values{\terms}}{p(\rs) \in I}} p(\rs).
  \end{equation}
  We denote this formula by $F$.
  It is the supporting formula of $p(\rs)$ for the well-supportedness of $I$ on $\Gamma$.
  We show that $F$ satisfies the properties of \cref{def:wsm-red-theory}:
  \begin{enumerate}
    \item $p(\rs) \in \head{F}$ by \cref{eq:wsm-red} as $p(\rs) \in I$,
    \item $I \models \body{F}$ as \cref{ass:wsm-prog-reduct-body} implies $I \models \glr{\tau(\body{R})}{I}$ by \cref{lem:sat-reduct} and $\body{F} = \glr{\tau(\body{R})}{I}$,
    \item for every conjunct $\bigvee_{\rs^{\prime} \in \values{\terms^{\prime}}, q(\rs^{\prime}) \in I} q(\rs^{\prime})$ we have some $\rs^{\prime}$ such that $q(\rs^{\prime}) \in I$ and $q(\rs^{\prime}) \prec p(\rs)$ by \cref{ass:wsm-prog-reduct-order}.
  \end{enumerate}
  Thus, $I$ is a well-supported model of $\Gamma$ with the same relation $\prec$ that shows the well-supportedness of $I$ on $\Pi$.
\end{proof}

Furthermore, we can also prove the converse statement.

\begin{lemma}\label{lem:wsm-reduct-prog}
  Let $\Pi$ be a \mg program and let $I$ be a set of precomputed atoms.
  We denote by $\Gamma$ the reduct of $\tau(\Pi)$ with respect to $I$.
  If $I$ is a well-supported model of $\Gamma$, then $I$ is also a well-supported model of $\Pi$.
\end{lemma}

Intuitively, from a supporting formula $F \in \Gamma$ we can just go back to the rule $R \in \inst{\Pi}$ with $F = \glr{\tau(R)}{I}$.
The reduct operation essentially removes the negative body of a rule and replaces it by $\true$ or $\false$ depending on whether $I$ satisfies the negative body.
Properties 1~and~3 of \cref{def:wsm} immediately follow as they only talk about the head and positive body.
For property~2, it is clear that $I$ is a model of the positive part of $\tau(\body{R})$ as $I \models \body{F}$.
The reduct of the negative body (including single and double negation) can either be $\true$ or $\false$.
We can rule out the $\false$ case as then $I$ could not be a model of $\body{F}$.
But then, as the reduct of the negative body is $\true$, it means that $I$ satisfies the negative body.
Thus, we get that $I$ satisfies $\tau(\body{R})$.

\begin{proof}
  As $I$ is a model of $\Gamma = \glr{\tau(\Pi)}{I}$ we also have that $I$ is a model of $\tau(\Pi)$ by \cref{lem:sat-reduct}.
  To show that $I$ is a well-supported interpretation of $\Pi$, we use the same order $\prec$ that shows the well-supportedness on $\Gamma$.
  Now let $p(\rs) \in I$ and let $F \in \Gamma$ be the supporting formula of $p(\rs)$, i.e., we have
  \begin{assumptions}
    \item\label{ass:wsm-reduct-prog-head} $p(\rs) \in \head{F}$,
    \item\label{ass:wsm-reduct-prog-body} $I \models \body{F}$,
    \item\label{ass:wsm-reduct-prog-order} for every conjunct $p_{i}(\rs_{i,1}) \lor \dots \lor p_{i}(\rs_{i,m_{i}})$ in $\body{F}$ there is some $r_{i,j}$ such that $p_{i}(\rs_{i,j}) \in I$ and $p_{i}(\rs_{i,j}) \prec p(\rs)$.
  \end{assumptions}

  Let $R \in \inst{\Pi}$ such that $\glr{\tau(R)}{I} = F$.
  We show that $R$ is the supporting rule of $p(\rs)$ regarding the well-supportedness on $\Pi$, i.e., that the properties of \cref{def:wsm} hold.

  First, we have $\headlit{R} = p(\terms)$ and $\rs \in \values{\terms}$ by \cref{ass:wsm-reduct-prog-head}.

  Next, we show $I \models \tau(\body{R})$.
  We know that $I \models F = \glr{\tau(R)}{I}$ as $I$ is a model of $\Gamma$.
  Thus, the reduct of $\tau(R)$ is $\glr{\tau(R)}{I} = \glr{\tau(\body{R})}{I} \imp \glr{\tau(\head{R})}{I}$.
  Therefore, $\body{F} = \body{\glr{\tau(R)}{I}} = \glr{\tau(\body{R})}{I}$, i.e., we can change the order of the body and reduct operations.
  By \cref{lem:sat-reduct} we then have that $I \models \tau(\body{R})$ as $I \models \glr{\tau(\body{R})}{I}$ (i.e., \cref{ass:wsm-reduct-prog-body}) holds.

  Finally, we need to show that for every $q(\terms^{\prime}) \in \bodylitp{R}$ there is some $\rs^{\prime} \in \values{\terms}$ with $q(\rs^{\prime}) \in I$ and $q(\rs^{\prime}) \prec p(\rs)$.
  For every $q(\terms^{\prime}) \in \bodylitp{R}$ we have $q(\rs_{1}) \lor \dots q(\rs_{m})$ as one of the conjuncts in $\body{F}$ where the $\rs_{i}$ are some of the values in $\values{\terms^{\prime}}$.
  By \cref{ass:wsm-reduct-prog-order} we then know that there is some $\rs_{j}$ such that $q(\rs_{j}) \in I$ and $q(\rs_{j}) \prec p(\rs)$, i.e., the $\rs^{\prime}$ that we need is $\rs_{j}$.

  Thus, we have shown that $I$ is a well-supported model of $\Pi$.
\end{proof}

Combining the two previous lemmata, we then obtain the following corollary stating that a program and its reduct have the same well-supported models.

\begin{corollary}
  Let $\Pi$ be a \mg program and let $I$ be a set of precomputed atoms.
  We denote by $\Gamma$ the reduct of $\tau(\Pi)$ with respect to $I$.
  Then $I$ is a well-supported model of $\Pi$ iff $I$ is a well-supported model of $\Gamma$.
\end{corollary}

\subsection{Equivalence of Well-Supported and Stable Models}
To prove that well-supported models are equivalent to stable models, we have to show that a well-supported model is the minimal model of the program after building the reduct.
Here we use the immediate consequence operator to talk about minimal models, as this enables us to extract information about the derivation order from the minimal model.
Thus, we first have to define the immediate consequence operator on reduct theories.

\begin{definition}[Immediate Consequence Operator]
  Let $\Gamma$ be a reduct theory.
  We define the immediate consequences of $\Gamma$ with respect to a set $I$ of precomputed atoms as follows
  \[
    \step{\Gamma}(I) = \{ p(\rs) \mid F \in \Gamma, p(\rs) \in \head{F}, I \models \body{F} \}.
  \]

  Next we define the repeated application $\stepi{\Gamma}{i}$ of the operator:
  \begin{alignat*}{2}
    \stepi{\Gamma}{0}(I)   & = I, \\
    \stepi{\Gamma}{n+1}(I) & = \stepi{\Gamma}{n}(\step{\Gamma}(I)).
  \end{alignat*}
\end{definition}

We will use the following shorthand notations: $\stepi{\Gamma}{n} = \stepi{\Gamma}{n}(\emptyset)$, $\stepf{\Gamma} = \bigcup_{i=0}^{\infty} \stepi{\Gamma}{i}$.
Note that the operator is monotonic, i.e., if $p(\rs) \in \stepi{\Gamma}{n}$ then also $p(\rs) \in \stepi{\Gamma}{m}$ for all $m > n$.
Also note that $\stepf{\Gamma}$ is exactly the minimal model of $\Gamma$.%
\footnote{We do not prove these properties here. However, note that we can convert reduct theories into sets of definite clauses/rules (i.e., formulas of the form $p_{1} \land \dots \land p_{n} \imp q$, where the $p_{i}$ and $q$ are propositional atoms) as follows. First, we can eliminate the conjunction in the head as $F \imp (q \land q^{\prime})$ is equivalent to the two formulas $F \imp q$ and $F \imp q^{\prime}$. Second, we can transform the body into a disjunction of conjunctions by properties of disjunction and conjunction. Then we can eliminate the disjunctions in the body as $F \lor F^{\prime} \imp q$ is equivalent to the two formulas $F \imp q$ and $F^{\prime} \imp q$. We can then refer to~\cite{VanEmdenKowalski1976,Lloyd1987} for proofs of monotonicity and equality to the minimal model for the immediate consequence operator on sets of definite clauses/rules.}

Before stating and proving the equivalence of well-supported models and stable models, we give some lemmata connecting the immediate consequence operator and well-supported models.
First, we consider how to build a strict well-founded partial order $\prec$ using the immediate consequence operator.

\begin{remark}
  \label{rem:ord-t}
  The immediate consequence operator provides an order for the precomputed atoms in the minimal model.
  I.e., let $\Gamma$ be a reduct theory and let $I = \stepf{\Gamma}$.
  We define an order on $I$ by setting $A \prec B$ iff $A$ is derived in an earlier step by $\step{\Gamma}$ than B.
  That is, given $A,B \in I$ we set $A \prec B$ iff for some $i,j$ with $i < j$, $A \in \stepi{\Gamma}{i}$ and $B \in \stepi{\Gamma}{j} \setminus \stepi{\Gamma}{i}$.
\end{remark}

The following lemmata state that $\prec$ constructed in this way is a strict well-founded partial order.

\begin{lemma}
  \label{lem:ord-t-spo}
  Let $\Gamma$ be a reduct theory and let $I = \stepf{\Gamma}$.
  The order $\prec$ as defined in \cref{rem:ord-t} is a strict partial order, i.e., $\prec$ is irreflexive and transitive.
\end{lemma}

Irreflexivity is immediate from the definition of $\prec$.
Transitivity follows from the monotonicity of $\step{\Gamma}$.

\begin{lemma}
  \label{lem:ord-t-wfo}
  Let $\Gamma$ be a reduct theory and let $I = \stepf{\Gamma}$.
  The order $\prec$ as defined in \cref{rem:ord-t} is well-founded.
\end{lemma}

Let $A \in I$ be an arbitrary precomputed atom.
As $A \in I$ we know there exists some $n$ such that $A \in \stepi{\Gamma}{n}$.
The maximum length for a descending chain starting at $A$ is thus $n-1$ and, in particular, finite.
Thus, $\prec$ does not contain infinite descending chains and is well-founded.

Next, we connect well-supported models to minimal models.

\begin{lemma}
  \label{lem:wsm-min}
  Let $\Gamma$ be a reduct theory and let $I$ be a well-supported model of $\Gamma$.
  Then $I \subseteq \stepf{\Gamma}$, i.e., $I$ is a subset of the minimal model of $\Gamma$.
\end{lemma}

\begin{proof}
  Let $I$ be a well-supported model of $\Gamma$ with the relation $\prec$.
  We make use of the notion of the rank of an atom as given by the relation $\prec$ on $I$.
  For each element $p(\rs) \in I$ we define the rank of that element with respect to the order $\prec$ as follows:
  the rank of $p(\rs)$ is the maximum of the numbers $n$ such that there exists a chain $e_{1} \prec e_{2} \prec \dots \prec e_{n} = p(\rs)$.\footnote{Note that the minimal rank according to this definition is 1.}

  We show by induction on the rank $n$ that for any $p(\rs) \in I$ there exists an $l$ such that $p(\rs) \in \stepi{\Gamma}{l}$.

  For the base case of the induction if the rank of $p(\rs)$ is 1, then the formula that supports $p(\rs)$ has an empty body (no atom can be used as the support of $p(\rs)$ as no element smaller than $p(\rs)$).
  Thus, $p(\rs)$ is a ``fact'' of $\Gamma$ (formally there is an $F \in \Gamma$ with $p(\rs) \in \head{F}$ and $\body{F} = \true$) and $p(\rs) \in \stepi{\Gamma}{1}$, that is $l=1$.

  Now let $p(\rs)$ be an atom at rank $n$.
  We assume by induction that for any atom $q(\rs^{\prime})$ with rank $n^{\prime} < n$ there exists a $j$ such that $q(\rs^{\prime}) \in \stepi{\Gamma}{j}$.

  By the assumption that $I$ is a well-supported model of $\Gamma$, the atom $p(\rs)$ is supported by a formula $F$ with:
  \begin{assumptions}
    \item $p(\rs) \in \head{F}$,
    \item $I \models \body{F}$,
    \item for every conjunct $p_{i}(\rs_{i,1}) \lor \dots \lor p_{i}(\rs_{i,m_{i}})$ in $\body{F}$ there is some $\rs_{i,j}$ such that $p_{i}(\rs_{i,j}) \in I$ and $p_{i}(\rs_{i,j}) \prec p(\rs)$.
  \end{assumptions}
  Then for every conjunct $p_{i}(\rs_{i,1}) \lor \dots \lor p_{i}(\rs_{i,m_{i}})$ we have a $p_{i}(\rs_{i,j})$ with a smaller rank than $p(\rs)$ (as $p_{i}(\rs_{i,j}) \prec p(\rs)$).
  By the induction hypothesis we then have an $l_{i}$ such that $p_{i}(\rs_{i,j}) \in \stepi{\Gamma}{l_{i}}$.
  Let $l_{m}$ be the maximum of the $l_{i}$.
  As $T_{\Gamma}$ is monotonic we have $p_{i}(\rs_{i,j}) \in \stepi{\Gamma}{l_{m}}$ for all $p_{i}(\rs_{i,j})$.
  In other words every conjunct of $\body{F}$ is satisfied by $\stepi{\Gamma}{l_{m}}$, i.e., $\stepi{\Gamma}{l_{m}} \models \body{F}$.
  Hence, we have $p(\rs) \in \stepi{\Gamma}{l_{m}+1}$ as $p(\rs) \in \head{F}$.

  Thus, we have shown that for any $p(\rs) \in I$ there exists an $l$ such that $p(\rs) \in \stepi{\Gamma}{l}$, i.e., $I \subseteq \stepf{\Gamma}$.
\end{proof}

Note that as well-supported models are in particular models of $\Gamma$, we have the reverse relation as well, i.e., $\stepf{\Gamma} \subseteq I$.

Finally, we are ready to state the equivalence of stable models and well-supported models.
We do so by considering the two directions of the equivalence separately in the following two lemmata.

\begin{lemma}\label{lem:sm-to-wsm}
  Let $\Pi$ be a \mg program and let $I$ be a set of precomputed atoms.
  If $I$ is a stable model of $\Pi$, then $I$ is also a well-supported model of $\Pi$.
\end{lemma}

\begin{proof}
  Let $I$ be a stable model of $\Pi$, i.e., $I$ is a stable model of $\tau(\Pi)$.
  In other words, $I$ is the minimal model of the reduct of $\tau(\Pi)$ with respect to $I$, which we denote by $\Gamma$.
  That is, $I$ is the fixpoint of applying the step-wise immediate consequence operator $\step{\Gamma}$, i.e., we have $I = \stepf{\Gamma}$.

  We show that $I$ is also a well-supported model of $\Pi$.
  By \cref{lem:wsm-reduct-prog} it is sufficient to show that $I$ is a well-supported model of $\Gamma$.
  As $I$ is the minimal model of $\Gamma$, it only remains to show $I$ is a well-supported interpretation of $\Gamma$.
  To do so, we need to construct a strict well-founded partial order $\prec$ on $I$.
  We construct $\prec$ as in \cref{rem:ord-t}.
  \cref{lem:ord-t-spo,lem:ord-t-wfo} show that $\prec$ constructed in this way is indeed a strict well-founded partial order.

  The three conditions from the definition of a well-supported interpretation (\cref{def:wsm-red-theory}) remain to be shown.
  We take some $p(\rs) \in I$ and show that there is a supporting formula $F \in \Gamma$ for $p(\rs)$.

  First, we need to show that $p(\rs) \in \head{F}$.
  We know that $p(\rs) \in \stepi{\Gamma}{i}$ for some $i$.
  We take the smallest such $i$, i.e., such that $p(\rs) \not\in \stepi{\Gamma}{i-1}$.
  As $p(\rs) \in \stepi{\Gamma}{i}$ we know that there is some $F \in \Gamma$ with $p(\rs) \in \head{F}$ and $\stepi{\Gamma}{i-1} \models \body{F}$.
  This $F$ is the supporting formula of $p(\rs)$.
  Thus, we have shown that the first condition, namely $p(\rs) \in \head{F}$, holds.

  Next, we show that $I \models \body{F}$.
  As noted before, we know that $\stepi{\Gamma}{i-1}$ is a model of $\body{F}$.
  As $\stepi{\Gamma}{i-1} \subseteq I$ we thus also have $I \models \body{F}$.

  It remains to show that for every conjunct
  \[
    p_{i}(\rs_{i,1}) \lor \dots \lor p_{i}(\rs_{i,m_{i}}),
  \]
  of $\body{F}$ there is some $\rs_{i,j}$ such that $p_{i}(\rs_{i,j}) \in I$ and $p_{i}(\rs_{i,j}) \prec p(\rs)$.
  We know that $\stepi{\Gamma}{i-1} \models \body{F}$ and thus in particular
  \[
    \stepi{\Gamma}{i-1} \models p_{i}(\rs_{i,1}) \lor \dots \lor p_{i}(\rs_{i,m_{i}}).
  \]
  This means that there is some $p_{i}(\rs_{i,j}) \in \stepi{\Gamma}{i-1}$ and therefore also $p_{i}(\rs_{i,j}) \in I$.
  By construction of $\prec$ we also have $p_{i}(\rs_{i,j}) \prec p(\rs)$.

  Thus, $I$ is a well-supported model of $\Gamma$.
  By \cref{lem:wsm-reduct-prog} we then obtain that $I$ is a well-supported model of $\Pi$ as well.
\end{proof}

\begin{lemma}\label{lem:wsm-to-sm}
  Let $\Pi$ be a \mg program and let $I$ be a set of precomputed atoms.
  If $I$ is a well-supported model of $\Pi$, then $I$ is also a stable model of $\Pi$.
\end{lemma}

\begin{proof}
  Let $I$ be a well-supported model of $\Pi$ and let $\prec$ be the relation on $I$ establishing the well-supportedness.
  We need to show that $I$ is the minimal model of the reduct of $\tau(\Pi)$ with respect to $I$, this reduct is again denoted by $\Gamma$.
  As $I$ is a well-supported model of $\Pi$, it is also a well-supported model of $\Gamma$ (\cref{lem:wsm-prog-reduct}).
  By \cref{lem:wsm-min} we then know that $I \subseteq \stepf{\Gamma}$.

  This is sufficient to show that $I$ is the minimal model of $\Gamma$.
  As $I$ is a model of $\tau(\Pi)$ and thus also of $\Gamma = \glr{\tau(\Pi)}{I}$ (by \cref{lem:sat-reduct}), we have that $\stepf{\Gamma} \subseteq I$ as $\stepf{\Gamma}$ is the minimal model of $\Gamma$.
  Thus, as $I \subseteq \stepf{\Gamma}$ and $\stepf{\Gamma} \subseteq I$ we have that $I = \stepf{\Gamma}$ which shows that $I$ is a stable model of $\Pi$.
\end{proof}

Combining the two lemmata, we obtain the main theorem of this chapter.

\begin{theorem}[Well-Support Semantics for \mg]\label{thm:sm-wsm}
  Let $\Pi$ be a \mg program.
  A set of precomputed atoms $I$ is a stable model of $\Pi$ iff it is a well-supported model of $\Pi$.
\end{theorem}

Thus, well-supported models give us an alternative to the reduct based semantics for defining stable models of \mg programs.
Importantly, they give us direct access to the derivation order~$\prec$.

\chapter{Completions for \texorpdfstring{\mg}{mini-gringo}}
\label{chp:completion}

In this chapter, we introduce ordered completion for \mg.
First, we review the ordinary completion for \mg in \cref{sec:completion}.
We then define ordered completion by modifying the normal completion in \cref{sec:oc}.
But this first version of ordered completion has a shortcoming: it only captures finite stable models.
We then introduce another modification, the ordered completion with level mapping, in \cref{sec:oc-lvl}, that also captures infinite models.
Finally, we finish the chapter by considering a simplification of the definition of ordered completion that produces simpler formulas in \cref{sec:oc-simp}.

\section{Review: Completion for \texorpdfstring{\mg}{mini-gringo}}
\label{sec:completion}

This definition of completion follows the definitions in the current literature on completion for \mg~\cite{FandinnoEtAl2024a,FandinnoEtAl2023,LifschitzEtAl2020,FandinnoEtAl2020,HarrisonEtAl2017a}.
Some notation is slightly altered to provide a better basis for defining ordered completion in the next section.

\begin{definition}[Completion of a Predicate]
  Let $\Pi$ be a \mg program.
  The completion of a predicate $p \in \lang{\Pi}$ is the following formula
  \begin{equation*}
    \forall \xs (p(\xs) \equi \bigvee_{\twosub{R \in \Pi}{\headlit{R} = p(\terms)}} (\exists \ys\, \form{R}{\xs})),
  \end{equation*}
  where $\xs$ are fresh program variables that match the arity of $p$ and $\ys$ are the free variables in $\form{R}{\xs}$.
\end{definition}

Alternatively, we can also split up the equivalences of the completion into two implications:

\begin{definition}[Completion in Parts]
  Let $\Pi$ be a \mg program.
  The completion of a predicate $p \in \lang{\Pi}$ is the conjunction of the following two formulas.
  First, are the rules of the program
  \begin{equation*}
    \comprule{p} = \forall\xs (\bigvee_{\twosub{R \in \Pi}{\headlit{R} = p(\terms)}} (\exists \ys\, \form{R}{\xs}) \imp p(\xs)),
  \end{equation*}
  where $\xs$ are fresh program variables that match the arity of $p$ and $\ys$ are the free variables in $\form{R}{\xs}$.
  Second, the part added by the completion
  \begin{equation*}
    \compdef{p} = \forall\xs (p(\xs) \imp \bigvee_{\twosub{R \in \Pi}{\headlit{R} = p(\terms)}} (\exists \ys\, \form{R}{\xs})),
  \end{equation*}
  where $\xs$ are fresh program variables that match the arity of $p$ and $\ys$ are the free variables in $\form{R}{\xs}$.
\end{definition}

The formula $\comprule{p}$ ensures that all the rules of $p$ in $\Pi$ are satisfied.
The addition of the formula $\compdef{p}$ makes sure that the rules are treated as the definition of the predicate: $p$ can only be true if its truth is supported by one of its rules.
Note that this provides us with a link to supported interpretations.
A set of precomputed atoms $I$ is a supported interpretation of $\Pi$ if and only if it satisfies the formulas $\compdef{p}$ for every $p$ in $\Pi$.

For the constraints in a program, we define the following formula.

\begin{definition}[Constraints]
  For a \mg program $\Pi$, we define the formula $\cons{\Pi}$ as
  \[
    \bigwedge_{\twosub{R \in \Pi}{\headlit{R} = \bot}} \forall \xs\, \lnot \taub(\body{R}),
  \]
  where $\xs$ are the free variables in $\taub(\body{R})$.
\end{definition}

Combining the formulas $\comprule{p}$ and the formula $\cons{\Pi}$ we essentially get the theory $\tau^{*}(\Pi)$.
For any constraint $R$ we know that $\tau^{*}(R)$ is the formula $\forall \xs (\lnot \taub{\body{R}})$ which is one of the formulas in the conjunction $\cons{\Pi}$.
The conjunction over all $\tau^{*}(R)$ for all basic and choice rules $R$ is exactly the formula $\bigwedge_{p \in \lang{\Pi}} \comprule{p}$ after applying the following simplification:
\[
  \forall \xs (F \imp \xs) \land \forall \xs (G \imp p(\xs)) \equi \forall \xs (F \lor G \imp p(\xs)),
\]
note that this is an equivalent transformation.

We state this relationship in the following lemma and also note the relationship to $\tau(\Pi)$.

\begin{lemma}\label{lem:tau-comprule}
  Let $\Pi$ be a \mg program and let $I$ be a set of precomputed atoms.
  We then have the following equivalent statements:
  \begin{enumerate}
    \item $I \models \tau(\Pi)$,
    \item $\ext{I}{\sigsort} \models \tau^{*}(\Pi)$,
    \item $\ext{I}{\sigsort} \models \cons{\Pi} \land \bigwedge_{p \in \lang{\Pi}} \comprule{p}$.
  \end{enumerate}
\end{lemma}

The equivalence of (1) and (2) is exactly \cref{lem:tau-tau*}.
The equivalence between (2) and (3) is justified by the above explanation.

Combining the above definitions, we can define the completion of a \mg program.

\begin{definition}[Completion]
  The completion of a \mg program $\Pi$, denoted by $\comp{\Pi}$, is the following formula
  \[
    \cons{\Pi} \land \bigwedge_{p \in \lang{\Pi}} \comprule{p} \land \compdef{p}.
  \]
\end{definition}

Every stable model of a program is then also a model of its completion as stated in the following lemma.
This is shown, for example, in~\cite{HarrisonEtAl2017a,LifschitzEtAl2020}.
Note that the completions defined in~\cite{HarrisonEtAl2017a,LifschitzEtAl2020} are slightly different from the one given here, but they are almost the same syntactically and semantically equivalent in classical logic.

\begin{lemma}\label{lem:sm-comp}
  Let $\Pi$ be a \mg program and let $I$ be a set of precomputed atoms.
  If $I$ is a stable model of $\Pi$, then $\ext{I}{\sigsort}$ is a model of its completion~$\comp{\Pi}$.
\end{lemma}

But not every model of the completion is a stable model.
This only holds for the case of tight logic programs, as stated in the following lemma.
A logic program is \emph{tight} if it does not contain positive recursion.
For more details see~\cite{ErdemLifschitz2003,HarrisonEtAl2017a,LifschitzEtAl2020}.

\begin{lemma}
  Let $\Pi$ be a \emph{tight} \mg program and let $I$ be a set of precomputed atoms.
  $I$ is a stable model of $\Pi$ if and only if $\ext{I}{\sigsort}$ is a model of its completion~$\comp{\Pi}$.
\end{lemma}

This is shown in~\cite{HarrisonEtAl2017a}.
\cite{FandinnoEtAl2024a}~proves a more general result for io-programs: \mg programs with input and output.
This more general result shows that the exact correspondence holds for \emph{locally-tight} io-programs.

Before turning to the definition of ordered completion in the next section, we want to consider an example that will highlight the issue that ordered completion needs to solve.
In \cref{chp:introduction} we already considered a simple non-tight program for which the completion had undesirable models.
We now want to discuss the example of a transitive closure program as in \cref{eq:transitive}.

\begin{example}\label{ex:transitive-completion}
  The following \mg program $\Pi_{TC}$ computes the transitive closure of a graph given by its edges
  \[
    t(X,Y) \revimp e(X,Y), \qquad t(X,Y) \revimp e(X,Z) \land t(Z,Y).
  \]

  Here we want to consider a simple finite graph containing three nodes $a_{1}, a_{2}$ and $b$.
  The nodes $a_{1}$ and $a_{2}$ are connected to each other while $b$ does not have any edges.
  We add the following corresponding facts to our logic program from above
  \[
    e(a_{1},a_{2}), \qquad e(a_{2},a_{1}).
  \]

  The only stable model of our program is the following set (where we omit the facts of the program)
  \[
    \{t(a_{1},a_{1}), t(a_{1},a_{2}), t(a_{2},a_{2}), t(a_{2},a_{1})\}.
  \]
  This (together with the facts) is also a model of the completion $\comp{\Pi_{TC}}$.
  However, the completion also has a second model that additionally contains the following atoms: $t(a_{1},b)$ and $t(a_{2},b)$.
  The fact that there is a model of the completion that is not a stable model is, of course, not surprising as our program $\Pi_{TC}$ is not (locally) tight.

  We take a closer look at why $t(a_{1},b)$ and $t(a_{2},b)$ can be true in a model of the completion.
  It is clear that even with those additional atoms, the rules of the program are still satisfied, i.e., the formula $\comprule{t}$ holds.
  More interesting is the formula $\compdef{t}$:
  \[
    \compdef{t} = \forall X Y (t(X,Y) \imp e(X,Y) \lor \exists Z (e(X,Z) \land t(Z,Y))).
  \]
  When this formula is instantiated by setting $X = a_{1}$ and $Y = b$ either $e(a_{1},b)$ or the formula
  \[
    \exists Z (e(a_{1},Z) \land t(Z,b))
  \]
  needs to hold.
  The former is obviously false.
  The latter, however, is true: we can choose $Z = a_{2}$, both $e(a_{1},a_{2})$ and $t(a_{2},b)$ are true.
  In the same way the formula $\compdef{t}$ is satisfied for $X = a_{2}$ and $Y = b$ by setting $Z = a_{1}$.
  But this is exactly the problem of the completion: the truth of $t(a_{1},b)$ and $t(a_{2},b)$ is justified circularly.
  $t(a_{1},b)$ is true because of $t(a_{2},b)$ and vice versa.
\end{example}

From this example, we can conclude the following.
While completion ensured that there is some justification for each true atom (due to the formula $\comprule{p}$), it fails to guarantee that this justification is not circular.
However, for a (locally) tight program, there can be no circular justification as the rules of the program do not contain any circularity.

\section{Ordered Completion}
\label{sec:oc}

Next, we define the ordered completion for \mg programs.
The key idea is that we modify the $\compdef{p}$ formulas to include newly introduced comparison predicates $<_{pq}$ to talk about the derivation order of atoms.
In this way, we include the idea of well-supported models into the formulas of the ordered completion and ensure an equivalence to stable models even for non-tight programs.

Compared to the definition in~\cite{AsuncionEtAl2012}, there are two main differences in our definition below.
First, we generalise their definition to the more complex language of \mg.

Second, the comparison predicates we use for encoding the derivation order are slightly different.
We discuss this difference below.

\subsection{Comparison Predicates and their Axioms}
To express restrictions on the derivation order of atoms, we have to introduce new comparison predicates for every existing pair of predicates in our program.

\begin{definition}[Comparison Predicates]
  For each pair of predicates $p$ and $q$, we introduce a new predicate $<_{pq}$ whose arity is the sum of arities of $p$ and $q$.
  In the following, we use infix notation for the comparison predicates, i.e., we write $<_{pq}(\xs, \ys)$ as $\xs <_{pq} \ys$.
\end{definition}

These newly introduced predicates result in a different logical signature used for expressing the ordered completion.
Let $\sigord$ denote the extension of the signature $\sigsort$ by the new predicates $<_{pq}$ as above.
We introduce some notations for moving between the signatures $\sigord$ and $\sigsort$.

\begin{definition}[Extensions and Reducts on Signatures]\label{def:ext-red}
  If $J$ is an interpretation of the signature $\sigord$, then $\red{J}{\sigsort}$ is the interpretation of signature $\sigsort$ that is obtained by removing the interpretations of the new predicates $<_{pq}$.
  We call an interpretation $J$ of $\sigord$ an \emph{extension} of an interpretation $I$ of $\sigsort$ if $\red{J}{\sigsort} = I$, $I$~is then called the \emph{reduct} of $J$ to the signature $\sigsort$.

  For a set $I$ of precomputed terms, the interpretation $\ext{I}{\sigord}$ denotes an extension of the standard interpretation $\ext{I}{\sigsort}$ to the signature $\sigord$.

  Let $I$ be an interpretation over the signature $\sigord$.
  We denote by $\at{I}$ the set of precomputed atoms $J$ that are true in $I$.
  I.e., we have that the reduct of $I$ to $\sigsort$ is exactly the standard interpretation $\ext{J}{\sigsort}$.
\end{definition}

Note that a set of precomputed atoms $I$ can be uniquely extended to a standard interpretation of the signature $\sigsort$ denoted by $\ext{I}{\sigsort}$.
On the other hand, the extension $\ext{I}{\sigord}$ to the signature $\sigord$ is not unique.
There are many ways in which we can extend $\ext{I}{\sigsort}$ to an interpretation of $\sigord$.

To ensure that the comparison predicates $<_{pq}$ encode a sensible derivation order, we have to add some helper axioms expressing the irreflexivity and transitivity of the encoded relation.

\begin{definition}[Irreflexivity Axioms]
  For a \mg program $\Pi$, we define the formula $\irref{\Pi}$ as follows
  \[
    \bigwedge_{p \in \lang{\Pi}} \forall \xs \lnot (\xs <_{pp} \xs),
  \]
  where $\xs$ are fresh program variables matching the arity of $p$.
\end{definition}

\begin{definition}[Transitivity Axioms]
  For a \mg program $\Pi$, we define the formula $\trans{\Pi}$ as follows
  \[
    \bigwedge_{p,q,r \in \lang{\Pi}} \forall \xs\ys\zs (\xs <_{pq} \ys \land \ys <_{qr} \zs \imp \xs <_{pr} \zs),
  \]
  where $\xs,\ys,\zs$ are fresh program variables matching the arity of $p,q,r$ respectively.
\end{definition}

In~\cite{AsuncionEtAl2012} the comparison predicates are defined as $\leq_{pq}$ instead of $<_{pq}$.
They also add transitivity axioms but do not add irreflexivity axioms.
However, in the formulas of the ordered completion, they always use subformulas $\leq_{pq} \land \lnot\!\leq_{qp}$, i.e., state that the relation is asymmetric.
As discussed in \cref{chp:support}, for a transitive relation irreflexivity and asymmetry are equivalent.
Thus, we do not have to state the asymmetry in our formulas of the ordered completion but instead use irreflexivity axioms.
We use the $<_{pq}$ notation here as it more closely resembles the order from the definition of well-supported models (\cref{def:wsm}).

We consider two example programs that will act as our running examples for the definition of the ordered completion in the following.

\begin{example}\label{ex:oc-basic}
  Let $\Pi_{1}$ be the following logic program
  \[
    p(A) \revimp q(A-1).
  \]
  For $\Pi_{1}$ we introduce the following four predicates
  \[
    <_{pp}/2 \quad <_{pq}/2 \quad <_{qp}/2 \quad <_{qq}/2.
  \]
\end{example}

\begin{example}\label{ex:oc-choice}
  Let $\Pi_{2}$ be the following logic program
  \[
    \{p(A)\} \revimp r(A,B).
  \]
  For $\Pi_{2}$ we introduce the following four predicates
  \[
    <_{pp}/2 \quad <_{pr}/3 \quad <_{rp}/3 \quad <_{rr}/4.
  \]
\end{example}

Using these newly introduced comparison predicates, we can then express in the formulas of the ordered completion that any positive atom $q$ that is used to justify $p$ has to be derived earlier, with respect to the relation encoded by the comparison predicates, by adding the atom $<_{qp}$.

\subsection{Definition of Ordered Completion}
\label{sec:oc-def}

We now turn to formally defining the ordered completion of a predicate using the derivation order predicates.

\begin{definition}[Ordered Completion of a Predicate]
  Given a \mg program $\Pi$, the ordered completion of a predicate $p \in \lang{\Pi}$ are the following two formulas.
  The first is the same as in the ordinary completion
  \begin{equation*}
    \comprule{p} = \forall \xs (\bigvee_{\twosub{R \in \Pi}{\headlit{R} = p(\terms)}} (\exists \ys\, \form{R}{\xs}) \imp p(\xs))
  \end{equation*}
  where $\xs$ are fresh program variables that match the arity of $p$ and $\ys$ are the free variables in $\form{R}{\xs}$.
  The second formula is where ordered completion differs
  \begin{equation*}
    \ocomp{p} =
    \forall \xs (p(\xs) \imp \bigvee_{\twosub{R \in \Pi}{\headlit{R} = p(\terms)}} \exists \ys (\form{R}{\xs} \land {}
    \bigwedge_{\nospace{q(\terms^{\prime}) \in \bodylitp{R}}} \exists \zs (\val{\terms^{\prime}}{\zs} \land q(\zs) \land \zs <_{qp} \xs)))
  \end{equation*}
  where $\xs$ and $\zs$ are fresh program variables that match the arity of $p$ and $q$, respectively, and $\ys$ are the free variables in $\form{R}{\xs}$.
  Note that free variables in $\terms^{\prime}$ are already free variables in $\form{R}{\xs}$ and thus already quantified.
\end{definition}

For a predicate $p$, we have the following relationship: $\ocomp{p} \models \compdef{p}$.
We return to our two example programs from above.

\begin{continuedexample}{ex:oc-basic}
  Continuing with $\Pi_{1}$ (i.e., $p(A) \revimp q(A-1)$) as defined above.
  First we compute $\comprule{p}$ and simplify it:
  \begin{alignat*}{3}
    \comprule{p}
    &= \forall X (\exists A (&&\val{A}{X} \land \taub(q(A-1))) \imp p(X)) \\
    &= \forall X (\exists A (&&X = A \land \exists V (\val{A-1}{V} \land q(V))) \imp p(X)) \\
    &= \forall X (\exists A (&&X = A \land {} \\
    &            &&\exists V (\exists I,J (I = A \land J = 1 \land V = I - J) \land q(V))) \imp p(X)) \\
    &= \forall X (\exists A (&&X = A \land \exists V (V = A - 1 \land q(V))) \imp p(X)) \\
    &= \forall A (q(A  && - 1) \imp p(A)).
  \end{alignat*}

  Next we compute $\ocomp{p}$ and simplify:
  \begin{alignat*}{2}
    \ocomp{p}
    &= \forall X (p(X) \imp {} \exists A (&&\val{A}{X} \land \taub(q(A-1)) \land {} \\
    &                         &&\exists Z (\val{A-1}{Z} \land q(Z) \land Z <_{qp} X))) \\
    &= \forall X (p(X) \imp {} \exists A (&&\val{A}{X} \land \exists V (\val{A-1}{V} \land q(V)) \land {} \\
    &                         &&\exists Z (\val{A-1}{Z} \land q(Z) \land Z <_{qp} X))) \\
    &= \forall X (p(X) \imp {} \exists A (&&X = A \land \exists V (V = A - 1 \land q(V)) \land {} \\
    &                         &&\exists Z (Z = A - 1 \land q(Z) \land Z <_{qp} X))) \\
    &= \forall A (p(A) \imp {} q(A  && - 1) \land A-1 <_{qp} A).
  \end{alignat*}
\end{continuedexample}

\begin{continuedexample}{ex:oc-choice}
  Continuing with $\Pi_{2}$ (i.e., $\{p(A)\} \revimp r(A,B)$) as defined above.
  First $\comprule{p}$ is the following formula:
  \begin{alignat*}{2}
    \comprule{p}
    &= \forall X (\exists A,B ( &&\val{A}{X} \land \taub(r(A,B)) \land \lnot \lnot p(X)) \imp p(X)) \\
    &= \forall X (\exists A,B ( &&X = A \land \exists V_{1},V_{2} (V_{1} = A \land V_{2} = B \land r(V_{1},V_{2})) \land {} \\
    &               &&\lnot \lnot p(X)) \imp p(X)) \\
    &= \forall X (\exists A,B ( &&X = A \land r(A,B) \land \lnot \lnot p(X)) \imp p(X)) \\
    &= \forall A (\exists B (r(A&&,B) \land \lnot \lnot p(A)) \imp p(A)).
  \end{alignat*}

  Next we compute $\ocomp{p}$:
  \begin{alignat*}{2}
    \ocomp{p}
    &= \forall X (p(X) \imp {} \exists A,B ( &&\val{A}{X} \land \exists V_{1},V_{2} (\val{A,B}{V_{1},V_{2}} \land r(V_{1},V_{2})) \land {} \\
    &                            &&\lnot \lnot p(X) \land \exists Z_{1},Z_{2} (\val{A,B}{Z_{1},Z_{2}} \land r(Z_{1},Z_{2}) \land {} \\
    &                            &&(Z_{1},Z_{2}) <_{rp} X))) \\
    &= \forall X (p(X) \imp {} \exists A,B ( &&X = A \land \exists V_{1},V_{2} (V_{1} = A \land V_{2} = B \land r(V_{1},V_{2})) \land \\
    &                            &&\lnot \lnot p(X) \land \exists Z_{1},Z_{2} (Z_{1} = A \land Z_{2} = B \land {} \\
    &                            &&r(Z_{1},Z_{2}) \land (Z_{1},Z_{2}) <_{rp} X))) \\
    &= \forall X (p(X) \imp {} \exists A,B ( &&X = A \land r(A,B) \land \lnot \lnot p(X) \land (A,B) <_{rp} X)) \\
    &= \forall A (p(A) \imp {} \exists B (r(A&&,B) \land \lnot \lnot p(A) \land (A,B) <_{rp} A)).
  \end{alignat*}
\end{continuedexample}

\begin{example}
  Now let $\Pi_{3} = \Pi_{1} \cup \Pi_{2}$, i.e., we combine the programs from \cref{ex:oc-basic,ex:oc-choice}.
  We get the following completions for $p$:
  \begin{alignat*}{2}
    \comprule{p}
    &= \forall X (&\exists A,B (&X = A \land \exists V_{1},V_{2} (V_{1} = A \land V_{2} = B \land r(V_{1},V_{2})) \land {} \\
                      & &&\lnot \lnot p(X)) \lor {} \\
    &       &\exists A (  &X = A \land \exists V (V = A - 1 \land q(V))) \imp p(X)),
  \end{alignat*}
  \vspace{-0.7cm}
  \begin{alignat*}{2}
    \ocomp{p}
    &= \forall X (p(X) \imp {}&\exists A,B (&X = A \land \exists V_{1},V_{2} (V_{1} = A \land V_{2} = B \land r(V_{1},V_{2})) \land {} \\
    &                   &       &\lnot \lnot p(X) \land \exists Z_{1},Z_{2} (Z_{1} = A \land Z_{2} = B \land {} \\
    &                   &       &r(Z_{1},Z_{2}) \land (Z_{1},Z_{2}) <_{rp} X)) \lor {} \\
    &                   &\exists A (  &X = A \land \exists V (V = A - 1 \land q(V)) \land {} \\
    &                   &       &\exists Z (Z = A - 1 \land q(Z) \land Z <_{qp} X))).
  \end{alignat*}
\end{example}

Combining the ordered completion of each predicate with the constraint formula and the helper axioms, we can define the ordered completion of \mg programs.

\begin{definition}[Ordered Completion]\label{def:oc}
  The ordered completion of a \mg program $\Pi$, denoted by $\oc{\Pi}$, is the following formula
  \[
    \cons{\Pi} \land \ocax{\Pi} \land \bigwedge_{p \in \lang{\Pi}} \comprule{p} \land \ocomp{p}.
  \]
\end{definition}

For a logic program $\Pi$, we have the following relationship to the ordinary completion: $\oc{\Pi} \models \comp{\Pi}$.

\subsection{The Ordered Completion Theorem}
\label{sec:oc-theorem}

Before considering the relationship between ordered completion models and stable models formally, we want to give some intuition about it.
In the following, we will not directly prove the equivalence of ordered completion models and stable models, but rather we will show that models of the ordered completion are equivalent to well-supported models.
By combining this equivalence with \cref{thm:sm-wsm}, we will obtain the equivalence to stable models as well.

The reason for proving equivalence to well-supported models is that both ordered completion and well-supported models use the same concepts.
A set of precomputed atoms $I$ is a well-supported model of a \mg program $\Pi$ if it: (1) is a model of the program (or rather its formula representation), and (2) it is a well-supported interpretation of $\Pi$.
First, as shown before in \cref{lem:tau-comprule}, the formula
\[
  \cons{\Pi} \land \bigwedge_{p \in \lang{\Pi}} \comprule{p},
\]
is equivalent to $\tau^{*}(\Pi)$.
Thus, the above formula, which is contained in $\oc{\Pi}$, ensures that we have a model of the program.

Second, the formula $\bigwedge_{p \in \lang{\Pi}} \ocomp{p}$ ensures that we have a well-supported interpretation.
We recall that if $I$ is a well-supported interpretation of $\Pi$, then for every $p(\rs)$ that is true in $I$ we need to have a supporting rule whose head matches $p(\rs)$, whose body is satisfied, and all positive body literals $q(\rs^{\prime})$ need to be derived before $p(\rs)$.
The formula $\ocomp{p}$ expresses exactly that:
\[
  \ocomp{p} =
    \forall \xs (p(\xs) \imp \bigvee_{\twosub{R \in \Pi}{\headlit{R} = p(\terms)}} \exists \ys (\form{R}{\xs} \land {}
    \bigwedge_{\nospace{q(\terms^{\prime}) \in \bodylitp{R}}} \exists \zs (\val{\terms^{\prime}}{\zs} \land q(\zs) \land \zs <_{qp} \xs))).
\]
If $p(\rs)$ is true then we need a rule $R$ whose head matches $p(\rs)$ (the head literal is $q(\terms)$ and $\form{R}{\xs}$ includes the formula $\val{\terms}{\rs}$) such that the body of the rule is satisfied ($\form{R}{\xs}$ includes $\tau^{*}(\body{R})$) and any positive body literal of $R$ is derived before $p(\rs)$ ($\val{\terms^{\prime}}{\zs} \land q(\zs) \land \zs <_{qp} \xs$, where the formula~$\ocax{\Pi}$ ensures that the $<_{qp}$ predicates encode a sensible order).
So conceptually, ordered completion and well-supported models use the same idea.
The difference is that ordered completion encodes this idea as a first-order formula, whereas the well-supported semantics only states this idea on a ``meta-level''.

However, there is one other important difference between the two definitions: well-supported models make use of the $\tau$ transformation, whereas ordered completion uses the $\tau^{*}$ transformation.
I.e., well-support deals with ground rules while ordered completion works on rules with variables.
To overcome this difference we make use of the quantification in $\ocomp{p}$: the disjunction over all the rules matching $p(\rs)$ includes the quantification $\exists \ys$ for each rule $R$, where $\ys$ are the free variables in $\form{R}{\xs}$, i.e., essentially the free variables in $R$.

From the perspective of well-supported models, we have an instance $R^{\prime}$ of some rule $R \in \Pi$.
This instantiation tells us how to replace the variables in $R$ to obtain a ground rule with the properties we need.
But this is exactly what the meaning of the existentially quantified formula $\exists \ys \form{R}{\xs}$ is: we need to know how to instantiate the free variables of $R$ in order to obtain a ground instance of $R$ that is satisfied.

These ideas (i.e., (1) that $\ocomp{p}$ encodes the well-support of $p$, and (2) the transition from ground rules to rules with variables) are the key concepts to prove the equivalence of well-supported and ordered completion models.
We now turn towards expressing this formally.
We start by proving that well-supported models are models of the ordered completion.
Before we do so, we have to consider how to construct an interpretation of the signature $\sigord$ from a stable model.

\begin{remark}\label{rem:wsm-sigord}
  Let $\Pi$ be a \mg program and let $I$ be a set of precomputed atoms such that $I$ is a well-supported model (with order $\prec$) of $\Pi$.
  Then we can extend $\ext{I}{\sigsort}$ to an interpretation $J$ of signature $\sigord$ as follows: for any $p(\terms) \in I$ and $q(\terms^{\prime}) \in I$ if $p(\terms) \prec q(\terms^{\prime})$ we set $\terms <_{pq} \terms^{\prime}$ to true in $J$, for all other values we set $<_{pq}$ to false.
  As we only add predicates of the signature $\sigord$ to $J$ that do not belong to $\sigsort$ we have that $\red{J}{\sigsort} = \ext{I}{\sigsort}$.
\end{remark}

An interpretation of $\sigord$ constructed in this way satisfied the helper axioms expressing irreflexivity and transitivity of the relation encoded by the order predicates.

\begin{lemma}\label{lem:wsm-sigord}
  Let $\Pi$ be a \mg program and let $I$ be a set of precomputed atoms such that $I$ is a well-supported model (with order $\prec$) of $\Pi$.
  We construct the interpretation $J$ of the signature $\sigord$ according to \cref{rem:wsm-sigord}.
  Then ${J \models \ocax{\Pi}}$.
\end{lemma}

This follows by construction of $J$ as $\prec$ is irreflexive and transitive.
Next, we can prove that stable models of a program $\Pi$ can be extended to a model of its ordered completion $\oc{\Pi}$.

\begin{lemma}\label{lem:sm-oc}
  Let $\Pi$ be a \mg program and let $I$ be a set of precomputed atoms.
  If $I$ is a stable model of $\Pi$ then there is an extension $J$ of $\ext{I}{\sigsort}$ such that $\red{J}{\sigsort} = \ext{I}{\sigsort}$ and $J$ is a model of $\oc{\Pi}$.
\end{lemma}

\begin{proof}
  Let $I$ be a stable model of $\Pi$.
  Then $I$ is a well-supported model of $\Pi$, and let $\prec$ be the order establishing the well-supportedness.

  We construct a model $J$ of $\oc{\Pi}$ that fulfils the condition $\red{J}{\sigsort} = \ext{I}{\sigsort}$.
  We do so according to \cref{rem:wsm-sigord}.

  It remains to show that $J$ constructed in this way is a model of $\oc{\Pi}$.
  Recall that
  \[
    \oc{\Pi} = \cons{\Pi} \land \ocax{\Pi} \land \bigwedge_{p \in \lang{\Pi}} \comprule{p} \land \ocomp{p}.
  \]
  As $I$ is a stable model of $\Pi$, $J$ is obviously a model of $\cons{\Pi}$.
  By \cref{lem:wsm-sigord} $J$ is a model of $\ocax{\Pi}$.
  The conjunction $\bigwedge_{p \in \lang{\Pi}} \comprule{p}$ is satisfied by $J$ as $\ext{I}{\sigsort}$ is a model of $\comp{\Pi}$ (by \cref{lem:sm-comp}).

  It remains to show that $J$ is a model of $\bigwedge_{p \in \lang{\Pi}} \ocomp{p}$.
  Let $p$ be an arbitrary predicate in $\lang{\Pi}$.
  We check whether
  \[
    \ocomp{p} = \forall \xs (p(\xs) \imp \bigvee_{\twosub{R \in \Pi}{\headlit{R} = p(\terms)}} \exists \ys (\form{R}{\xs} \land \bigwedge_{\nospace{q(\terms^{\prime}) \in \bodylitp{R}}} \exists \zs (\val{\terms^{\prime}}{\zs} \land q(\zs) \land \zs <_{qp} \xs)))
  \]
  is satisfied.
  If $p(\rs) \not\in I$ for some $\rs$, then $\ocomp{p}$ is trivially satisfied.
  Otherwise, we assume $p(\rs) \in I$.
  We know that there is a instance $R^{\prime}$ of a rule $R \in \Pi$ by the definition of well-supported models (\cref{def:wsm}) such that
  \begin{assumptions}
    \item\label{itm:sm-oc-head} $\head{R^{\prime}} = p(\terms)$ and $\rs \in \values{\terms}$,
    \item\label{itm:sm-oc-body} $I \models \tau(\body{R^{\prime}})$,
    \item\label{itm:sm-oc-ord} for all $q(\terms^{\prime}) \in \bodylitp{R^{\prime}}$, there is some $\rs^{\prime} \in \values{\terms^{\prime}}$ with $q(\rs^{\prime}) \in I$ and $q(\rs^{\prime}) \prec p(\rs)$.
  \end{assumptions}
  We show that the disjunct
  \[
    \exists \ys (\form{R}{\xs} \land \bigwedge_{\nospace{q(\terms^{\prime}) \in \bodylitp{R}}} \exists \zs (\val{\terms^{\prime}}{\zs} \land q(\zs) \land \zs <_{qp} \rs))
  \]
  holds for this specific $R$.
  We set the variables $\ys$ according to the instantiation of $R$ given by $R^{\prime}$.
  Therefore, in the following, the quantification $\exists \ys$ is dropped and $R$ is replaced by $R^{\prime}$.
  Note that also the variables in $\terms^{\prime}$ are instantiated according to $R^{\prime}$.
  Then our remaining formula is
  \[
    \form{R^{\prime}}{\xs} \land \bigwedge_{\nospace{q(\terms^{\prime}) \in \bodylitp{R^{\prime}}}} \exists \zs (\val{\terms^{\prime}}{\zs} \land q(\zs) \land \zs <_{qp} \rs).
  \]

  First, we consider the case that $R^{\prime}$ is a basic rule.
  Then we have that $\form{R^{\prime}}{\xs} = \val{\terms}{\rs} \land \taub(\body{R^{\prime}})$.
  By~\cref{itm:sm-oc-head} we have $\rs \in \values{\terms}$ and thus $\val{\terms}{\rs}$ is satisfied (as it is equivalent to $\true$ by \cref{lem:val-values}).
  $\taub(\body{R^{\prime}})$ is satisfied as \cref{itm:sm-oc-body} holds which is equivalent to $\ext{I}{\sigsort} \models \taub(\body{R^{\prime}})$ by \cref{lem:tau-taub}.

  Second, we consider the case that $R^{\prime}$ is a choice rule.
  Then $\form{R^{\prime}}{\xs}$ is the formula $\val{\terms}{\rs} \land \taub(\body{R^{\prime}}) \land \lnot \lnot p(\rs)$.
  The first two conjuncts are satisfied by the same reasoning as the above case of $R^{\prime}$ being a basic rule.
  As $p(\rs) \in I$, $\lnot \lnot p(\rs)$ is also satisfied.

  It remains to show that
  \[
    \bigwedge_{\nospace{q(\terms^{\prime}) \in \bodylitp{R^{\prime}}}} \exists \zs (\val{\terms^{\prime}}{\zs} \land q(\zs) \land \zs <_{qp} \rs)
  \]
  is satisfied.
  For any $q(\terms^{\prime}) \in \bodylitp{R^{\prime}}$ we have a $q(\rs^{\prime}) \in I$ for some $\rs^{\prime} \in \values{\terms^{\prime}}$ and $q(\rs^{\prime}) \prec p(\rs)$ by \cref{itm:sm-oc-ord}.
  We instantiate $\zs$ with $\rs^{\prime}$.
  As $\rs^{\prime} \in \values{\terms^{\prime}}$ the formula $\val{\terms^{\prime}}{\rs^{\prime}}$ is satisfied by \cref{lem:val-values}.
  $q(\rs^{\prime})$ holds as $q(\rs^{\prime}) \in I$.
  As we have $q(\rs^{\prime}) \prec p(\rs)$ we also have that $\rs^{\prime} <_{qp} \rs$ holds in $J$ by construction of $J$.

  Thus, we have shown that for an arbitrary $p(\rs) \in I$ the instance of $\ocomp{p}$ with $\xs$ instantiated to $\rs$ is satisfied by $J$.
  In other words, $\ocomp{p}$ is satisfied by~$J$.
  With that, we have shown that every conjunct of $\oc{\Pi}$ is satisfied by~$J$.
  I.e., that $J$ is a model of $\oc{\Pi}$ and by construction of $J$ (\cref{rem:wsm-sigord}) we have $\red{J}{\sigsort} = \ext{I}{\sigsort}$.
\end{proof}

Before proving the other direction, we show how to construct a strict well-founded partial order $\prec$ from a model of the ordered completion.

\begin{remark}\label{rem:oc-ord}
  A model $I$ of the ordered completion $\oc{\Pi}$ of a \mg program $\Pi$ encodes an order $\prec$ on the set $\at{I}$ of precomputed atoms that are true in $I$.
  For atoms $p(\rs) \in \at{I}$ and $q(\rs^{\prime}) \in \at{I}$ we set $p(\rs) \prec q(\rs^{\prime})$ iff $\rs <_{pq} \rs^{\prime}$ holds in $I$.
\end{remark}

The order $\prec$ constructed in this way is indeed a strict partial order.

\begin{lemma}\label{lem:oc-ord-spo}
  Let $\Pi$ be a \mg program, $J$ be a model of $\oc{\Pi}$ and ${I = \at{J}}$ be the set of precomputed atoms true in $J$.
  The relation $\prec$ on $I$ as defined in \cref{rem:oc-ord} is a strict partial order.
\end{lemma}

To show that $\prec$ is a strict partial order, we only need to verify that it is irreflexive and transitive.
As the helper axioms $\ocax{\Pi}$ are part of $\oc{\Pi}$, it is immediate that $\prec$ is a strict partial order.

\begin{lemma}\label{lem:oc-ord-wfo}
  Let $\Pi$ be a \mg program and $J$ be a model of $\oc{\Pi}$.
  Furthermore, we assume that the set of precomputed atoms $I = \at{J}$ that are true in $J$ is \emph{finite}.
  The relation $\prec$ on $I$ as defined in \cref{rem:oc-ord} is well-founded.
\end{lemma}

As $\prec$ is a strict partial order (\cref{lem:oc-ord-spo}) on a finite set, it is necessarily well-founded (see \cref{lem:fin-well-founded}).
Note that we here gain a restriction for our below lemma: we can only prove that models of the ordered completion in which a finite set of precomputed atoms is true will correspond to stable models.
For models of the ordered completion where an infinite set of precomputed atoms is true, we can not prove that the order $\prec$ constructed according to \cref{rem:oc-ord} is well-founded, and thus we can not obtain a well-supported model.
Next, we show that a model of the ordered completion in which only a finite set of precomputed atoms is true corresponds to a stable model.

\begin{lemma}\label{lem:oc-sm}
  Let $\Pi$ be a \mg program and let $I$ be a \emph{finite} set of precomputed atoms.
  Furthermore, let $J$ be an interpretation of the signature $\sigord$ such that $\red{J}{\sigsort} = \ext{I}{\sigsort}$.
  If $J$ is a model of $\oc{\Pi}$, then $I$ is a stable model of $\Pi$.
\end{lemma}

\begin{proof}
  Let $J$ be a model of $\oc{\Pi}$.
  We show that $I$ is a well-supported model of $\Pi$ and thus a stable model of $\Pi$.
  It is clear that $I$ is a model of $\tau(\Pi)$ (by \cref{lem:tau-comprule}), so it only remains to show that $I$ is a well-supported interpretation of $\Pi$.

  To do so, we first construct the order $\prec$ on $I$.
  We do this according to \cref{rem:oc-ord}.
  Then by \cref{lem:oc-ord-spo,lem:oc-ord-wfo} this is a strict well-founded partial order.

  It remains to show that for every atom $p(\rs) \in I$ there exists an instance $R^{\prime}$ of a rule $R \in \Pi$ such that
  \begin{assumptions}
    \item\label{itm:oc-sm-head} $\head{R^{\prime}} = p(\terms)$ and $\rs \in \values{\terms}$,
    \item\label{itm:oc-sm-body} $I \models \tau(\body{R^{\prime}})$,
    \item\label{itm:oc-sm-ord} for all $q(\terms^{\prime}) \in \bodylitp{R^{\prime}}$, for some $\rs^{\prime} \in \values{\terms^{\prime}}$ with $q(\rs^{\prime}) \in I$, $q(\rs^{\prime}) \prec p(\rs)$.
  \end{assumptions}

  As $J$ is a model of $\oc{\Pi}$ we know that the following formula (i.e., $\ocomp{p}$ where $\xs$ is instantiated to $\rs$ and simplified as $p(\rs)$ is true by assumption) holds
  \[
    \bigvee_{\twosub{R \in \Pi}{\headlit{R} = p(\terms)}} \exists \ys (\form{R}{\xs} \land \bigwedge_{\nospace{q(\terms^{\prime}) \in \bodylitp{R}}} \exists \zs (\val{\terms^{\prime}}{\zs} \land q(\zs) \land \zs <_{qp} \rs)).
  \]
  Thus, there is some $R \in \Pi$ whose head literal is $p(\terms)$ such that
  \[
    \exists \ys (\form{R}{\xs} \land \bigwedge_{\nospace{q(\terms^{\prime}) \in \bodylitp{R}}} \exists \zs (\val{\terms^{\prime}}{\zs} \land q(\zs) \land \zs <_{qp} \rs))
  \]
  is satisfied.
  We first consider the case that $R$ is a basic rule.
  Then the formula is the following
  \[
    \exists \ys (\val{\terms}{\rs} \land \taub(\body{R}) \land \bigwedge_{\nospace{q(\terms^{\prime}) \in \bodylitp{R}}} \exists \zs (\val{\terms^{\prime}}{\zs} \land q(\zs) \land \zs <_{qp} \rs)).
  \]
  Then, by the values of $\ys$ for which the above formula holds, we have an instantiation $R^{\prime}$ of the rule $R$.
  Plugging those values into the formula, we have
  \[
    \val{\terms}{\rs} \land \taub(\body{R^{\prime}}) \land \bigwedge_{\nospace{q(\terms^{\prime}) \in \bodylitp{R^{\prime}}}} \exists \zs (\val{\terms^{\prime}}{\zs} \land q(\zs) \land \zs <_{qp} \rs).
  \]
  As the first conjunct ($\val{\terms}{\rs}$) is satisfied we know that $\rs \in \values{\terms}$ holds by \cref{lem:val-values} and thus \cref{itm:oc-sm-head} is satisfied.
  \cref{itm:oc-sm-body} is satisfied by the second conjunct: $J \models \taub(\body{R^{\prime}})$ is equivalent to $\ext{I}{\sigsort} \models \taub(\body{R^{\prime}})$ (as the formula is over the signature $\sigsort$) which is equivalent to $I \models \tau(\body{R^{\prime}})$ by \cref{lem:tau-taub}.

  It remains to verify \cref{itm:oc-sm-ord}.
  Let $q(\terms^{\prime}) \in \bodylitp{R^{\prime}}$.
  Then we know that the following is true
  \[
    \exists \zs (\val{\terms^{\prime}}{\zs} \land q(\zs) \land \zs <_{qp} \rs).
  \]
  Now let $\rs^{\prime}$ be the instantiation of $\zs$ for which the inner formula holds.
  I.e., we have the following formula, which is satisfied by $J$:
  \[
    \val{\terms^{\prime}}{\rs^{\prime}} \land q(\rs^{\prime}) \land \rs^{\prime} <_{qp} \rs.
  \]
  With this we can verify \cref{itm:oc-sm-ord}: for $q(\terms^{\prime}) \in \bodylitp{R^{\prime}}$ we have an $\rs^{\prime} \in \values{\terms^{\prime}}$ (by \cref{lem:val-values} as $\val{\terms^{\prime}}{\rs^{\prime}}$ holds) and $q(\rs^{\prime}) \in I$ (as $q(\rs^{\prime})$ holds) and it indeed holds that $q(\rs^{\prime}) \prec p(\rs)$ by construction of $\prec$ as $\rs^{\prime} <_{qp} \rs$ holds in $J$.

  Thus, we have verified all conditions for a well-supported interpretation and have shown that $I$ is a well-supported model and, equivalently, also a stable model of $\Pi$.
\end{proof}

By \cref{lem:sm-oc,lem:oc-sm} we then obtain the ordered completion theorem for \mg.
Note that while \cref{lem:sm-oc} holds for arbitrary sets of precomputed atoms, \cref{lem:oc-sm} only holds for \emph{finite} sets of precomputed atoms.
Thus, the ordered completion theorem only holds for \emph{finite} sets of precomputed atoms.

\begin{theorem}[Ordered Completion for \mg]
  Let $\Pi$ be a \mg program and let $I$ be a \emph{finite} set of precomputed atoms.
  Then $I$ is a stable model of $\Pi$ if and only if there is an extension $J$ of $\ext{I}{\sigsort}$ such that $\red{J}{\sigsort} = \ext{I}{\sigsort}$ and $J$ is a model of $\oc{\Pi}$.
\end{theorem}

We will conclude this section by returning to the transitive closure example from the previous section (\cref{ex:transitive-completion}).

\begin{example}\label{ex:oc-trans-finite}
  Recall the \mg program $\Pi_{TC}$ defining the transitive closure of a graph:
  \[
    t(X,Y) \revimp e(X,Y), \qquad t(X,Y) \revimp e(X,Z) \land t(Z,Y).
  \]

  We again want to consider the graph with nodes $a_{1}$ and $a_{2}$ that are connected to each other and the unconnected node $b$.
  This graph is given by the following facts that we add to $\Pi_{TC}$
  \[
    e(a_{1},a_{2}), \qquad e(a_{2},a_{1}).
  \]

  Recall that the only stable model has the following transitivity edges
  \[
    t(a_{1},a_{1}), t(a_{1},a_{2}), t(a_{2},a_{2}), t(a_{2},a_{1})
  \]
  while the completion of $\Pi_{TC}$ has an additional model with the transitivity edges
  \[
    t(a_{1},b), t(a_{2},b).
  \]
  In the completion, this was possible as these two atoms justify themselves circularly (see \cref{ex:transitive-completion}).

  We now consider the ordered completion of $\Pi_{TC}$ (note that the following theory is already simplified):
  \begin{alignat*}{3}
    &\forall X Y (e(X,Y) \equi   &(X = {} &a_{1} \land Y = a_{2}) \lor (X = a_{2} \land Y = a_{1})) \\
    &\forall X Y (e(X,Y) \lor {} &\exists Z (e( &X,Z) \land t(Z,Y)) \imp t(X,Y)) \\
    &\forall X Y (t(X,Y) \imp    &( e(X,  &Y) \land (X,Y)  <_{et} (X,Y)) \lor {} \\
    &                      &\exists Z (   &e(X,Z) \land (X,Z) <_{et} (X,Y) \land {} \\
    &                      &        &t(Z,Y) \land (Z,Y) <_{tt} (X,Y))).
  \end{alignat*}
  As the $e/2$ predicate is given as facts, we can always satisfy $<_{et}$ predicates.
  (We can set the $e(X,Y)$ atoms as the smallest in our order as there is no order predicate of the form $<_{ee}$ or $<_{te}$.)
  This means we can drop the $<_{et}$ from the last formula.

  The atoms $t(a_{1},b)$ and $t(a_{2},b)$ satisfy the first two formulas.
  We instantiate the last formula of $\oc{\Pi_{TC}}$ by setting $X = a_{1}$ and $Y = b$:
  \[
    t(a_{1},b) \imp (e(a_{1},b) \lor \exists Z (e(a_{1},Z) \land t(Z,b) \land (Z,b) <_{tt} (a_{1},b))).
  \]
  Clearly $e(a_{1},b)$ does not hold.
  Thus, the only way for the formula to be true is if the formula
  \[
    \exists Z (e(a_{1},Z) \land t(Z,b) \land (Z,b) <_{tt} (a_{1},b))
  \]
  holds.
  The only value of $Z$ for which this is possible is $a_{2}$:
  \[
    e(a_{1},a_{2}) \land t(a_{2},b) \land (a_{2},b) <_{tt} (a_{1},b).
  \]
  This formula is satisfied if $t(a_{2},b)$ is derived before $t(a_{1},b)$.

  We now consider the same formula for $t(a_{2},b)$.
  By instantiating, we obtain the formula
  \[
    e(a_{2},b) \lor \exists Z (e(a_{2},Z) \land t(Z,b) \land (Z,b) <_{tt} (a_{2},b)).
  \]
  Clearly, $e(a_{2},b)$ is false again, and the second disjunct can only hold for $Z = a_{1}$.
  Then the remaining formula
  \[
    e(a_{2},a_{1}) \land t(a_{1},b) \land (a_{1},b) <_{tt} (a_{2},b)
  \]
  needs to be satisfied.
  While both $e(a_{2},a_{1})$ and $t(a_{1},b)$ are satisfied we can not satisfy $(a_{1},b) <_{tt} (a_{2},b)$.
  If this atom were to hold we would have that $(a_{1},b) <_{tt} (a_{1},b)$ holds by transitivity as $(a_{2},b) <_{tt} (a_{1},b)$ holds from above.
  But this is in contradiction to the irreflexivity axioms of ordered completion.

  Thus, we have verified that the atoms $t(a_{1},b)$ and $t(a_{2},b)$ can not be true in any model of the ordered completion of $\Pi_{TC}$.
\end{example}

This example illustrates how ordered completion ensures that justifications are non-circular.
We now want to consider an infinite graph as the input to our transitive closure program in order to illustrate why some models of the ordered completion (in which infinitely many precomputed atoms are true) are not stable models.

\begin{example}\label{ex:oc-trans-infinite}
  Our infinite graph contains nodes $a_{i}$ for $i \geq 1$ as well as a node $b$.
  Every node $a_{i}$ is connected to the node $a_{i+1}$.
  I.e., the $a$ nodes are an infinite chain.
  As before, $b$ is an unconnected node.
  We add $e(a_{i},a_{i+1})$ for every $i \geq 1$ as a fact to our program $\Pi_{TC}$.

  The only stable model of our program consists of the following transitivity edges:
  \[
    t(a_{i},a_{j}) \quad \text{for } i,j \geq 1 \text{ and } i < j.
  \]
  We, of course, have a model of the ordered completion $\oc{\Pi_{TC}}$ with these same transitivity edges.
  However, we additionally have a model of the ordered completion that contains the following
  \[
    t(a_{i},b) \quad \text{for } i \geq 1.
  \]
  Note that this is an infinite set of precomputed atoms.
  Remember that in this case, not every model of the ordered completion corresponds to a stable model.
  We investigate why this is the case for this specific example.

  Again, the interesting formula in this case is $\ocomp{t}$
  \[
    \forall X Y (t(X,Y) \imp (e(X,Y) \lor \exists Z (e(X,Z) \land t(Z,Y) \land (Z,Y) <_{tt} (X,Y))).
  \]
  Note that we already applied the simplification of dropping the $<_{et}$ predicates as discussed in the previous example.
  For all other formulas of $\oc{\Pi_{TC}}$, it is easy to see that they are still satisfied.

  As $t(a_{1},b)$ is true we need some $Z$ such that
  \[
    e(a_{1},Z) \land t(Z,b) \land {} (Z,b) <_{tt} (a_{1},b)
  \]
  holds (clearly the other disjunct, $e(a_{1},b)$, does not hold).
  The only option here is $Z = a_{2}$.
  Indeed, $e(a_{1},a_{2})$ and $t(a_{2},b)$ hold.
  We have to set
  \[
    (a_{2},b) <_{tt} (a_{1},b)
  \]
  to true.
  Then $\ocomp{t}$ is satisfied for $t(a_{1},b)$.

  We continue with $t(a_{2},b)$.
  By the same reasoning as above for $t(a_{1},b)$ we get that
  \[
    (a_{3},b) <_{tt} (a_{2},b)
  \]
  needs to hold in order for $\ocomp{t}$ to be satisfied for $t(a_{2},b)$.

  Continuing this reasoning, we get the following infinitely decreasing chain
  \[
    (a_{1},b), (a_{2},b), (a_{3},b), \dots
  \]
  where $(a_{i+1},b) <_{tt} (a_{i},b)$ for all $i \geq 1$.
  However, we know from the definition of well-supported models (\cref{def:wsm}) that the derivation of a stable model is not allowed to have such an infinitely decreasing chain.
  This is why this model of the ordered completion is not a stable model.
  As noted in the discussion of \cref{lem:oc-ord-wfo}, we can only guarantee that the derivation order is well-founded, i.e., it does not contain an infinitely decreasing chain, when we only have finitely many precomputed atoms that are true.
\end{example}

\section{Ordered Completion with Level Mapping}
\label{sec:oc-lvl}

In this section, we will define a modification of the ordered completion in order to avoid the shortcomings of the ordered completion, in that it only captures finite stable models, as seen in the above example.

The idea in this section is to make use of the relation $<$ on natural numbers.
This is a well-founded relation on an infinite set.
Therefore, by modifying the ordered completion, we will be able to establish well-foundedness even in the case that we have an infinite set of precomputed atoms that are true.
However, using this relation means that we necessarily need a first-order logic with arithmetic.%
\footnote{Alternatively, instead of a first-order logic with arithmetic, we may consider a normal first-order logic but restrict our following theorems to standard interpretations, i.e., interpretations that interpret numbers, arithmetic operations, and comparisons in the usual way.}
For the kind of logic program considered in~\cite{AsuncionEtAl2012}, arithmetic is not necessary, and thus changing to a stronger logic may be considered undesirable.
However, as we consider \mg as the language of our logic programs, this is not a drawback: we need arithmetic anyway to capture the semantics of \mg programs.

The key idea is to replace the literals $\xs <_{pq} \ys$ by a relation on natural numbers.
To do so, we need two functions $\lvl{p}$ and $\lvl{q}$ that have the same arity as $p$ and $q$ and that map to natural numbers.
Then we can just write $\lvl{p}(\xs) < \lvl{q}(\ys)$ using the usual $<$ relation on natural numbers.

This then requires a first-order logic that has \emph{natural numbers} and the $<$~relation built in.
However, in the \mg context, we use a logic with \emph{integers} instead.
Therefore, we have to add helper axioms that express that our $\lvl{p}$ functions only map to values greater than or equal to $0$.

As stated above, using $<$~on natural numbers allows us to establish well-foundedness of the derivation order even for infinite sets of precomputed atoms that are true in a model of the ordered completion.
Additionally, it is no longer necessary to add irreflexivity and transitivity axioms.

\subsection{Derivation Level Functions}
We introduce the derivation level functions needed to define the ordered completion with level mapping.
\begin{definition}[Level Mapping Functions]
  For each predicate $p$ of arity $n$, we introduce a new function $\lvl{p}$ of the same arity $n$ that maps to integers.
\end{definition}

Now let $\siglvl$ denote the extension of the signature $\sigsort$ by the new functions $\lvl{p}$ of arity $n$, where $p/n$ is a predicate of the signature $\sigsort$.

We use the same notations as for $\sigord$ to describe reducts and extensions for this new signature (see \cref{def:ext-red}).

We furthermore define axioms that state that the level mapping functions only map to natural numbers.
\begin{definition}
  Let $\Pi$ be a \mg program.
  We define the formula $\nat{\Pi}$ as follows
  \[
    \bigwedge_{p \in \lang{\Pi}} \forall \xs (\lvl{p}(\xs) \geq 0)
  \]
  where $\xs$ are fresh program variables matching the arity of $p$.
\end{definition}

\subsection{Definition of Ordered Completion With Level Mapping}

Using the level mapping function, we define a modification of $\ocomp{p}$.
Note that we do not modify the definition of $\comprule{p}$.

\begin{definition}[Ordered Completion With Level Mapping of a Predicate]
  Let $\Pi$ be a \mg program and let $p \in \lang{\Pi}$ be a predicate of this program.
  We define the formula $\ncomp{p}$ as follows
  \begin{alignat*}{2}
    \ncomp{p} =
    \forall \xs (p(\xs) \imp \bigvee_{\twosub{R \in \Pi}{\headlit{R} = p(\terms)}} \exists \ys (\form{R}{\xs} \land {}
    \bigwedge_{\nospace{q(\terms^{\prime}) \in \bodylitp{R}}} \exists \zs (&\val{\terms^{\prime}}{\zs} \land q(\zs) \land {} \\
    &\lvl{q}(\zs) < \lvl{p}(\xs)))),
  \end{alignat*}
  where $\xs$ and $\zs$ are fresh program variables that match the arity of $p$ and $q$, respectively, and $\ys$ are the free variables in $\form{R}{\xs}$.
  Note that the free variables in $\terms^{\prime}$ are already free variables in $\form{R}{\xs}$ and thus already quantified.
\end{definition}

Finally, we define the ordered completion with level mapping of \mg programs.

\begin{definition}[Ordered Completion With Level Mapping]
  The ordered completion with level mapping of a \mg program $\Pi$, denoted by $\nc{\Pi}$, is the following formula
  \[
    \nat{\Pi} \land \cons{\Pi} \land \bigwedge_{p \in \lang{\Pi}} \comprule{p} \land \ncomp{p}.
  \]
\end{definition}

\subsection{The Ordered Completion With Level Mapping Theorem}
The proof for the ordered completion with level mapping follows the same ideas as the proof of the ordered completion theorem in \cref{sec:oc-theorem}.
The only difference is that for the ordered completion with level mapping, we can show that the relation encoded with the derivation level functions is well-founded, whether we have a finite or infinite set of precomputed atoms.
Almost all the lemmata from \cref{sec:oc-theorem} continue to hold.
We only have to modify two constructions and their corresponding lemmata.

First, we modify the construction of the interpretation $J$ of signature $\siglvl$ given a well-supported model.

\begin{remark}\label{rem:wsm-siglvl}
  Let $\Pi$ be a \mg program and let $I$ be a set of precomputed atoms such that $I$ is a well-supported model (with order $\prec$) of $\Pi$.
  Then we can extend $\ext{I}{\sigsort}$ to an interpretation $J$ of signature $\siglvl$ as follows: for any $p(\terms) \in I$ we set $\lvl{p}(\terms)$ to the rank of $p(\terms)$ with respect to $\prec$.
  For terms $\terms^{\prime}$ for which $p(\terms^{\prime}) \not\in I$ we can set $\lvl{p}(\terms^{\prime})$ arbitrarily.
  As we only add components of the signature $\siglvl$ to $J$ that do not belong to $\sigsort$, we have that $\red{J}{\sigsort} = \ext{I}{\sigsort}$.
\end{remark}

\begin{lemma}\label{lem:wsm-siglvl}
  Let $\Pi$ be a \mg program and let $I$ be a set of precomputed atoms such that $I$ is a well-supported model (with order $\prec$) of $\Pi$.
  We construct the interpretation $J$ of signature $\siglvl$ according to \cref{rem:wsm-siglvl}.
  Then for any $p(\terms) \in I$ and $q(\terms^{\prime}) \in I$ we have $p(\terms) \prec q(\terms^{\prime})$ iff $J \models \lvl{p}(\terms) < \lvl{q}(\terms)$.
\end{lemma}

This follows by construction of $J$.
If $p(\terms) \prec q(\terms^{\prime})$ then the rank $n$ of $p(\terms)$ is smaller than the rank $m$ of $q(\terms^{\prime})$.
By construction, we have $\lvl{p}(\terms) = n$ and $\lvl{q}(\terms^{\prime}) = m$.

\begin{lemma}
  Let $\Pi$ be a \mg program and let $I$ be a set of precomputed atoms.
  If $I$ is a stable model of $\Pi$ then there is an extension $J$ of $\ext{I}{\sigsort}$ such that $\red{J}{\sigsort} = \ext{I}{\sigsort}$ and $J$ is a model of $\nc{\Pi}$.
\end{lemma}

The proof of this lemma is the same as the proof of \cref{lem:sm-oc}, where we replace the construction of $J$ with the construction in \cref{rem:wsm-siglvl}.

Second, we modify the construction of an order $\prec$ given a model of the ordered completion with level mapping.

\begin{remark}\label{rem:nc-ord}
  A model $I$ of the ordered completion with level mapping $\nc{\Pi}$ of a \mg program $\Pi$ encodes an order $\prec$ on the set $\at{I}$ of precomputed atoms that are true in $I$.
  For atoms $p(\rs) \in \at{I}$ and $q(\rs^{\prime}) \in \at{I}$ we set $p(\rs) \prec q(\rs^{\prime})$ iff $\lvl{p}(\rs) < \lvl{q}(\rs^{\prime})$ holds in $I$.
\end{remark}

\begin{lemma}
  Let $\Pi$ be a \mg program and $I$ be a model of $\nc{\Pi}$.
  The relation $\prec$ on $\at{I}$ as defined in \cref{rem:nc-ord} is a strict well-founded partial order.
\end{lemma}

We get irreflexivity and transitivity of $\prec$ as $<$ is, by definition, irreflexive and transitive.
Furthermore, we can obtain the well-foundedness of $\prec$ as the relation $<$ on natural numbers is well-founded.
In contrast to $\oc{\Pi}$, this holds in both the \emph{finite} and \emph{infinite} case.
With that, we can prove the following lemma along the same way as the proof of \cref{lem:oc-sm}, where we replace the construction of $\prec$ by the construction in \cref{rem:nc-ord} and drop the restriction that only a finite set of precomputed atoms is true.

\begin{lemma}
  Let $\Pi$ be a \mg program and let $I$ be a set of precomputed atoms.
  Furthermore, let $J$ be an interpretation of the signature $\siglvl$ such that ${\red{J}{\sigsort} = \ext{I}{\sigsort}}$.
  If $J$ is a model of $\nc{\Pi}$, then $I$ is a stable model of $\Pi$.
\end{lemma}

Using these two lemmata, we then obtain the ordered completion with level mapping theorem for \mg.
Compared to the ordered completion from \cref{sec:oc}, models of the level mapping approach are equivalent to finite and infinite stable models.

\begin{theorem}[Ordered Completion with Level Mapping for \mg]
  Let $\Pi$ be a \mg program and let $I$ be a set of precomputed atoms.
  Then $I$ is a stable model of $\Pi$ if and only if there is an extension $J$ of $\ext{I}{\sigsort}$ such that $\red{J}{\sigsort} = \ext{I}{\sigsort}$ and $J$ is a model of $\nc{\Pi}$.
\end{theorem}

We conclude this section by revisiting the example of transitive closure on an infinite graph (\cref{ex:oc-trans-infinite}).
Recall that with the ordered completion as defined in \cref{sec:oc}, we had a model that did not correspond to a stable model.
The reason for this was that the derivation order contained an infinitely decreasing chain.
We now reconsider this example for the ordered completion with level mapping to see how the level mapping allows us to fix this issue.

\begin{example}
  Recall our \mg program $\Pi_{TC}$
  \[
    t(X,Y) \revimp e(X,Y), \qquad t(X,Y) \revimp e(X,Z) \land t(Z,Y), \\
  \]
  supplemented with facts $e(a_{i},a_{i+1})$ for every $i \geq 1$.
  The stable model only contains transitivity edges $t(a_{i},a_{j})$ for every $i,j \geq 1$ and $i < j$.
  The ordered completion $\oc{\Pi_{TC}}$ also had a model with additional transitivity edges $t(a_{i},b)$ for every $i \geq 1$ (see \cref{ex:oc-trans-infinite}).
  We check whether this can also be true for a model of $\nc{\Pi_{TC}}$.

  Again the formulas $\compdef{e}$, $\ncomp{e}$, and $\compdef{t}$ are satisfied by an interpretation containing the $t(a_{i},b)$ atoms.
  We consider the formula $\ncomp{t}$:
  \begin{alignat*}{3}
    &\forall X Y (t(X,Y) \imp (&e(X, &Y) \land \lvl{e}(X,Y)  < \lvl{t}(X,Y)) \lor {} \\
    &                    &\exists Z (&e(X,Z) \land \lvl{e}(X,Z) < \lvl{t}(X,Y) \land {} \\
    &                    &     &t(Z,Y) \land \lvl{t}(Z,Y) < \lvl{t}(X,Y))),
  \end{alignat*}
  Again, we can drop all the comparisons involving the function $\lvl{e}(X,Y)$ as we can just set $\lvl{e}$ to be $0$ for every argument.
  Our remaining formula then is
  \[
    \forall X Y (t(X,Y) \imp e(X,Y) \lor \exists Z (e(X,Z) \land t(Z,Y) \land \lvl{t}(Z,Y) < \lvl{t}(X,Y))).
  \]

  We first consider this formula for $t(a_{1},b)$.
  Clearly the first disjunct ($e(a_{1},b)$) is not satisfied.
  In order to satisfy the second disjunct, we can only choose $Z = a_{2}$ (for all other values $e(a_{1},Z)$ is not satisfied).
  Then both $e(a_{1},a_{2})$ and $t(a_{2},b)$ are satisfied.
  It remains to check that $\lvl{t}(a_{2},b) < \lvl{t}(a_{1},b)$ is satisfiable.
  $\lvl{t}(a_{1},b)$ is some natural number $m$.
  Then $\lvl{t}(a_{2},b)$ can at most be $m - 1$.

  We continue with checking $\ncomp{t}$ for $t(a_{2},b)$.
  Repeating the same reasoning as above, we need to check that $\lvl{t}(a_{3},b) < \lvl{t}(a_{2},b)$ is satisfied.
  As $\lvl{t}(a_{2},b)$ can at most be $m - 1$ we get that $\lvl{t}(a_{3},b)$ can be at most $m - 2$.

  Overall to justify $t(a_{1},b)$ we again get the infinite chain
  \[
    t(a_{1},b), t(a_{2},b), t(a_{3},b), \dots
  \]
  However, now it is not possible to satisfy $\lvl{t}(a_{i+1},b) < \lvl{t}(a_{i},b)$ for all $i$: $\lvl{t}(a_{1},b)$ is some natural number $m$.
  If we now assume that $\lvl{t}(a_{2},b)$ is $m - 1$, $\lvl{t}(a_{3},b)$ is $m - 2$, and so on we get that $\lvl{t}(a_{m+1},b)$ is $0$.
  But then it is not possible to satisfy $\lvl{t}(a_{m+2},b) < \lvl{t}(a_{m+1},b)$ as $\lvl{t}$ needs to be greater or equal to $0$ for every value.
  (If the values of $\lvl{t}$ do not go down in steps of $1$ as assumed, we reach the value $0$ earlier, but the same reasoning still holds.)

  Thus, the interpretation containing the $t(a_{i},b)$ atoms does not satisfy the formula $\ncomp{t}$ and thus also not $\nc{\Pi_{TC}}$.
  By mapping the derivation order to an order on natural numbers, we ensure that it is well-founded: it can not have any infinitely decreasing chain.
\end{example}

In the following, we will drop the special notation $\nc{\Pi}$ for the ordered completion with level mapping and use the same notation $\oc{\Pi}$.
It will be clear from context which version we are using.

\section{Simplifying the Ordered Completion}
\label{sec:oc-simp}

The definition of ordered completion in its current form often contains essentially repeated subformulas.
For example let us recall $\ocomp{p}$ from \cref{ex:oc-basic} (in the level mapping version):
\begin{alignat*}{2}
  \ocomp{p}
  &= \forall X (p(X) \imp {} \exists A (&&\val{A}{X} \land \exists V (\val{A-1}{V} \land q(V)) \land {} \\
  &                         &&\exists Z (\val{A-1}{Z} \land q(Z) \land \lvl{q}(Z) < \lvl{p}(X)))).
\end{alignat*}
Notice that we have the subformula
\[
  \exists V (\val{A-1}{V} \land q(V))
\]
which is contained (with renaming of the quantified variable) in the subformula
\[
  \exists Z (\val{A-1}{Z} \land q(Z) \land \lvl{q}(Z) < \lvl{p}(X)).
\]
We can just remove the contained formula (as it is subsumed by the larger formula) to obtain a simplified $\ocomp{p}$:
\[
  \forall X (p(X) \imp {} \exists A (\val{A}{X} \land \exists Z (\val{A-1}{Z} \land q(Z) \land \lvl{q}(Z) < \lvl{p}(X)))).
\]

We use this idea to introduce a simplified version of $\ocomp{p}$, which we will denote by $\ocomps{p}$.
Below, we will show that this simplification is equivalent to the definitions from \cref{sec:oc,sec:oc-lvl}.
Thus, all the theorems from before will continue to hold.

We will first describe the idea of this simplification before formally defining it.
We have that $\ocomp{p}$ is the following formula:
\[
  \forall \xs (p(\xs) \imp \bigvee_{\twosub{R \in \Pi}{\headlit{R} = p(\terms)}} \exists \ys (\form{R}{\xs} \land \bigwedge_{\nospace{q(\terms^{\prime}) \in \bodylitp{R}}} \exists \zs (\val{\terms^{\prime}}{\zs} \land q(\zs) \land \ord{q(\zs)}{p(\xs)}))),
\]
where $\ord{q(\zs)}{p(\xs)}$ is the formula expressing the derivation order of $q(\zs)$ and $p(\xs)$, i.e., $\zs <_{qp} \xs$ for the version of ordered completion defined in \cref{sec:oc} and $\lvl{q}(\zs) < \lvl{p}(\xs)$ for the version defined in \cref{sec:oc-lvl}.
We use this notation here to treat both versions of the ordered completion uniformly, as this simplification applies to both.

The idea is to pull the subformula $\ord{q(\zs)}{p(\xs)}$ into the formula $\form{R}{\xs}$ as the latter contains the formula $\val{\terms^{\prime}}{\zs} \land q(\zs)$ for each $q(\terms^{\prime}) \in \bodylitp{R}$.
I.e., we will modify $\form{R}{\xs}$ such that for each positive body literal of $R$ it will contain the formula $\val{\terms^{\prime}}{\zs} \land q(\zs) \land \ord{q(\zs)}{p(\xs)}$ instead.
The formulas in $\form{R}{\xs}$ for negated literals (whether that is a single or double negation) as well as for comparisons will remain unchanged.

We will denote this modification of $\form{R}{\xs}$ by $\oform{R}{\xs}$, which we define as follows.
If $R$ is a basic rule, then
\[
  \oform{R}{\xs} = \val{\terms}{\xs} \land \otaub(\body{R}).
\]
If $R$ is a choice rule, then
\[
  \oform{R}{\xs} = \val{\terms}{\xs} \land \otaub(\body{R}) \land \lnot \lnot p(\xs).
\]
If $\body{R}$ is the conjunction $B_{1} \land \dots \land B_{n}$ then $\otaub(\body{R})$ is the conjunction
\[
  \otaub(B_{1}) \land \dots \land \otaub(B_{n}),
\]
where we define $\otaub(B_{i})$ as
\begin{itemize}
  \item $\exists \vars (\val{\terms}{\vars} \land q(\vars) \land \ord{q(\vars)}{p(\xs)})$ if $B_{i}$ is $q(\terms)$,
  \item $\taub(B_{i})$ otherwise.
\end{itemize}

Using these two definitions, we can then define $\ocomps{p}$.

\begin{definition}[Simplified Ordered Completion of a Predicate]
  Given a \mg program $\Pi$ and a predicate $p$ we define the formula $\ocomps{p}$ as follows:
  \[
    \ocomps{p} = \forall \xs (p(\xs) \imp \bigvee_{\twosub{R \in \Pi}{\headlit{R} = p(\terms)}} \exists \ys (\oform{R}{\xs})).
  \]
\end{definition}

\begin{lemma}\label{lem:ocomps}
  Let $\Pi$ be a \mg program and let $p$ be a predicate in the language of $\Pi$.
  Then the formulas $\ocomp{p}$ and $\ocomps{p}$ are equivalent, i.e., we have $\ocomp{p} \equi \ocomps{p}$.
\end{lemma}

\begin{proof}
  Both formulas have the same outer structure
  \[
    \forall \xs (p(\xs) \imp \bigvee_{\twosub{R \in \Pi}{\headlit{R} = p(\terms)}} \phi_{R}),
  \]
  only the inner formulas $\phi_{R}$ differ.
  It is thus sufficient to show that for each rule $R$, the inner formulas $\phi_{R}$ are equivalent.

  We start with the case that $R$ is a basic rule.
  The inner formula of $\ocomp{p}$ then is
  \[
    \exists \ys (\val{\terms}{\xs} \land \taub(\body{R}) \land \bigwedge_{\nospace{q(\terms^{\prime}) \in \bodylitp{R}}} \exists \zs (\val{\terms^{\prime}}{\zs} \land q(\zs) \land \ord{q(\zs)}{p(\xs)})).
  \]

  Now let $\body{R} = B_{1} \land \dots B_{n}$, then we unfold $\taub(\body{R})$
  \[
    \exists \ys (\val{\terms}{\xs} \land \taub(B_{1}) \land \dots \land \taub(B_{n}) \land \bigwedge_{\nospace{q(\terms^{\prime}) \in \bodylitp{R}}} \exists \zs (\val{\terms^{\prime}}{\zs} \land q(\zs) \land \ord{q(\zs)}{p(\xs)})).
  \]
  Now let $B_{i} = q(\terms^{\prime})$, i.e., $B_{i}$ is a positive body literal.
  Then $\taub(B_{i})$ is
  \[
    \exists \vars (\val{\terms^{\prime}}{\vars} \land q(\vars)).
  \]
  Furthermore, the big conjunction (over all $q(\terms^{\prime}) \in \bodylitp{R}$) contains
  \[
    \exists \zs (\val{\terms^{\prime}}{\zs} \land q(\zs) \land \ord{q(\zs)}{p(\xs)}).
  \]
  as $q(\terms^{\prime})$ is a positive body literal.
  Of course, the conjunction of these two formulas (as it occurs in $\ocomp{p}$ after applying associativity and commutativity of conjunction)
  \[
    \exists \vars (\val{\terms^{\prime}}{\vars} \land q(\vars)) \land \exists \zs (\val{\terms^{\prime}}{\zs} \land q(\zs) \land \ord{q(\zs)}{p(\xs)})
  \]
  is equivalent to just the second formula
  \[
    \exists \zs (\val{\terms^{\prime}}{\zs} \land q(\zs) \land \ord{q(\zs)}{p(\xs)}),
  \]
  which is exactly $\otaub(B_{i})$.
  We can repeat this for all positive body literals and, of course, for all non-positive body literals and comparisons we can just replace $\taub(B_{i})$ by $\otaub(B_{i})$ as they are, by definition, equal.
  We then obtain exactly
  \[
    \exists \ys (\val{\terms}{\xs} \land \otaub(B_{1}) \land \dots \land \otaub(B_{n})),
  \]
  which after folding the definition of $\oform{R}{\xs}$ is
  \[
    \exists \ys (\oform{R}{\xs}).
  \]
  This is exactly the inner formula of $\ocomps{p}$.
  Thus, we have shown that for all basic rules $R$ the inner formulas of $\ocomp{p}$ and $\ocomps{p}$ are equivalent.

  If we now consider the case that $R$ is a choice rule, note that the only change is that both $\form{R}{\xs}$ and $\oform{R}{\xs}$ contain an additional conjunct $\lnot \lnot p(\xs)$.
  Thus, the same reasoning from above shows that the inner formulas of $\ocomp{p}$ and $\ocomps{p}$ are equivalent in the case that $R$ is a choice rule.

  As the inner formulas of $\ocomp{p}$ and $\ocomps{p}$ are equivalent, we can conclude that the formulas themselves are equivalent.
\end{proof}

Next, we use $\ocomps{p}$ to define a simplified ordered completion as follows.

\begin{definition}[Simplified Ordered Completion] \label{def:oc-simp}
  The simplified ordered completion of a \mg program $\Pi$, denoted by $\ocs{\Pi}$, is the following formula
  \[
    \cons{\Pi} \land Axioms(\Pi) \land \bigwedge_{p \in \lang{\Pi}} \comprule{p} \land \ocomps{p},
  \]
  where $Axioms(\Pi)$ is the formula $\irref{\Pi} \land \trans{\Pi}$ for the normal ordered completion and $\nat{\Pi}$ for the ordered completion with level mapping.
\end{definition}

By \cref{lem:ocomps} we then obtain the following lemma.

\begin{lemma}
  Let $\Pi$ be a \mg program.
  Then the ordered completion according to \cref{def:oc} is equivalent to the simplified definition according to \cref{def:oc-simp}, i.e., we have $\oc{\Pi} \equi \ocs{\Pi}$.
\end{lemma}

This means that all the lemmata and the main theorems from \cref{sec:oc,sec:oc-lvl} continue to hold for this simplified version of the ordered completion.
In the following, when we talk about the ordered completion of a program, we will implicitly refer to this simplified version, and we will drop the change of notation for the simplification.

\chapter{Verification using Ordered Completion}
\label{chp:verification}

Now that we have the ordered completion transformation to capture the stable model semantics of \mg programs, we want to use it in some practical verification scenarios.
To do so, we implemented ordered completion as a translation as well as a verification mode using it in the \anthem system.
\anthem is a system for formally verifying Answer Set Programs.
We give a quick review of the \anthem system focusing on the translating and verifying functionalities below and explain the newly implemented features in \cref{sec:implementation}.
We then consider some example verification problems that we can verify thanks to the ordered completion in \cref{sec:experiments}.
Afterwards, we examine some examples that show us the limitations of how we can use ordered completion in verification in \cref{sec:limitations}.
Finally, in \cref{sec:summary} we summarise our results from the different verification experiments.

\section{Implementation}
\label{sec:implementation}

First, \anthem can translate \mg programs or first-order theories by applying one of the following transformations: $\tau^{*}$, $\gamma$, $Comp$.
As introduced in \cref{sec:tau-star}, $\tau^{*}$~transforms \mg programs into first-order theories~\cite{LifschitzEtAl2019,FandinnoEtAl2024a}.
The transformation~$\gamma$ is used in the verification of strong equivalence~\cite{Heuer2020,Heuer2023,FandinnoLifschitz2023}.
$Comp$ is the completion as defined in~\cite{FandinnoEtAl2024a}.
This is a more general version of completion than defined in \cref{sec:completion}: it is a transformation on so-called completable first-order theories.
To compute the completion of a \mg program as in \cref{sec:completion}, it is necessary to first apply $\tau^{*}$ to the program and then apply the $Comp$ transformation.

Second, \anthem provides a mode to verify certain equivalence properties of \mg programs and/or first-order theories.
The first equivalence that can be verified is strong equivalence~\cite{LifschitzEtAl2019,Heuer2020,Heuer2023}.
Two programs are strongly equivalent if they can be exchanged in any context.
I.e., combined with any other logic program, they will produce the same stable models.
The second equivalence is external equivalence~\cite{FandinnoEtAl2023}.
External equivalence is a form of equivalence under certain inputs to the two logic programs.
The models of the two programs do not have to be equal under these inputs, only the presence of output atoms in the models is of interest.

The verification of external equivalence makes use of the completion transformation.
Thus, external equivalence can only be verified for (locally) tight programs.
Tightness is checked by the \anthem system; however, as local tightness is not a simple syntactic property, it can not be easily checked.
But it is possible for the user of the \anthem system to manually state that a program is locally tight.

For the verification of both properties, \anthem first applies some of the above-mentioned transformations and then produces problem files in the \tptp language~\cite{Sutcliffe2009,SutcliffeEtAl2012} which are given as the input to the automated theorem prover \vampire~\cite{KovacsVoronkov2013}.
For more information on the \anthem system refer to its documentation.\footnote{\url{https://github.com/potassco/anthem}}

We now turn toward the newly implemented features concerning ordered completion.
A version of \anthem with these new features is available online.\footnote{\url{https://github.com/janheuer/anthem/tree/oc-prototype}}
First, we implemented ordered completion as a new translation.
Analogously to how ordinary completion is implemented in \anthem, ordered completion is also implemented as a transformation on first-order theories.
Applying first $\tau^{*}$ and then the ordered completion to a logic program using \anthem results in the ordered completion as defined in \cref{sec:oc-simp}.
To do so, we can invoke \anthem using the following command(s).

\begin{lstlisting}
anthem translate --with tau-star program.lp | anthem translate --with ordered-completion
\end{lstlisting}

For example, we use the following as our \mg program (from \cref{ex:oc-basic}) saved in the file \code{program.lp}.

\begin{lstlisting}[language=mg]
p(A) :- q(A-1).
\end{lstlisting}

\anthem then produces the following formulas using the above command.

\begin{lstlisting}[language=formula]
forall V1 (p(V1) <- exists A (V1 = A and exists Z (exists I$i J$i (Z = I$i - J$i and I$i = A and J$i = 1) and q(Z)))).
forall V1 (p(V1) -> exists A (V1 = A and exists Z (exists I$i J$i (Z = I$i - J$i and I$i = A and J$i = 1) and (q(Z) and less_q_p(Z, V1))))).
\end{lstlisting}

Note that this command does not produce the axioms of the comparison predicates (i.e., irreflexivity and transitivity axioms).
Furthermore, the implementation only computes the ordered completion of predicates that are defined in the program (i.e., predicates that occur in some rule head).
The ordered completion with level mapping (\cref{sec:oc-lvl}) is currently not implemented as \anthem does not support functions in the logical language yet.

Second, we implemented a mode for verifying first-order statements of the form
\[
  \oc{\Pi} \imp T,
\]
where $\Pi$ is a \mg program and $T$ is a first-order theory.
This mode applies $\tau^{*}$ and ordered completion to the program $\Pi$ and then produces problem files that contain these formulas as axioms and the formulas in $T$ as conjectures.
These problems are then given to the theorem prover \vampire.
To use this mode, we can invoke \anthem with the following command.

\begin{lstlisting}
anthem verify --equivalence ordered-completion --direction forward program.lp theory.spec
\end{lstlisting}

Using this verification mode, we can prove that a (non-locally tight) logic program implies certain properties specified as a first-order theory.%
\footnote{We assume here that the theory $T$ does not contain any of the comparison predicates, i.e., $T$ only contains the original vocabulary of the program. Otherwise the statement $\oc{\Pi} \imp T$ does not necessarily imply that the stable models of $\Pi$ imply the theory $T$.}
We consider some examples of this kind of verification problem below in \cref{sec:experiments}.
Note that in this mode, \anthem always applies the ordered completion and does not check the program for tightness.
It is possible to add the option \code{--bypass-tightness} to the above command to force \anthem to use the normal completion instead.

Alternatively, instead of supplying a theory $T$ we can also use a second program $\Pi^{\prime}$ and instead verify the formula
\[
  \oc{\Pi} \imp \oc{\Pi^{\prime}}.
\]

In addition to just proving the ``forward'' direction of these problems, we can prove the ``backward'' direction (e.g., $\oc{\Pi} \revimp T$) or the ``universal'' direction (e.g., $\oc{\Pi} \equi T$).
To do so, we just change the value of the direction flag in the above command accordingly.
(Or for the universal direction, just omit the direction flag.)

However, note that these verification problems (where we use a direction different from ``forward'' and/or a second \mg program instead of a theory) are somewhat limited in the conclusion we can draw from them.
We discuss these limitations in \cref{sec:limitations}.

\section{Verification Experiments}
\label{sec:experiments}

The below examples can be found in the respective example directories of the implementation at \code{res/examples/ordered_completion}.

\begin{example}[Tight Logic Program]
  We start with a simple tight program, \cref{eq:tight} from the introduction.
  Let \code{tight.lp} be the following \mg program

  \begin{lstlisting}[language=mg]
p(X) :- q(X).
p(X) :- not r(X).
r(1).
q(1).
\end{lstlisting}

  We would like to show that the predicate \code{p} holds for every value.
  We express this in the following conjecture (in the file \code{tight.spec}).

  \begin{lstlisting}[language=formula]
forall X p(X).
\end{lstlisting}

  We invoke \anthem as follows to verify that the ordered completion of our program implies this conjecture.

  \begin{lstlisting}
anthem verify --equivalence ordered-completion --direction forward tight.lp tight.spec
\end{lstlisting}

  \anthem produces the following output, where we omit some of the axioms produced.

  \begin{lstlisting}[language=formula,numbers=left]
> Proving forward_0...
Axioms:
    forall X1 not less_q_q(X1, X1)
    forall X1 not less_p_p(X1, X1)
    forall X1 not less_r_r(X1, X1)
    forall X1 X2 X3 (less_q_q(X1, X2) and less_q_q(X2, X3) -> less_q_q(X1, X3))
    forall X1 X2 X3 (less_q_p(X1, X2) and less_p_r(X2, X3) -> less_q_r(X1, X3))
    ...
    forall V1 (q(V1) or (r(V1) -> #false) -> p(V1))
    forall V1 (V1 = 1 -> r(V1))
    forall V1 (V1 = 1 -> q(V1))
    forall V1 (p(V1) -> q(V1) and less_q_p(V1, V1) or (r(V1) -> #false))
    forall V1 (r(V1) -> V1 = 1)
    forall V1 (q(V1) -> V1 = 1)

Conjectures:
    forall X p(X)

> Proving forward_0 ended with a SZS status
Status: Theorem (81 ms)

> Success! Anthem found a proof of the forward theorem. (82 ms)
\end{lstlisting}

  The output starts by indicating which sub-problem is being verified.
  As our claim only consists of a single formula, we only have one sub-problem.
  If our claim had more formulas, there would be one sub-problem for each formula.

  \anthem then starts by listing the axioms in lines~3~to~14.
  Lines~3~to~5 are the irreflexivity axioms.
  From line~6 we have the transitivity axioms.
  We only show two of them here and omit the rest.
  In lines~9~to~11 we have the formulas $\comprule{p}$, $\comprule{r}$ and $\comprule{q}$.
  In lines~12~to~14 we have the formulas $\ocomp{p}$, $\ocomp{r}$ and $\ocomp{q}$.
  The conjecture is shown in line~17.
  Note that all the formulas are simplified using the simplification algorithm implemented in \anthem and may thus be slightly different from the definitions given in this thesis (but they are equivalent).

  Finally, line~20 indicates the status of the sub-problem, and in line~22 we have the overall result: \anthem successfully verified our conjecture from the axioms in $\qty{82}{ms}$.
  By the results of \cref{sec:oc-theorem}, we then also know that every stable model of our program implies the conjecture.
  Of course, here it would have been sufficient to use the normal completion as the axioms, as our program is tight.
  We can test this by adding the option \code{--bypass-tightness} to the above command.
\end{example}

\begin{example}[Non-Tight Logic Program]
  Next, we consider the non-tight program \cref{eq:non-tight} from the introduction.
  I.e., we take \code{non_tight.lp} to be the following \mg program.

  \begin{lstlisting}[language=mg]
p(X) :- q(X).
q(X) :- p(X).
\end{lstlisting}

  We want to verify that in every stable model of this program, both predicates \code{p} and \code{q} are false for every value.
  We state this in the following first-order formula in the file \code{non_tight.spec}.

  \begin{lstlisting}[language=formula]
forall X (not p(X) and not q(X)).
\end{lstlisting}

  Using these two files, we call \anthem to verify that the ordered completion of our program implies our claim.

  \begin{lstlisting}
anthem verify --equivalence ordered-completion --direction forward non_tight.lp non_tight.spec
\end{lstlisting}

  \anthem successfully verified this problem in around $\qty{28}{ms}$.
  If instead we were to try to prove that our claim follows from the completion of our program, this would not be successful, as our program is not (locally) tight, and it is thus necessary to use ordered completion.
  (This can be tested by adding the option \code{--bypass-tightness} to the above command, which forces \anthem to use ordinary completion instead of ordered completion.)
\end{example}

\begin{example}[Locally-Tight Logic Program]
  We now consider a program that is not tight but locally tight.
  Again, here it is not necessary to use ordered completion as the normal completion would be sufficient.
  However, as \anthem does not currently have any method to check for local tightness, we would have to prove local tightness manually first.
  For our example below, this is simple, but in practice, it is in general not trivial to check whether a program is locally tight or not.

  We use the following example \mg program (\cref{eq:locally-tight} from the introduction).

  \begin{lstlisting}[language=mg]
p(X+1) :- p(X), X > 0.
p(1).
\end{lstlisting}

  We want to verify that the predicate \code{p} holds for every integer greater than or equal to $1$.
  We express this in the file \code{locally_tight.spec}.

  \begin{lstlisting}[language=formula]
forall X$ (X$ >= 1 -> p(X$)).
\end{lstlisting}

  We verify this claim using the following command.

  \begin{lstlisting}
anthem verify --equivalence ordered-completion --direction forward locally_tight.lp locally_tight.spec
\end{lstlisting}

  \anthem verifies this claim in around $\qty{32}{s}$.

\end{example}

\begin{example}[Transitive Closure]
  Next, we consider the example of computing the transitive closure of a (finite) graph (as in \cref{ex:oc-trans-finite}).
  To do so, we use the following \mg program corresponding to \cref{eq:transitive}, which is saved in the file \code{transitive.lp}.

  \begin{lstlisting}[language=mg]
t(X,Y) :- e(X,Y).
t(X,Y) :- e(X,Z), t(Z,Y).
\end{lstlisting}

  We add facts corresponding to the edges of the graph to our logic program.

  \begin{lstlisting}[language=mg]
e(a1,a2).
e(a2,a1).
\end{lstlisting}

  The transitive closure of this graph is the set
  \[
    \{t(a_{1},a_{2}),t(a_{1},a_{1}),t(a_{2},a_{1}),t(a_{2},a_{2})\}.
  \]
  However, the completion of this program has an additional model with the transitivity edges $t(a_{1},b)$ and $t(a_{2},b)$ (see \cref{ex:transitive-completion}).
  We want to verify that the stable model of this program includes the actual transitive closure and that the additional transitivity edges possible in the completion of the program are not in the stable model.
  We thus have the following first-order formula as our conjecture (in the file \code{transitive.spec}).

  \begin{lstlisting}[language=formula]
t(a1,a2) and t(a1,a1) and t(a2,a1) and t(a2,a2) and not t(a1,b) and not t(a2,b).
\end{lstlisting}

  We again invoke \anthem using the following command.

  \begin{lstlisting}
anthem verify --equivalence ordered-completion --direction forward transitive.lp transitive.spec
\end{lstlisting}

  \anthem is able to verify that our claim follows from the ordered completion of our logic program in around $\qty{4}{s}$.
  Or equivalently, our claim holds in every stable model of the logic program.
  If we instead tried to verify the claim using the completion of the program, this would fail.
  (Running \anthem with the added \code{--bypass-tightness} option results in a timeout of the theorem prover.)
  Indeed, as noted above, the transitive edges \code{t(a1,b)} and \code{t(a2,b)} are true in a model of the completion, which is why we can not prove the claim using completion.
\end{example}

\section{Limitations}
\label{sec:limitations}

So far, we have only considered verification problems of the form
\[
  \oc{\Pi} \imp T.
\]
We will now consider some other kinds of problems and explore the limitations of ordered completion for them.
Note that below we use tight logic programs as our examples, but the limitations that these examples highlight can also occur for non-tight examples.
We only use the tight examples here for simplicity.

\begin{example}[Backward Direction]
  First, we want to consider a problem of the form
  \[
    T \imp \oc{\Pi},
  \]
  i.e., the reverse direction of the problems we have considered so far.
  Let our \mg program be the following (\code{backward.lp}).

  \begin{lstlisting}[language=mg]
p.
q :- p.
\end{lstlisting}

  Our theory will consist of the following formula (\code{backward.spec}).

  \begin{lstlisting}[language=formula]
p and q.
\end{lstlisting}

  If we try to verify that this theory is implied by the ordered completion of our program (i.e., $\oc{\Pi} \imp T$), \anthem will report a successful verification of the theorem.
  Instead, we want to try verifying the backward direction.
  We do so with the following command.

  \begin{lstlisting}
anthem verify --equivalence ordered-completion --direction backward backward.lp backward.lp
\end{lstlisting}

  This will result in a timeout.
  However, increasing the time limit will not help us here.
  Let us look at the ordered completion of our program
  \[
    p, \qquad p \imp q, \qquad q \imp p \land {} <_{pq}.
  \]
  The first two formulas can easily be verified using $p \land q$ as the axiom.
  The last formula can not be verified as we can not prove that $<_{pq}$ holds given our axiom.

  Let us consider what this means for the stable models of our program $\Pi$.
  By verifying $T \imp \oc{\Pi}$ we show that every model of $T$ is also a model of $\oc{\Pi}$.
  If this implication holds, we could conclude that every model of $T$ is also a stable model of $\Pi$.

  But as $T \imp \oc{\Pi}$ does not hold, can we conclude that there are models of $T$ that are not stable models of $\Pi$?
  This is, of course, not the case.
  If we recall the theorem on ordered completion, we notice that it essentially says the following: for every stable model, there \emph{exists} an extension that is a model of the ordered completion.
  Thus, what we really should try to verify is the following formula
  \[
    \exists <_{pq} (p \land (p \imp q) \land (q \imp p \land {} <_{pq})).
  \]
  Note that this is a second-order formula: we existentially quantify the predicate~$<_{pq}$.%
  \footnote{In general, we have to use the following second-order formula: $\widetilde{\exists} \oc{\Pi}$ where $\widetilde{\exists}$ quantifies over all order predicates in $\oc{\Pi}$. In this example, we can drop the irreflexivity axioms and transitivity axioms and thus only quantify $<_{pq}$.}
  And indeed, this second-order formula is a logical consequence of our axiom $p \land q$.

  Our original formula (i.e., not the second-order version) would be the correct formula to verify if the ordered completion theorem stated: for every stable model, \emph{every} extension is a model of the ordered completion.
\end{example}

\begin{example}[Equivalence of Ordered Completions]
  Next, we want to consider a verification problem of the form
  \[
    \oc{\Pi_{1}} \equi \oc{\Pi_{2}},
  \]
  i.e., verifying that the ordered completions of two programs are equivalent.
  Let $\Pi_{1}$ be the program (\code{equivalence.1.lp}):
  \begin{lstlisting}[language=mg]
p.
q :- p.
\end{lstlisting}
  and let $\Pi_{2}$ be (\code{equivalence.2.lp}):
  \begin{lstlisting}[language=mg]
q.
p :- q.
\end{lstlisting}

  Of course, these two programs have the same stable model: $\{p,q\}$.
  We may try to verify that their ordered completions are equivalent by using the command

  \begin{lstlisting}
anthem verify --equivalence ordered-completion --direction universal equivalence.1.lp equivalence.2.lp
\end{lstlisting}

  This will again result in a timeout.
  Let us consider the two ordered completions to see why this equivalence does not hold.
  The ordered completion of $\Pi_{1}$ contains the formulas (omitting the irreflexivity and transitivity axioms)
  \[
    p, \qquad p \imp q, \qquad q \imp p \land {} <_{pq},
  \]
  and the ordered completion of $\Pi_{2}$ contains
  \[
    q, \qquad q \imp p, \qquad p \imp q \land {} <_{qp}.
  \]
  Taking $\oc{\Pi_{1}}$ as the axioms, we can verify the first two formulas of $\oc{\Pi_{2}}$, but we can not verify that
  \[
    p \imp q \land {} <_{qp}
  \]
  holds.
  While our two programs have the same stable model, the models of their ordered completion are not the same as the extensions (by adding order predicates) are different: we have to add $<_{pq}$ to get a model of $\oc{\Pi_{1}}$ while we have to add $<_{qp}$ to get a model of $\oc{\Pi_{2}}$.

  We may try a second-order formula similar to the previous example instead.
  I.e., we try to verify that $\oc{\Pi_{1}}$ implies the following formula:
  \[
    \exists \prec_{qp} (q \land (q \imp p) \land (p \imp q \land {} \prec_{qp}))
  \]
  which is just $\oc{\Pi_{2}}$ with the order predicate $\prec_{qp}$ existentially quantified.\footnote{As in the previous example this formula is in general $\widetilde{\exists} \oc{\Pi_{2}}$. We can simplify the formula for this specific example.}
  (Note that we renamed the predicate $<_{qp}$ to $\prec_{qp}$ here for clarity.)
  Indeed, we can verify that this second-order formula holds.
\end{example}

\section{Summary}
\label{sec:summary}

To summarise this chapter, we have the following three verification scenarios using ordered completion and their relation to the stable models of the program(s):
\begin{enumerate}
  \item $\oc{\Pi} \imp T$ if and only if every stable models of $\Pi$ is a model of $T$,
  \item $T \imp \oc{\Pi}$ implies that every model of $T$ is a stable model of $\Pi$,
  \item $\oc{\Pi_{1}} \equi \oc{\Pi_{2}}$ implies that $\Pi_{1}$ and $\Pi_{2}$ have the same stable models.
\end{enumerate}
For all three scenarios, we have good results for the positive verification scenario: if we can prove our claim, we can draw conclusions about the stable model of our program(s).
Only in the first scenario can we also draw conclusions in the negative case.
In order to be able to draw conclusions in the negative case for the other two scenarios, we would have to replace our claim with a second-order formula.

Furthermore, the implementation is currently restricted as it does not implement the level mapping variant of ordered completion.
The equivalence in verification scenario (1) only holds for the level mapping variant.
With the currently implemented variant of ordered completion, we only have an implication in this first case.
The reason for this is that the formula $\oc{\Pi} \imp T$ may not hold because of a model of $\oc{\Pi}$ (that is not a model of $T$) that contains infinitely many precomputed atoms.
This model of the ordered completion does not necessarily correspond to a stable model of $\Pi$.

\chapter{Conclusion}
\label{chp:conclusion}

In this chapter, we first summarise the contributions of this thesis: well-support semantics for \mg, ordered completion for \mg, and first practical verification experiments using ordered completion.
Afterwards, we conclude the thesis by considering related (\cref{sec:related}) and future work (\cref{sec:future}).

\section{Contributions}
\label{sec:contributions}

In this thesis, we introduced two new characterisations of stable models for \mg programs.
First, we defined well-supported models of \mg programs.
Well-supported models have been considered in the literature for simpler languages~\cite{Fages1991,Elkan1990}.
Intuitively, the truth of every atom needs to be supported by one of the rules of the logic program.
I.e., for every true atom we need a rule whose head contains this atom and whose body is satisfied.
Additionally, for well-supported models, we require that positive elements of the body have to be derived earlier with respect to the derivation order of the model.
We showed that this characterisation is equivalent to the usual definition of stable models of \mg programs (in terms of stable models of propositional theories).

Second, we introduced ordered completion for \mg programs.
This is based on the definitions from~\cite{AsuncionEtAl2012} that define ordered completion on a more restricted language.
For ordered completion, we split up the equivalences of the normal completion: the ``$\revimp$'' direction ensures that we have a model of the program, while the ``$\imp$'' direction ensures that we have a supported model.
For the definition of ordered completion, we then supplement the formulas of the ``$\imp$'' direction by adding formulas talking about the derivation order.
We have defined two different approaches for expressing the derivation order: (1) using predicates $<_{pq}$, and (2) using functions $\lvl{p}$ and $\lvl{q}$ mapping to natural numbers (called the \emph{level mapping} variant).
With the former approach, we only capture finite stable models, while the latter approach also captures infinite stable models.
We proved the equivalence to stable models by showing that models of the ordered completion and well-supported models are equivalent.
Ordered completion and well-supported model use the same ideas but express them in different ways.
The well-support semantics provides criteria to pick the desirable models from the set of all classical models of the program.
On the other hand, ordered completion encodes these criteria into formulas in first-order logic so that the classical models are exactly our desired models.

Besides providing two new characterisations of stable models for the \mg language, we also considered ordered completion for the verification of non-tight \mg programs.
We implemented ordered completion in the ASP verification tool \anthem, as a new translation and as part of a new verification mode.
The level mapping variant of ordered completion is not yet implemented.
We successfully verified statements of the form ${\oc{\Pi} \imp T}$ where $\Pi$ is a \mg program and $T$ is a first-order theory, i.e., a formal specification.
Using our theoretical results, we can conclude that if this statement holds, then all stable models of $\Pi$ also imply the theory $T$.
With the level mapping variant, we have an equivalence of the two statements (i.e., the theory is implied by the ordered completion if and only if it is implied by the stable models).
For other verification problems (i.e., $T \imp \oc{\Pi}$ and $\oc{\Pi_{1}} \equi \oc{\Pi_{2}}$), we only have that if these statements hold the corresponding statement on the stable models holds as well (even with the level mapping variant).
The reason for this is that ordered completion makes the derivation order explicit, while it is only implicit in a stable model.
In order to get equivalences for these types of verification problems, we would have to make use of existential second-order quantification.

\section{Related Work}
\label{sec:related}

This thesis provides the first approach to capturing the semantics of arbitrary \mg programs in first-order logic.
Related work on \mg is restricted to tight or locally-tight programs.
However, there are some other works on capturing the stable model semantics in first-order logic for more restricted languages.
Specifically, the related work investigated in this section considers logic programs with rules of the form
\begin{equation}\label{eq:lp}
  p(\terms) \revimp L_{1} \land \dots \land L_{n},
\end{equation}
where $p(\terms)$ is an atom and each $L_{i}$ is an atom possibly preceded by one or two occurrences of negation~$\pnot$.
A term~$t$ is a symbolic constant or a variable.
Compared to \mg, we have the following main restrictions: terms can not contain any numbers and arithmetic, bodies can not contain comparisons, and only basic rules are allowed (i.e., we can not have choices or constraints).

We separate the related work into two classes: \emph{modifications of completion}, and \emph{preprocessing approaches}.
First, modifications of completion are exactly like the ordered completion presented in this thesis: they take completion as the basis and add or modify it to capture the semantics of arbitrary programs.
Second, preprocessing approaches transform an arbitrary program~$\Pi$ into a program~$\Pi^{\prime}$ such that the stable models of~$\Pi^{\prime}$ correspond to the completion models of~$\Pi^{\prime}$.
In this way, we can view the transformation from $\Pi$ to $\Pi^{\prime}$ as a preprocessing step to applying completion.

\subsection{Modifications of Completion}

First, we have the idea of loop formulas.
As mentioned in the introduction of this thesis, propositional loop formulas~\cite{LinZhao2004} are used in the design of modern ASP solvers.
Loop formulas have been generalised to the first-order case~\cite{ChenEtAl2006}.
However, as noted before, the set of loop formulas of a logic program with variables can, in general, be infinite and is thus not suited for automated verification.

Second, there are two approaches~\cite{Ben-EliyahuDechter1994,Niemela2008} closely related to ordered completion.
Both approaches only consider propositional logic programs (i.e., programs without variables).
In the case of propositional logic programs of the form \cref{eq:lp}, the ordered completion of a predicate~$p$ consists of the following formulas
\[
  \comprule{p} = \bigvee_{\twosub{R \in \Pi}{\headlit{R}=p}} \pform{R} \imp p, \qquad \ocomp{p} = p \imp \bigvee_{\twosub{R \in \Pi}{\headlit{R}=p}} \pform{R} \land \bigwedge_{\nospace{q \in \bodylitp{R}}} \ord{q}{p},
\]
where $\ord{q}{p}$ is either the formula $<_{qp}$ or $\lvl{q} < \lvl{p}$ depending on the version of ordered completion we use.
Both~\cite{Ben-EliyahuDechter1994,Niemela2008} use the same formulas, only the definition of $\ord{q}{p}$ is different.

In~\cite{Ben-EliyahuDechter1994} we have $\ord{q}{p} = \#p < \#q$.
This is the same idea as the level mapping version of ordered completion.
However, in~\cite{Ben-EliyahuDechter1994} the formula $\#p < \#q$ is encoded in purely propositional logic, i.e., no form of arithmetic is used.

On the other hand, \cite{Niemela2008} uses difference logic to express the formula $\ord{q}{p}$.
Difference logic is an extension of propositional logic by simple linear inequalities of the form $x + k \geq y$ where $x,y$ are integer variables and $k$ is an integer constant.
The formula $\ord{q}{p}$ is then defined as $x_{q} - 1 \geq x_{p}$ where $x_{q},x_{p}$ are variables associated with~$q$ and $p$~respectively.
The use of difference logic makes it possible to compute stable models by using SMT (Satisfiability Modulo Theories) solvers.

\subsection{Preprocessing Approaches}

An important approach in this class is the so-called tightening transformation~\cite{Wallace1993}.
Tightening transforms arbitrary (first-order) logic programs into locally-tight logic programs.\footnote{In~\cite{Wallace1993}, different terminology is used as local tightness is a more recent concept.}
This is done by extending each predicate occurring in the program by an additional argument that counts the number of rule applications needed to derive it.
For example, if we consider the following logic program
\begin{equation}\label{eq:related}
  p \revimp q, \qquad q \revimp p,
\end{equation}
its tightening contains the following rules
\[
  p(N+1) \revimp q(N) \land N \geq 0, \qquad q(N+1) \revimp p(N) \land N \geq 0,
\]
as well as projection rules
\[
  p \revimp p(N) \land N \geq 0, \qquad q \revimp q(N) \land N \geq 0.
\]
As the tightening of a program~$\Pi$ is a locally-tight program~$\Pi^{\prime}$, the stable models of~$\Pi^{\prime}$ are fully captured by the model of its completion.
Furthermore, the stable models of~$\Pi$ and~$\Pi^{\prime}$ are the same after projection to the vocabulary of~$\Pi$.
Note that the tightening of a program contains some arithmetic expressions and comparisons.
Alternatively, it would be possible to express them using Herbrand semantics.

Thus, in the same way that using ordered completion enables us to capture the stable models of arbitrary programs, we can also do so by using tightening and ordinary completion.

In \cref{sec:limitations} we noted that even if two programs have the same stable models, their ordered completions are not necessarily equivalent.
The same holds for the tightening of the programs.
If we recall the example programs~$\Pi_{1}$
\[
  p, \qquad q \revimp p,
\]
and~$\Pi_{2}$
\[
  q, \qquad p \revimp q,
\]
from \cref{sec:limitations}, their tightenings are
\[
  p(N) \revimp N \geq 0, \qquad q(N+1) \revimp p(N) \land N \geq 0,
\]
and
\[
  q(N) \revimp N \geq 0, \qquad p(N+1) \revimp q(N) \land N \geq 0,
\]
respectively.
(Note that we omit the projection rules here.)
The stable model of the tightening of~$\Pi_{1}$ contains~$p(N)$ for every $N \geq 0$ and~$q(N)$ for every $N \geq 1$.
On the other hand, the stable model of the tightening of~$\Pi_{2}$ contains~$p(N)$ for every $N \geq 1$ and~$q(N)$ for every $N \geq 0$.
Note the difference that while the stable model of $\Pi_{1}$ contains~$p(0)$ the stable model of $\Pi_{2}$ does not (and vice versa for~$q(0)$).
Thus, the stable models of the tightenings (and also the models of their completions) are not the same even though the stable models are the same.

As tightening produces locally tight programs, it could be used in the verification of external equivalence.
The procedure for verifying external equivalence from~\cite{FandinnoEtAl2023} is correct for locally tight programs~\cite{FandinnoEtAl2024a}.
Furthermore, external equivalence includes private predicates: certain predicates can be marked as private, which has the effect that they are not taken into account for the equivalence.
I.e., in the above example, we could mark both $p/1$ and $q/1$ as private, as we are only interested in whether our models include the original predicates $p/0$ and $q/0$.

However, the external equivalence problem is, in general, a second-order problem and can only be reduced to the first-order case if there is no \emph{private recursion}.
A program contains private recursion if it contains recursion over private predicates or a choice rule over a private predicate.
The tightened program from above does not contain private recursion, so the external equivalence statement would be a first-order problem.
However, this is only the case because the original program we considered in the example was tight already.
If we instead consider the non-tight program \cref{eq:related}, its tightening contains private recursion because of the rules
\[
  p(N+1) \revimp q(N) \land N \geq 0, \qquad q(N+1) \revimp p(N) \land N \geq 0.
\]
Thus, the tightening of a non-tight program can not be verified with the external equivalence procedure from~\cite{FandinnoEtAl2023} as the external equivalence statement does not reduce to a first-order problem.

Besides tightening, there are some other approaches that are restricted to the propositional case~\cite{Janhunen2004,LinZhao2003,JanhunenNiemela2011,JanhunenEtAl2009}.

First, \cite{LinZhao2004} transform arbitrary propositional programs into programs that are \emph{inherently tight}.
A program~$\Pi$ is called inherently tight if it is inherently tight on all models of its completion.
A program~$\Pi$ is inherently tight on a set of atoms~$S$ if there is a program $\Pi^{\prime} \subseteq \Pi$ such that $\Pi^{\prime}$~is tight and $S$~is a stable model of~$\Pi^{\prime}$.
It is then shown that for an inherently tight program~$\Pi$, the stable models of~$\Pi$ are exactly the models of the completion of~$\Pi$.
An arbitrary program~$\Pi$ is transformed into an inherently tight program~$\Pi^{\prime}$ such that the stable models of both programs are the same after projecting to the vocabulary of~$\Pi$.

We look at an example of this transformation to understand its idea.
We again consider the non-tight program \cref{eq:related}.
It is transformed into the following inherently tight program
\begin{alignat*}{3}
p'        &\revimp \pnot less(p,q) \land q,\qquad &q'        &\revimp \pnot less(q,p) \land p, \\
p         &\revimp p',                            &q         &\revimp q', \\
less(q,p) &\revimp p',                            &less(p,q) &\revimp q'.
\end{alignat*}
The idea is to introduce for each atom~$p$ and each rule of~$p$ a new atom, here~$p^{\prime}$, as each atom only has one rule.
The rule $p \revimp q$ is then modified by replacing the head by~$p^{\prime}$ and adding $\pnot less(p,q)$ to the body.
The latter expresses that~$p$ can not be derived earlier than~$q$.
The atom~$p$ is true if its rule is true, i.e., if~$p^{\prime}$ is true.
Furthermore, if~$p^{\prime}$ is true, then~$q$ needs to be derived before~$p$, i.e., $less(q,p)$~needs to hold.
It is clear that the only stable model of this program is the empty model, as both~$p$ and~$q$ can not be true at the same time, as then both $less(p,q)$ and $\pnot less(p,q)$ would need to hold.
Again, the idea is essentially the same as with ordered completion: information is added to state that positive body atoms have to be derived earlier than the head.
The difference is that in~\cite{LinZhao2003} this information is already added on the program level and not when constructing the completion.

Second, \cite{Janhunen2004} transform arbitrary propositional programs into tight programs.
To do so, \cite{Janhunen2004} first introduces a variant of well-supported models.
However, instead of using a strict well-founded partial order as in \cref{sec:well-support}, a level mapping function similar to $\lvl{p}$ from \cref{sec:oc-lvl} is used.
The difference is that additional constraints are added to the level mapping function that ensure that this mapping is unique.
In \cite{Janhunen2004} levels are not only assigned to atoms but also to the supporting rules (i.e., all rules whose bodies are satisfied).
The level of a rule is then the maximal level of its positive body atoms plus one.
The level of an atom is the minimal level of its supporting rules.
For example, if we have the program
\[
  p, \qquad q \revimp p,
\]
the level of the fact $p$ is $1$ and thus $\lvl{p}$ is also $1$.
The level of the rule $q \revimp p$ is $\lvl{p} + 1$, i.e., $2$, and $\lvl{q}$ is also $2$.
In the level mapping as we defined it in \cref{sec:oc-lvl}, we only require that the level of $p$ needs to be smaller than the level of $q$.
Thus, there are infinitely many valid level mappings for the single stable model $\{p,q\}$, whereas with the version of a level mapping from \cite{Janhunen2004} there is a unique level mapping.

This idea of a unique level mapping is then encoded into a logic program.
We again consider the simple example program \cref{eq:related} to illustrate this transformation.
First, the rules of the program are transformed by preceding all positive body literals by two negations $\pnot$
\[
  p \revimp \pnot \pnot q, \qquad q \revimp \pnot \pnot p.
\]
Now the program is tight, but its stable models do not match the stable models of the original program.
Instead, the stable models of this program are exactly the supported models of the original program.
We thus have to encode the (unique) level mapping in the program.
We start by assigning levels to both $p$ and $q$
\[
  \{ lvl(p,1..2) \} \revimp p, \qquad \{ lvl(q,1..2) \} \revimp q,
\]
note that we only assign a level if the corresponding atom is true.
We furthermore have to add constraints to ensure that exactly one level is assigned to both atoms
\begin{alignat*}{3}
  & \revimp lvl(p,1) \land lvl(p,2) \land p, \qquad && \revimp lvl(q,1) \land lvl(q,2) \land q, \\
  & \revimp \pnot lvl(p,1) \land \pnot lvl(p,2) \land p, \qquad && \revimp \pnot lvl(q,1) \land \pnot lvl(q,2) \land q.
\end{alignat*}
Finally, we need to relate the levels of $p$ and $q$: the level of $p$ is $\lvl{p \revimp q} + 1$ where $\lvl{p \revimp q}$ is the level of $q$ and vice versa for the level of $q$.
This is expressed in the following constraints
\[
  \revimp p \land lvl(q,N) \land \pnot lvl(p,N+1), \qquad \revimp q \land lvl(p,N) \land \pnot lvl(q,N+1).
\]
Now it is no longer possible that both $p$ and $q$ are true in a stable model of this program.
Note that the actual transformation from \cite{Janhunen2004} is more complex as: (1) we used the \mg language here to represent the program whereas \cite{Janhunen2004} uses a propositional version of the language in \cref{eq:lp}, (2) the levels are encoded in a binary fashion, and (3) as we only have a single rule of the form $p \revimp q$ for each atom here the definition of levels is significantly simplified.%
\footnote{I.e., here we can set $\lvl{p \revimp q}$ to $\lvl{q}+1$ whereas with a rule that contains more positive body atoms we have to set the level of the rule to the maximum level of its positive body atoms plus one. Furthermore, the level of $p$ is exactly the level of its single rule, while in general it needs to be the minimum level of all its rules whose bodies are satisfied.}

Finally, \cite{JanhunenEtAl2009,JanhunenNiemela2011} combine the ideas from \cite{Janhunen2004} with ideas from \cite{Niemela2008}.
Furthermore, the translation is extended to logic programs including choice rules (in a more general form than in \mg), cardinality rules, and weight rules, however, it is still restricted to propositional programs.
The translation happens in a two-step process: first, programs with choice, cardinality, and weight rules are transformed into propositional programs as in \cref{eq:lp}, then these propositional programs are transformed into tight programs.
We do not consider an example for \cite{JanhunenEtAl2009,JanhunenNiemela2011} as the concepts are the same as in the above example, only the actual encoding differs.

\section{Future Work}
\label{sec:future}

First, while on the theoretical side we have the ordered completion with level mapping to capture finite and infinite stable models, this is not yet implemented.
So far, we only implemented the variant of ordered completion from \cref{sec:oc} in \anthem.
To implement the level mapping variant, the logical language of \anthem has to be extended by functions.
By having the level mapping version of ordered completion, we obtain better results for the verification: $\oc{\Pi} \imp Spec$ is equivalent to the statement that all stable models of $\Pi$ imply the specification $Spec$.
Without level mapping, we only have an implication between these two statements.

As noted in \cref{sec:limitations} for verification problems of the form $Spec \imp \oc{\Pi}$ or $\oc{\Pi_{1}} \equi \oc{\Pi_{2}}$ we only have an implication to the respective statement about the stable models of the program(s), even with the level mapping variant of ordered completion.
In order to obtain equivalences to the stable models statements, we have to use second-order logic.
Thus, it would be of interest to investigate the usage of higher-order automated theorem provers~\cite{SutcliffeBenzmuller2010,BhayatSuda2024} with \anthem.

The original work on ordered completion~\cite{AsuncionEtAl2012} outlines some ideas to optimise the theory produced by ordered completion so that fewer new predicates and fewer helper axioms are needed.
These ideas should also apply to the version of ordered completion from \cref{sec:oc}.
However, for the ordered completion with level mapping (\cref{sec:oc-lvl}) these optimisations can not be applied.

In \cref{sec:related} we have outlined related approaches to ordered completion.
Generalising these approaches to the \mg language could be another topic for future work.
However, most approaches only consider propositional languages and can thus not be directly generalised.
For them, the first question would be whether a generalisation to programs with variables is possible.
Only the tightening transformation~\cite{Wallace1993} is already defined on programs with variables.
We expect that a generalisation of tightening to the \mg language should be possible in the same way that this thesis generalised ordered completion.
But all the approaches (whether only propositional or not) have the same limitation, as shown for ordered completion in \cref{sec:limitations}, that programs with the same stable models will not necessarily have the same models after transforming them (whether that is a transformation to theories or programs).

Finally, as stated in the introduction of this thesis, the main motivation for investigating extensions of completion is the usage of completion in external equivalence~\cite{FandinnoEtAl2023}.
So far, the prototypical implementation of ordered completion in \anthem does not integrate ordered completion into external equivalence.
Only simple equivalence between programs or between a program and a theory can be verified.
The concepts of inputs and outputs from external equivalence are not taken into account.
Thus, we would like to investigate the integration of ordered completion into external equivalence in future work.
However, the limitations outlined in \cref{sec:limitations} also limit the results of external equivalence with ordered completion.

\appendix
\chapter{Reduct Theories}
\label{chp:red-theories}

Let us assume we have a \mg program $\Pi$ and some set $I$ of precomputed atoms.
We want to analyse the shape of the formulas of $\tau(\Pi)$ after building the reduct with respect to $I$ (\cref{def:reduct}), i.e., the shape of the formulas in $\glr{\tau(\Pi)}{I}$.
We will call theories that have this shape \emph{reduct theories}.
The motivation of this analysis is to then define the immediate consequence operator on reduct theories.

In the following, we only consider ground rules, as $\tau(\Pi)$ is the set $\tau(R)$ for all ground instances $R$ of the rules of $\Pi$.
We start by considering the case that $R$ is a ground constraint.
I.e., $R$ has the form $\revimp Body$.
We then have
\[
  \tau(R) = \tau(\revimp Body) = \lnot \tau(Body).
\]
To construct the reduct, we have to consider two cases depending on whether $\tau(Body)$ is satisfied by $I$ or not.
\begin{enumerate}
  \item $I \models \tau(Body)$: then $\glrp{\lnot \tau(Body)}{I} = \false$.
  \item $I \not\models \tau(Body)$: then $\glrp{\lnot \tau(Body)}{I} = \glr{Body}{I} \imp \glr{\false}{I} = \false \imp \false = \true$.
\end{enumerate}
Thus, we do not need to consider the case that $R$ is a constraint for determining the overall shape that our formulas can have.
We summarise this in the following lemma.

\begin{lemma}[Reduct of a Constraint]\label{lem:red-constraint}
  Let $R$ be a ground constraint and let $I$ be a set of precomputed atoms.
  The formula $\glr{\tau(R)}{I}$ is either
  \begin{enumerate}
    \item $\false$ or,
    \item $\true$.
  \end{enumerate}
\end{lemma}

In the first case, $I$ can not be a (stable) model of any program containing $R$.
In the second case, we can remove the formula without affecting the models.

For the remaining two cases (i.e., $R$ is a basic rule or a choice rule), we first start by considering the heads of the rules.
We start with the head of a basic rule, i.e., we have $Head = p(\terms)$.
Then $\tau(Head) = \bigwedge_{\rs \in \values{\terms}} p(\rs)$.
We build the reduct: $\glr{\tau(Head)}{I} = \glrp{\bigwedge_{\rs \in \values{\terms}} p(\rs)}{I}$ by considering two cases.
\begin{enumerate}
  \item $I \models \bigwedge_{\rs \in \values{\terms}} p(\rs)$, then $\glrp{\bigwedge_{\rs \in \values{\terms}} p(\rs)}{I} = \bigwedge_{\rs \in \values{\terms}} \glr{p(\rs)}{I} = \bigwedge_{\rs \in \values{\terms}} p(\rs)$.
  \item $I \not\models \bigwedge_{\rs \in \values{\terms}} p(\rs)$, then $\glrp{\bigwedge_{\rs \in \values{\terms}} p(\rs)}{I} = \false$.
\end{enumerate}
We obtain the following lemma.

\begin{lemma}[Reduct of a Basic Head]\label{lem:red-basic-head}
  Let $Head = p(\terms)$ be the head of a basic rule and let $I$ be a set of precomputed atoms.
  The formula $\glr{\tau(Head)}{I}$ is either
  \begin{enumerate}
    \item $\bigwedge_{\rs \in \values{\terms}} p(\rs)$ or,
    \item $\false$.
  \end{enumerate}
\end{lemma}

Next, we consider the case of a choice rules, i.e., we have $Head = \{ p(\terms) \}$.
Then $\tau(Head) = \bigwedge_{\rs \in \values{\terms}} (p(\rs) \lor \lnot p(\rs))$.
The reduct is then just the reduct of each inner disjunction $\glrp{\bigwedge_{\rs \in \values{\terms}} (p(\rs) \lor \lnot p(\rs))}{I} = \bigwedge_{\rs \in \values{\terms}} \glrp{p(\rs) \lor \lnot p(\rs)}{I}$.
We consider two cases for the inner reduct.
\begin{enumerate}
  \item $I \models p(\rs)$, then $\glr{p(\rs)}{I} = p(\rs)$ and $\glrp{\lnot p(\rs)}{I} = \false$, and thus the reduct of the disjunction is $\glrp{p(\rs \lor \lnot p(\rs))}{I} = p(\rs)$.
  \item $I \not\models p(\rs)$, then $\glr{p(\rs)}{I} = \false$ and $\glrp{\lnot p(\rs)}{I} = \glrp{p(\rs)}{I} \imp \glr{\false}{I} = \false \imp \false = \true$, and thus the reduct of the disjunction is $\glrp{p(\rs \lor \lnot p(\rs))}{I} = \true$.
\end{enumerate}
Thus, overall, we have a conjunction of atoms as stated in the following lemma.

\begin{lemma}[Reduct of a Choice Head]\label{lem:red-choice-head}
  Let $Head = \{ p(\terms) \}$ be the head of a choice rule and let $I$ be a set of precomputed atoms.
  The formula $\glr{\tau(Head)}{I}$ is
  \[
    \bigwedge_{\twosub{\rs \in \values{\terms}}{p(\rs) \in I}} p(\rs).
  \]
\end{lemma}

Note that structurally this is the same as the first case of the reduct of the head of a basic rule.
In both cases, the conjunction can also be empty.
In this case, the reduct of the head is $\true$ and thus the whole rule can be removed.

Next, we will consider the reduct of a rule body.
A rule body is a conjunction of literals and comparisons.
If we have a comparison $t_{1} \prec t_{2}$ then $\tau(t_{1} \prec t_{2})$ is either $\true$ or $\false$.
In the first case, we can remove $\tau(t_{1} \prec t_{2})$.
In the second case, we can remove the whole rule.
Thus, we do not need to consider comparisons below.

Let us now consider the results of constructing the reduct of a literal.
Let $p(\rs)$ be our atom.
We consider whether $I$ satisfies $p(\rs)$ and analyse each possible literal.
\begin{enumerate}
  \item $I \models p(\rs)$, then
        \begin{enumerate}
          \item $\glr{p(\rs)}{I} = p(\rs)$,
          \item $\glrp{\lnot p(\rs)}{I} = \false$,
          \item $\glrp{\lnot \lnot p(\rs)}{I} = \glrp{\lnot p(\rs)}{I} \imp \false = \false \imp \false = \true$.
        \end{enumerate}
  \item $I \not\models p(\rs)$, then
        \begin{enumerate}
          \item $\glr{p(\rs)}{I} = \false$,
          \item $\glrp{\lnot p(\rs)}{I} = \glr{p(\rs)}{I} \imp \false = \false \imp \false = \true$,
          \item $\glrp{\lnot \lnot p(\rs)}{I} = \false$.
        \end{enumerate}
\end{enumerate}
Thus, we get the following lemma for the reducts of literals.

\begin{lemma}[Reduct of a Literal]\label{lem:red-literal}
  Let $Lit$ be a literal and let $I$ be a set of precomputed atoms.
  The formula $\glr{\tau(Lit)}{I}$ is
  \begin{enumerate}
    \item if $Lit = p(\terms)$, then $\glr{\tau(Lit)}{I} = \glrp{\bigvee_{\rs \in \values{\terms}} p(\rs)}{I} = \bigvee_{\rs \in \values{\terms}, p(\rs) \in I} p(\rs)$.
    \item if $Lit = \lnot p(\terms)$, then $\glr{\tau(Lit)}{I} = \glrp{\bigvee_{\rs \in \values{\terms}} \lnot p(\rs)}{I} = \true$ if some $p(\rs) \not\in I$ otherwise $\false$.
    \item if $Lit = \lnot \lnot p(\terms)$, then $\glr{\tau(Lit)}{I} = \glrp{\bigvee_{\rs \in \values{\terms}} \lnot \lnot p(\rs)}{I} = \true$ if some $p(\rs) \in I$ otherwise $\false$.
  \end{enumerate}
\end{lemma}

For the reduct of a rule body, we then obtain the following lemma.

\begin{lemma}[Reduct of a Body]\label{lem:red-body}
  Let $Body$ be a rule body and let $I$ be a set of precomputed atoms.
  The formula $\glr{\tau(Body)}{I}$ is either
  \begin{enumerate}
    \item $(p_{1}(\rs_{1,1}) \lor \dots \lor p_{1}(\rs_{1,m_{1}})) \land \dots \land (p_{n}(\rs_{n,1}) \lor \dots \lor p_{n}(\rs_{n,m_{n}}))$ or,
    \item $\false$.
  \end{enumerate}
  Note that in the first case, the conjunction can also be empty, i.e., $\glr{\tau(Body)}{I}$ can also be the formula $\true$.
\end{lemma}

Using the above lemmata, we get the following result on the reduct of a \mg rule.

\begin{lemma}[Reduct of a Rule]
  Let $R$ be a \mg rule and let $I$ be a set of precomputed atoms.
  The formula $\glr{\tau(R)}{I}$ is a formula of the form $B \imp H$ where $B$ is either
  \begin{enumerate}
    \item $(p_{1}(\rs_{1,1}) \lor \dots \lor p_{1}(\rs_{1,m_{1}})) \land \dots \land (p_{n}(\rs_{n,1}) \lor \dots \lor p_{n}(\rs_{n,m_{n}}))$ or,
    \item $\false$,
  \end{enumerate}
  and $H$ is either
  \begin{enumerate}
    \item $q(\rs_{1}) \land \dots \land q(\rs_{m})$ or,
    \item $\false$.
  \end{enumerate}
\end{lemma}

Note that for the conjunction cases of $B$ and $H$, the conjunctions can also be empty, in which case $B$/$H$ is $\true$.
In the case that both $B$ and $H$ are $\false$, the formula $F$ is $\true$.
If only $B$ is $\false$, then $F$ is also $\true$.
If only $H$ is $\false$, then the formula $F$ can not be satisfied by $I$ as $I \models B$ by the definition of the reduct.

In order to summarise our analysis, we first introduce so-called \emph{reduct theories}.

\begin{recalldefinition}{def:red-theory}[Reduct Theory]
  We call a theory $\Gamma$ a \emph{reduct theory} if it only contains formulas of the following form:
  \[
    B_{1} \land \dots \land B_{n} \imp (q(\rs_{1}) \land \dots \land q(\rs_{m})),
  \]
  where each $B_{i} = (p_{i}(\rs_{i,1}) \lor \dots \lor p_{i}(\rs_{i,m_{i}}))$.
  If $F \in \Gamma$ is such a formula we define
  \begin{alignat*}{1}
    \head{F} &= \{ q(\rs_{1}), \dots, q(\rs_{m}) \}, \\
    \body{F} &= B_{1} \land \dots \land B_{n}.
  \end{alignat*}
  Note that both $\body{F}$ and the formula $q(\rs_{1}) \land \dots \land q(\rs_{m})$ can also be the empty conjunction, i.e., $\true$.
\end{recalldefinition}

We then obtain the following lemma based on our analysis from above.

\begin{recalllemma}{lem:red-theory}
  Let $\Pi$ be a \mg program and let $I$ be a set of precomputed atoms.
  The theory $\glr{\tau(\Pi)}{I}$ is either
  \begin{enumerate}
    \item a theory that is not satisfied by $I$ or,
    \item a reduct theory.
  \end{enumerate}
\end{recalllemma}

This theorem is sufficient for our intended purpose.
We wanted to find a form of theories that we can define an immediate consequence operator for.
This is the case for the theories in the second case of our theorem, i.e., for the reduct theories.
In the first case, there is no need for an immediate consequence operator as the theory is not satisfied by $I$ and thus $I$ can not be its minimal model.
Anyway, we only apply the immediate consequence operator to actual (stable) models of our programs to extract a derivation order.
Thus, the unsatisfiability case does not actually occur in our usage of the immediate consequence operator.

\backmatter
\bibliography{references}
\chapter{Declaration of Authorship}
I hereby declare that I have independently written the present thesis and have not used any sources other than those I have specified.
\bigskip

\noindent Hiermit erkläre ich, dass ich die vorliegende Arbeit selbstständig verfasst und keine anderen als die von mir angegebenen Quellen genutzt habe.
\vspace{2 cm}

\noindent\textsc{\textbf{\theauthor}}

\noindent \textsc{\theplace, \thedate}

\end{document}